\documentclass[a4paper,fleqn]{cas-sc}

\usepackage[numbers,sort&compress]{natbib}
\usepackage{subcaption}
\usepackage{algorithm}
\usepackage{algpseudocode}
\usepackage{float}

\usepackage{setspace}
\usepackage{makecell}

\usepackage{threeparttable}
\usepackage{lineno}

\usepackage{amsmath,amssymb}

\usepackage{xcolor}

\usepackage{adjustbox}
\usepackage{graphicx}
\algrenewcommand\algorithmicindent{0.5em}
\def\tsc#1{\csdef{#1}{\textsc{\lowercase{#1}}\xspace}}
\tsc{WGM}
\tsc{QE}
\tsc{EP}
\tsc{PMS}
\tsc{BEC}
\tsc{DE}

\begin{document}
\let\WriteBookmarks\relax
\def\floatpagepagefraction{1}
\def\textpagefraction{.001}


\shorttitle{Two-stage monitoring design and budgeted condition-based maintenance}


\title [mode = title]{Two-stage monitoring design and budgeted condition-based maintenance for LED lighting systems: a gamma-process-based ensemble Kalman filter approach}


\author[1]{Haohao Shi}
\cormark[1]
\ead{haohao.shi@hdr.qut.edu.au}
\credit{Writing -- original draft, Visualization, Software, Methodology, Conceptualization}
\affiliation[1]{organization={Faculty of Engineering, Queensland University of Technology (QUT)},
    city={Brisbane},
    postcode={4000},
    state={Queensland},
    country={Australia}}

\author[1]{Huy Truong-Ba}
\credit{Writing -- review \& editing, Supervision, Methodology, Conceptualization}

\author[1]{Michael E. Cholette}
\credit{Writing -- review \& editing, Supervision, Funding acquisition, Methodology, Conceptualization}

\author[2]{Brenden Harris}
\credit{Writing -- review \& editing, Funding acquisition, Formal analysis}
\author[2]{Juan Montes}
\credit{Writing -- review \& editing, Formal analysis}
\author[1]{Tommy H.T. Chan}
\credit{Writing -- review \& editing, Supervision, Funding acquisition}

\affiliation[2]{organization={Fredon Queensland},
    city={Brisbane},
    postcode={4119},
    state={Queensland},
    country={Australia}}

\cortext[cor1]{Corresponding author: Haohao Shi}

\begin{abstract}
Light-emitting diode (LED) luminaires degrade gradually during operation, reducing working plane (WP) illuminance and increasing the risk of inadequate indoor lighting. Because individual luminaire degradation states are difficult to measure directly after installation, this paper estimates latent luminaire degradation states from sparse in situ WP illuminance measurements. A two-stage monitoring design and budgeted condition-based maintenance (CBM) optimization framework is proposed. At the design stage, Radiance simulations are used to construct a linear-Gaussian observation model that maps latent luminaire degradation states to WP illuminance measurements. An identifiability-constrained D-optimal design then selects an informative reference layout of WP measurement points for inverse state estimation at a single epoch. At the runtime stage, the measurement design determines the visit schedule and the active reference measurement points for each visit, while an ensemble Kalman filter (EnKF) combines these sparse measurements with nonhomogeneous gamma process degradation predictions to update latent luminaire degradation state estimates. Posterior predictive failure probabilities are used as the risk metric for preventive replacement. Under a monitoring budget, the resulting runtime monitoring and CBM policy jointly selects the visit schedule, active reference measurement points, and risk threshold to minimize expected total downtime and replacement cost. A case study on an office lighting zone demonstrates design-stage measurement layout selection, runtime-stage measurement design, and budgeted CBM policy optimization.
\end{abstract}


\begin{highlights}
    \item A linear-Gaussian observation model enables degradation state inference from working plane illuminance.
    \item Identifiability-constrained D-optimal design constructs an informative reference layout.
    \item Sequential filtering enables sparse runtime monitoring below the single epoch identifiability limit.
    \item Visit schedules, measurement selection, and maintenance decisions are jointly optimized under a monitoring budget.
    \item A case study demonstrates the framework under varying degradation uncertainty and monitoring budgets.
\end{highlights}

\maketitle

\section{Introduction}

Light-emitting diode (LED) lighting systems are widely used in buildings because they offer high efficacy, long service life, and flexible control \citep{tan_chapter_2023}. In indoor lighting assessment, the working plane (WP) is a virtual task surface, typically located at desk height, on which illuminance requirements for visual tasks are evaluated. The WP illuminance field is collectively produced by all luminaires in a lighting zone and is therefore the main performance quantity for assessing indoor lighting quality. During operation, gradual LED luminaire degradation reduces light output and deteriorates the WP illuminance field, thereby increasing the risk of inadequate lighting. Reliable estimates of luminaire degradation are therefore important for planning maintenance that preserves adequate WP illuminance. In large-scale lighting systems, monitoring and maintenance costs further require that such state information be obtained efficiently.

Stochastic degradation models can predict LED degradation trajectories under uncertainty, but their predictions for individual installed luminaires may be inaccurate because actual operating conditions and unit-to-unit degradation variability are difficult to fully characterize. Direct measurement of each luminaire degradation state after installation would provide more reliable information for maintenance decision-making. However, such direct measurements are generally impractical in real buildings because they require luminaire removal and laboratory photometric testing, such as integrating sphere or goniophotometer measurements. In contrast, in situ WP illuminance measurements provide practical indirect observations for inferring luminaire degradation states. Existing lighting analysis and maintenance studies commonly rely on physics- or photometry-based simulation tools, including Radiance \citep{shakespeare_rendering_2021}, DIALux evo \citep{dialux_evo_2026}, ReluxDesktop \citep{reluxdesktop_2026}, and AGi32 \citep{agi32_2026}, to perform forward illuminance calculations from specified luminaire photometry and scene descriptions \citep{shi_gamma_2026}. Although their underlying calculation engines and implementation details differ, these tools are primarily designed for forward illuminance evaluation; the resulting simulator is typically an implicit input--output map rather than an explicit state-to-measurement observation model for inverse state estimation from sparse WP measurements. For condition monitoring that supports condition-based maintenance, this limitation motivates an observation model that quantifies both state-to-measurement sensitivities and measurement and model uncertainties.

Given such an explicit observation model, latent luminaire degradation states can in principle be inferred from WP illuminance measurements. This task, however, is an inverse state estimation problem rather than a conventional local sensing or field reconstruction problem commonly considered in sensor and monitoring placement studies \citep{NIRESI2025112153,tian_efficient_2025}. A WP measurement is not a local observation of a nearby luminaire; instead, it reflects the superposed contributions of many, and in principle all, luminaires, whereas the quantity of interest is the high-dimensional vector of luminaire degradation states. Consequently, an informative measurement layout must provide sufficiently independent and informative observations to distinguish degradation patterns across luminaires. This motivates formulating WP measurement point selection at the design stage as an optimal experimental design problem for inverse state estimation \cite{yang_interval_2024, tian_efficient_2025}. In this formulation, a subset of feasible WP measurement points is selected to provide informative observations for estimating luminaire degradation states and reducing their generalized uncertainty \citep{pukelsheim_optimal_2006}. However, design optimality cannot overcome basic identifiability limits. In this inverse setting, reliable recovery of component states is fundamentally constrained by linear algebraic identifiability conditions: the number of independent measurements must be at least as large as the number of latent variables; otherwise, the problem becomes rank deficient and highly sensitive to measurement noise \cite{hansen_rank-deficient_2005, aster_parameter_2018}. These structural constraints impose a lower bound on the number of required WP measurements for any standalone inverse estimate.

For large LED lighting systems, repeating a standalone inverse estimate at every monitoring visit can be economically burdensome and time-consuming, because each visit would require enough WP measurements to satisfy the single-epoch identifiability requirement. Sequential filtering provides a way to reduce runtime measurement effort by carrying state information forward across visits \cite{niu_remaining_2025,niu_degradation_2026}. Rather than estimating luminaire degradation states independently at each visit epoch, a filter propagates prior state estimates through the degradation model and assimilates newly collected sparse WP measurements \cite{truong-ba_sensor_2025}. The propagated prior information can partly compensate for limited instantaneous observability. It can therefore help preserve state estimation accuracy even when each visit collects fewer measurements than would be required for a standalone identifiable estimate. This reduction in per-visit measurement effort introduces a runtime-stage measurement design problem: which WP measurement points should be measured when only sparse observations are collected. Sequential filtering alone does not address this measurement selection problem. Recent filtering-based monitoring and prognostics studies assimilate data through prespecified observation channels or sensor layouts \citep{niu_remaining_2025,niu_degradation_2026,truong-ba_sensor_2025}, rather than treating the measurement locations used at each monitoring epoch as design variables. Directly optimizing visit-specific measurement subsets over all WP grid points across the full monitoring horizon would, however, create a large combinatorial search space and complicate field implementation.

Taken together, the design-stage identifiability requirement for standalone inverse estimation and the runtime-stage measurement design problem introduced by sparse sequential measurements motivate a two-stage monitoring design strategy. The first stage addresses design-stage measurement layout optimization: an informative reference layout of WP measurement points is optimized over the full WP grid subject to the minimum measurement count required for identifiability. This reference layout provides an informative reference set for standalone inverse estimation and defines a compact candidate set for runtime monitoring. In the transition to runtime operation, however, the second stage no longer enforces single-epoch identifiability at every monitoring visit. Instead, it addresses runtime-stage measurement design, which determines both when monitoring visits are conducted and which subset of the reference layout is measured at each visit to support reliable sequential state estimation. The points in this measured subset are referred to as the active reference measurement points for that visit. The sequential filter then combines the resulting sparse observations with prior information propagated from previous visits. This separation between the design and runtime stages avoids optimizing over all WP grid points across the full horizon and reduces operational measurement effort while maintaining reliable degradation state estimates.

The estimated luminaire degradation states are ultimately used not only for monitoring but also for maintenance decision-making. The posterior degradation state distribution updated by the sequential filter can be used to predict the probability that each luminaire will exceed a failure threshold before the next monitoring visit. This posterior predictive failure probability is used as the risk metric in a condition-based maintenance (CBM) rule: preventive replacement is triggered when the probability exceeds a prescribed risk threshold. The runtime-stage measurement design is therefore coupled with CBM decision-making. The measurement design component determines how much information to collect through the monitoring visit schedule and active reference measurement points, which affects posterior degradation state estimation and the uncertainty in the resulting posterior predictive failure probabilities. The CBM component then uses these posterior predictive failure probabilities and a prescribed risk threshold to determine preventive replacement actions. More frequent monitoring visits and more active reference measurement points can improve posterior degradation state estimation and reduce uncertainty in posterior predictive failure probabilities, but they also increase monitoring cost. A lower risk threshold triggers earlier preventive replacement and can reduce failed-luminaire downtime, whereas a higher threshold delays replacement and can reduce unnecessary replacement cost. These tradeoffs motivate a budgeted runtime monitoring and CBM policy that jointly selects the monitoring visit schedule, active reference measurement points, and risk threshold to minimize expected total downtime and replacement cost subject to a monitoring budget.

Accordingly, this study proposes a two-stage monitoring design and budgeted condition-based maintenance optimization framework for LED lighting systems. The framework integrates design-stage WP measurement layout optimization, runtime-stage measurement design, ensemble Kalman filter (EnKF)-based latent degradation state estimation, and risk-based preventive replacement. The main contributions are organized around this two-stage monitoring design framework as follows:
\begin{itemize}
    \item \textbf{A state-to-measurement observation model for inverse luminaire degradation state estimation.} A Radiance-based linear-Gaussian surrogate is developed to convert forward illuminance simulations into an explicit observation model linking latent luminaire degradation states to WP illuminance measurements, with measurement and model uncertainties quantified.

    \item \textbf{Design-stage measurement layout optimization with identifiability.} WP measurement point selection is formulated as an optimal experimental design problem for inverse estimation of latent luminaire degradation states. The layout is constrained by the minimum measurement count required for single-epoch identifiability and optimized using a D-optimal criterion to form an informative reference layout.

    \item \textbf{Runtime-stage measurement design enabled by sequential EnKF degradation state estimation.} The design-stage reference layout is used as a compact candidate set for runtime-stage measurement design. Instead of enforcing single-epoch identifiability at every monitoring visit, the runtime stage selects the visit schedule and active reference measurement points, while an EnKF assimilates sparse WP measurements with gamma process degradation predictions to recursively update latent luminaire degradation state estimates.

    \item \textbf{Budgeted runtime monitoring and CBM policy optimization.} Posterior predictive failure probabilities are used as the risk metric for preventive replacement decisions. Under a monitoring budget, the runtime-stage measurement design is coupled with CBM decision-making to jointly optimize the visit schedule, active reference measurement points, and risk threshold so as to minimize expected total downtime and replacement cost.
\end{itemize}

\begin{figure}[pos=!htbp]
    \centering
    \includegraphics[width=0.55\textwidth]{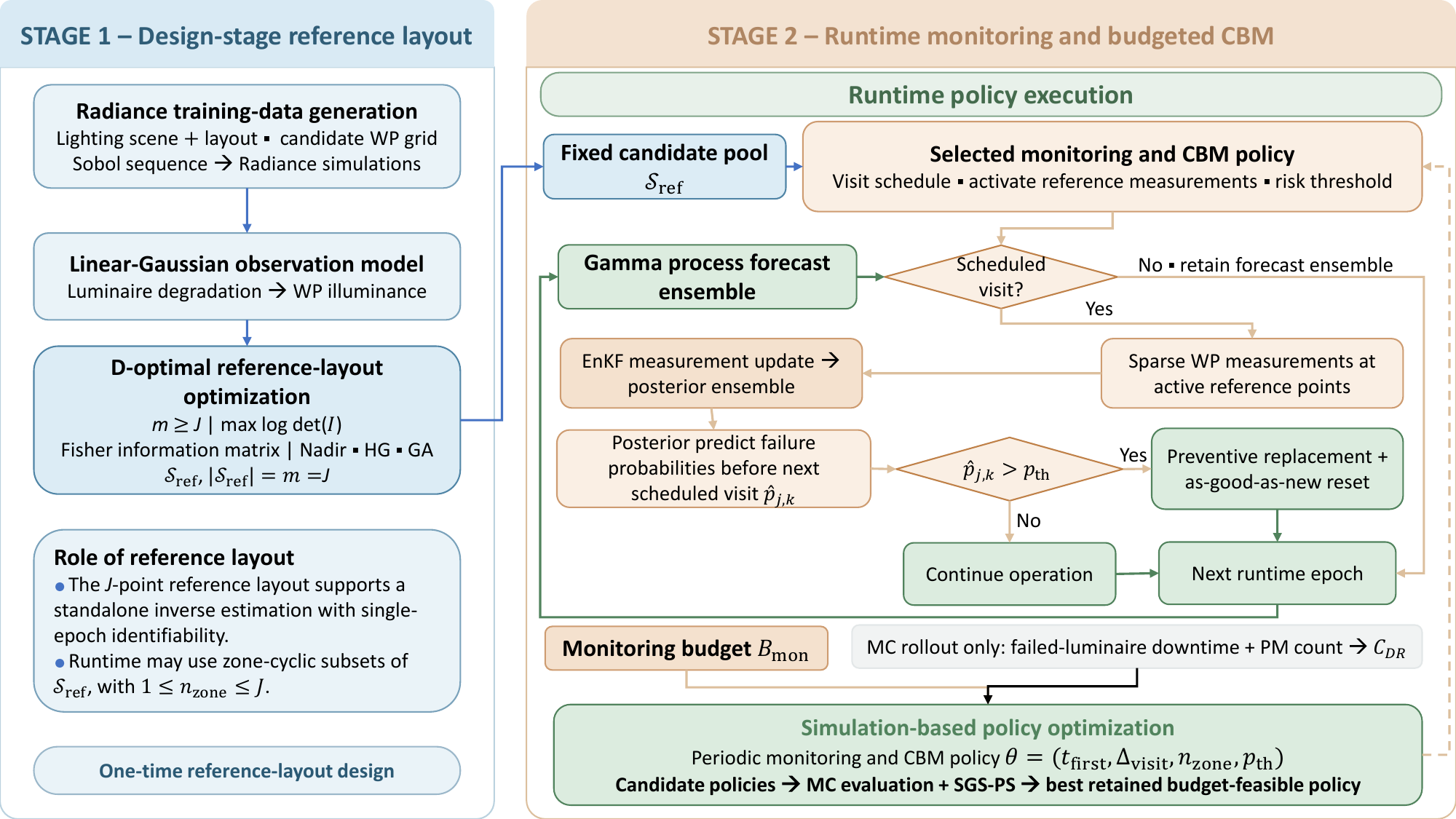}
    \caption{Overview of the proposed two-stage monitoring design and budgeted condition-based maintenance framework.}
    \label{fig:F1_two_stage_framework}
\end{figure}

An overview of the proposed two-stage monitoring design and budgeted CBM framework is shown in Fig.~\ref{fig:F1_two_stage_framework}.

The remainder of this paper is organized as follows. Section~\ref{sec:design_sensor_opt} develops the degradation state model, the state-to-illuminance mapping, the sparse WP measurement model, and the design-stage D-optimal reference measurement layout selection method. Section~\ref{sec:online_tracking} formulates the runtime-stage measurement design and CBM optimization problem, including the joint visit--measurement schedule, EnKF-based degradation tracking, risk-based preventive replacement rule, monitoring budget constraint, and sequential grid search with paired screening. Section~\ref{sec:case_study_settings} presents the office-zone case study, including the baseline settings, design-stage layout selection results, runtime-stage CBM performance, and sensitivity analyses for hitting time uncertainty and monitoring budget. Section~\ref{sec:conclusion} concludes the paper and outlines future research directions. Table~\ref{tab:symbols_abbreviations} summarizes the main symbols and abbreviations used throughout the paper.

\begin{table}[width=1\linewidth,cols=2,pos=h]
    \caption{Main symbols and abbreviations used in this paper.}
    \label{tab:symbols_abbreviations}
    \footnotesize
    \begin{tabular}{@{}p{0.24\linewidth}p{0.70\linewidth}@{}}
        \toprule
        \textbf{Symbol/Abbreviation} & \textbf{Description} \\
        \midrule
        \multicolumn{2}{@{}l}{\textit{Abbreviations}} \\
        LED & Light-emitting diode \\
        WP & Working plane \\
        CBM & Condition-based maintenance \\
        EnKF & Ensemble Kalman filter \\
        SGS-PS & Sequential grid search with paired screening \\
        HG & Hybrid greedy search \\
        GA & Genetic algorithm \\
        MC & Monte Carlo simulation \\
        PM & Preventive maintenance (implemented as preventive replacement in this study) \\
        \addlinespace
        \multicolumn{2}{@{}l}{\textit{Symbols}} \\
        $J$ & Number of luminaires; dimension of the latent degradation state vector \\
        $N$ & Number of candidate WP grid points \\
        $m$ & Number of selected WP measurement points in the design-stage layout \\
        $m_k$ & Number of active reference measurement points at runtime epoch $t_k$ \\
        $X_j(t)$ & Degradation state of luminaire $j$ at time $t$ \\
        $P_{j,\mathrm{out}}(t)$ & Effective normalized luminous output of luminaire $j$ \\
        $\mathbf{x}(t)$, $\mathbf{x}_k$ & Luminaire degradation state vector at time $t$ and runtime epoch $t_k$ \\
        $x_{\mathrm{th}}$ & Degradation failure threshold; $x_{\mathrm{th}}=0.3$ under the $L_{70}$ criterion \\
        $\mathbf{E}(t)$ & WP illuminance field vector over all candidate grid points \\
        $\mathbf{C}$, $\mathbf{b}$ & Surrogate sensitivity matrix and intercept in the linear-Gaussian state-to-illuminance model \\
        $\boldsymbol{\Sigma}$ & Surrogate residual covariance \\
        $\mathbf{y}(t)$, $\mathbf{y}_k$ & Sparse WP measurement vector at time $t$ and runtime epoch $t_k$ \\
        $\mathbf{S}(t)$, $\mathbf{S}_{\mathrm{ref}}$ & Active measurement selection matrix and reference-layout selection matrix \\
        $\mathcal{S}_{\mathrm{ref}}$ & Design-stage reference WP measurement layout \\
        $\mathbf{H}(t)$, $\mathbf{H}_k$ & Effective observation matrix for selected WP measurements \\
        $\mathbf{R}(t)$, $\boldsymbol{\Omega}_k$ & Photometer measurement-noise covariance and runtime aggregated observation covariance \\
        $t_k$, $K$, $\delta$ & Runtime grid epoch, final grid index, and grid step size \\
        $v_k$, $u_{k,r}$ & Visit indicator and indicator that reference point $r$ is measured at $t_k$ \\
        $\mathbf{Z}$ & Joint visit--measurement schedule matrix \\
        $p_{\mathrm{th}}$ & Risk threshold for preventive replacement decisions \\
        $\hat p_{j,k}$ & Posterior predictive failure probability for luminaire $j$ at visit epoch $t_k$ \\
        $d_{j,k}$ & Replacement decision for luminaire $j$ at visit epoch $t_k$ \\
        $\mathrm{DT}_{j,k}$ & Failed-luminaire downtime over $(t_{k-1},t_k]$ \\
        $\mathrm{PM}_{j,k}$ & Preventive replacement indicator at visit epoch $t_k$ \\
        $C_{\mathrm{DR}}$ & Total downtime and replacement cost \\
        $C_{\mathrm{insp}}$ & Total visit--measurement cost \\
        $B_{\mathrm{mon}}$ & Total monitoring budget \\
        $\boldsymbol{\theta}$ & Periodic policy parameter vector $(t_{\mathrm{first}},\Delta_{\mathrm{visit}},n_{\mathrm{zone}},p_{\mathrm{th}})$ \\
        $n_{\mathrm{zone}}$ & Number of runtime reference-layout zones and active reference measurements per visit \\
        \bottomrule
    \end{tabular}
\end{table}

\section{State-to-illuminance modeling and design-stage measurement layout selection}
\label{sec:design_sensor_opt}
This section develops the state-to-illuminance modeling formulation and the design-stage measurement layout selection method. Because direct measurement of individual luminaire degradation states after installation is impractical in real buildings, the proposed monitoring framework infers latent luminaire degradation states from in situ WP illuminance measurements. The section first defines the luminaire degradation state and models its stochastic evolution using a nonhomogeneous gamma process. Given the resulting degradation state vector, it then defines the WP illuminance field and the physics-based Radiance mapping from luminaire degradation states to this field. A linear-Gaussian surrogate model is introduced to provide an explicit quantitative relationship between luminaire degradation states and WP illuminance values.

The surrogate model is then combined with WP measurement selection matrices to form a sparse inverse observation model for estimating latent luminaire degradation states. Specifically, a Fisher information measure is derived to evaluate the informativeness of selected WP measurement points. A D-optimal criterion is then adopted to formulate the design-stage measurement layout optimization problem: determining how many WP measurement points are required for single-epoch inverse identifiability and which WP grid points should be selected to form an informative reference layout.

\subsection{Luminaire degradation and state-to-illuminance modeling}
\label{subsec:state_illum_model}

In this work, luminaire failure is defined from a photometric-performance perspective. Specifically, photometric failure occurs when the effective luminous output of a luminaire decreases below 70\% of its initial value, corresponding to the commonly used $L_{70}$ endpoint in luminous-flux-maintenance assessment \citep{ies_lm84_2020}. The degradation state therefore represents the effective luminaire-level loss of luminous output. Define the degradation state of luminaire $j$ as
\begin{equation}
    X_{j}(t) = 1 - P_{j,\text{out}}(t), \quad j=1,2,\ldots,J,
    \label{eq:LO}
\end{equation}
where $0 \leq P_{j,\text{out}}(t) \leq 1$ denotes the effective normalized luminous output of luminaire $j$ (with $P_{j,\text{out}}(0)=1$). With this definition, the photometric-failure threshold is $P_{j,\text{out}}(t) \leq 0.7$, equivalently $X_j(t) \geq 0.3$.

Let the latent luminaire degradation state vector at time $t$ be
\begin{equation}
    \mathbf{x}(t) = \left[ X_1(t),\, X_2(t),\, \dots,\, X_J(t)\right]^{\top},
    \label{eq:state_vec}
\end{equation}
where $J$ is the total number of luminaires.

For each luminaire $j$, the degradation state process $\{X_j(t)\}_{t\ge 0}$ is modeled as a nonhomogeneous gamma process with independent, non-negative increments. This choice is consistent with the monotonic accumulation of lumen depreciation and with common stochastic degradation modeling practice for LED degradation processes \citep{sun_stochastic_2018, ling_bayesian_2019, fan_gamma_2021, fan_prognostics_2021}. More formally, for $t>s$,
\begin{equation}
    X_j(t)-X_j(s) \sim \mathrm{Ga}\bigl(\alpha(t)-\alpha(s), \beta\bigr),
\end{equation}
where $\beta>0$ is the rate parameter and the shape function is specified as $\alpha(t)=A\exp(\gamma_{\alpha}t)$, with $A>0$ and $\gamma_{\alpha}>0$ so that $\alpha(t)$ is increasing. The exponential shape function is consistent with the projection form used in TM-28 \citep{ies_tm28_2020}.

Because LEDs typically have long service lives, degradation parameters are inferred from accelerated degradation testing (ADT) data and extrapolated to service conditions. Following the linear acceleration assumption \citep{elsayed_reliability_2020}, $A$ and $\gamma_{\alpha}$ are treated as stress-invariant, whereas the rate parameter is temperature-dependent and modeled by the Arrhenius relationship $\beta(T)=\beta_0\exp\left(E_a/(k_B T)\right)$ \citep{tobias_applied_2012,jiang_inference_2019,rocchetta_uncertainty_2024}, where $\beta_0$ is a pre-exponential factor, $E_a$ is the activation energy (eV), $k_B$ is the Boltzmann constant ($8.62\times 10^{-5}$ eV/K), and $T$ is the absolute temperature (K). The posterior distribution of the transformed degradation model parameters $\boldsymbol{\psi}=(\ln A,\,\gamma_{\alpha},\,\ln \beta_0,\,E_a)$ is inferred using the Bayesian framework described in our previous work \citep{shi_gamma_2026}.

Given this latent degradation state representation, the next step is to map luminaire degradation states to the WP illuminance field. In indoor lighting analysis, illuminance is evaluated on the WP, a virtual task surface typically located 0.7--0.85~m above the floor to represent desk height for common visual tasks. The WP is discretized into grid points that quantify the illuminance field and serve as candidate locations for subsequent measurements.

Following our previous work \citep{shi_gamma_2026}, Radiance is used as the high-fidelity forward model to compute the WP illuminance field from luminaire degradation states \citep{shakespeare_rendering_2021}.

Given the degradation state vector $\mathbf{x}(t)$ defined in Eq.~\eqref{eq:state_vec}, the Radiance mapping produces illuminance values at the $N$ WP grid points:
\begin{equation}
    \mathbf{E}(t) = \left[ E_1(t),\, E_2(t),\, \dots,\, E_N(t) \right]^{\top}\in \mathbb{R}^{N},
    \label{eq:IMWP}
\end{equation}
where $E_i(t)$ is the illuminance at WP grid point $i$ (lux), and the full candidate set is denoted by $\mathcal{C}=\{1,\ldots,N\}$. Conceptually, this implicit forward mapping is written as $\mathbf{E}(t)=\mathcal{F}_{\mathrm{Rad}}\!\left(\mathbf{x}(t)\right)$. Because Radiance does not directly provide an explicit observation model for inverse state estimation, this mapping is approximated by the linear-Gaussian surrogate
\begin{align}
    \mathbf{E}(t) = \mathbf{C}\mathbf{x}(t) + \mathbf{b} + \boldsymbol{\varepsilon}(t),
    \qquad
    \boldsymbol{\varepsilon}(t) \sim \mathcal N(\mathbf 0,\boldsymbol{\Sigma}),
    \label{eq:sr1}
\end{align}
where $\mathbf{C}\in\mathbb{R}^{N\times J}$ is the surrogate sensitivity matrix, $\mathbf{b}\in\mathbb{R}^{N}$ is the intercept, and $\boldsymbol{\Sigma}\in\mathbb{S}^{N}_{++}$ captures surrogate residual uncertainty (e.g., model mismatch and numerical variability).

The sensitivity matrix and intercept are estimated by ordinary least squares (OLS), and the residual covariance is estimated from the OLS residuals. Details of the Radiance-generated training data, Sobol sampling, and parameter estimation are provided in Appendix~\ref{app:surrogate_est_details}. The resulting surrogate model provides the basis for constructing sparse WP measurement models for inverse luminaire degradation state estimation.

\subsection{Sparse working plane measurement model for luminaire degradation state estimation}
\label{subsec:sparse_obs_model}

Combining illuminance measurements collected at selected WP grid points with the surrogate model in Eq.~\eqref{eq:sr1} yields a sparse observation model for estimating latent luminaire degradation states. A key characteristic of this measurement setting is global coupling: the illuminance measured at a single WP grid point reflects the combined contribution of multiple, and in principle all, luminaires rather than a local one-to-one sensor--luminaire relation. Therefore, a WP grid point set should be evaluated not by the informativeness of individual points in isolation, but by how well the selected measurements collectively distinguish the latent luminaire degradation states in the inverse estimation problem. The following formulation establishes the sparse measurement model used to derive the subsequent information-based measurement point selection criterion.

For measurement-based luminaire degradation state estimation, a subset of WP grid points is measured using a handheld photometer at a generic observation time $t$. If $m$ WP grid points are selected, the measured illuminance vector $\mathbf{y}(t)\in\mathbb{R}^{m}$ is
\begin{equation*}
    \mathbf{y}(t) = \mathbf{S}(t) \mathbf{E}(t) + \boldsymbol{\eta}(t),
    \qquad
    \boldsymbol{\eta}(t) \sim \mathcal{N}(\mathbf{0},\,\mathbf{R}(t)),
\end{equation*}
where $\mathbf{S}(t)\in\{0,1\}^{m\times N}$ (one ``1'' per row) is a binary active measurement selection matrix that encodes the subset of WP grid points measured at time $t$, and $\mathbf{R}(t)\in\mathbb{S}^{m}_{++}$ is the measurement noise covariance. This covariance represents uncertainty introduced by the in situ photometer measurements, including instrument noise and repeatability error at the selected WP grid points. When independent photometer reading errors with a common standard deviation are assumed, $\mathbf{R}(t)=\sigma_{\mathrm{read}}^2\mathbf{I}_m$. The measurement noise covariance is distinct from the surrogate residual covariance $\boldsymbol{\Sigma}$, which represents the approximation error in mapping latent luminaire degradation states to the full WP illuminance field.

Substituting the surrogate model in Eq.~\eqref{eq:sr1} into the measurement relation above gives the sparse inverse observation model
\begin{equation}
\begin{aligned}
    \mathbf{y}(t)
    &= \mathbf{S}(t)\mathbf{E}(t)+\boldsymbol{\eta}(t) \\
    &= \mathbf{H}(t)\mathbf{x}(t)+\mathbf{d}(t)+\boldsymbol{\nu}(t), \\
    \boldsymbol{\nu}(t)
    &\sim \mathcal{N}\!\left(\mathbf{0},\boldsymbol{\Omega}(t)\right),
\end{aligned}
    \label{eq:obs_lg_t}
\end{equation}
where $\mathbf{H}(t)=\mathbf{S}(t)\mathbf{C}\in\mathbb{R}^{m\times J}$ is the effective observation matrix, $\mathbf{d}(t)=\mathbf{S}(t)\mathbf{b}\in\mathbb{R}^{m}$ is the effective intercept, $\boldsymbol{\nu}(t):=\mathbf{S}(t)\boldsymbol{\varepsilon}(t)+\boldsymbol{\eta}(t)$ is the aggregated uncertainty, and $\boldsymbol{\Omega}(t)=\mathbf{S}(t)\boldsymbol{\Sigma}\mathbf{S}(t)^{\top}+\mathbf{R}(t)\in\mathbb{S}^{m}_{++}$ is its covariance when the surrogate residual and photometer measurement noise are treated as independent. The latent state $\mathbf{x}(t)\in\mathbb{R}^{J}$ evolves according to the gamma process specified in \autoref{subsec:state_illum_model}.

\subsection{Fisher information and D-optimal measurement point selection for inverse luminaire degradation state estimation}
\label{subsec:placement_opt}

The globally coupled observation model also creates an opportunity for measurement optimization: informative estimates of the latent luminaire degradation states may be obtained without measuring all WP grid points. At the design stage, the measurement design task has two coupled parts. The first is to determine how many WP measurements are required for single-epoch inverse identifiability. The second is to determine which WP grid points should be selected so that the measurements provide the most information for estimating the latent luminaire degradation states. To define this objective, consider a single observation epoch and suppress the time argument for clarity. Let $m$ be the number of selected measurement points and $\mathbf{S}$ denote the corresponding selection matrix. Equation~\eqref{eq:obs_lg_t} then gives
\begin{equation*}
    \mathbf{y}\mid\mathbf{x}
    \sim
    \mathcal{N}\!\left(\mathbf{H}\mathbf{x}+\mathbf{d},\boldsymbol{\Omega}\right),
\end{equation*}
where $\mathbf{H}=\mathbf{S}\mathbf{C}\in\mathbb{R}^{m\times J}$, $\mathbf{d}=\mathbf{S}\mathbf{b}\in\mathbb{R}^{m}$, and $\boldsymbol{\Omega}=\mathbf{S}\boldsymbol{\Sigma}\mathbf{S}^{\top}+\mathbf{R}\in\mathbb{S}^{m}_{++}$. For a fixed selected WP grid point set, this distribution defines a linear-Gaussian inverse estimation problem for the latent state vector $\mathbf{x}$. The likelihood of $\mathbf{x}$ given the measured illuminance vector $\mathbf{y}$ is
\begin{equation*}
    \mathcal{L}(\mathbf{x};\mathbf{y})
    =
    \frac{1}{\sqrt{(2\pi)^{m}\det(\boldsymbol{\Omega})}}
    \exp\!\left(
    -\frac{1}{2}(\mathbf{y}-\mathbf{d}-\mathbf{H}\mathbf{x})^{\top}\boldsymbol{\Omega}^{-1}(\mathbf{y}-\mathbf{d}-\mathbf{H}\mathbf{x})
    \right).
\end{equation*}

The maximum likelihood estimator (MLE) is
\begin{equation*}
    \hat{\mathbf{x}}
    =
    \left(\mathbf{H}^{\top}\boldsymbol{\Omega}^{-1}\mathbf{H}\right)^{-1}
    \mathbf{H}^{\top}\boldsymbol{\Omega}^{-1}(\mathbf{y}-\mathbf{d}),
\end{equation*}
provided that $\mathbf{H}^{\top}\boldsymbol{\Omega}^{-1}\mathbf{H}$ is invertible. Substituting $\mathbf{y}=\mathbf{H}\mathbf{x}+\mathbf{d}+\boldsymbol{\nu}$ into the estimator above gives
\begin{equation*}
    \hat{\mathbf{x}}
    =
    \mathbf{x}
    +
    \left(\mathbf{H}^{\top}\boldsymbol{\Omega}^{-1}\mathbf{H}\right)^{-1}\mathbf{H}^{\top}\boldsymbol{\Omega}^{-1}\boldsymbol{\nu}.
\end{equation*}

Hence the estimator is unbiased and Gaussian:
\begin{equation}
    \hat{\mathbf{x}}
    \sim
    \mathcal{N}\!\left(\mathbf{x},\,\mathbf{P}\right),
    \qquad
    \mathbf{P}
    =
    \left(\mathbf{H}^{\top}\boldsymbol{\Omega}^{-1}\mathbf{H}\right)^{-1}.
    \label{eq:cov_inv_fi}
\end{equation}

The matrix
\begin{equation}
    \mathcal{I}(\mathbf{x})
    =
    \mathbf{H}^{\top}\boldsymbol{\Omega}^{-1}\mathbf{H}
    \label{eq:fisher_general}
\end{equation}
is the Fisher information matrix for $\mathbf{x}$ under this linear-Gaussian observation model. The Fisher information does not depend on the value of $\mathbf{x}$, and when it is nonsingular, $\mathbf{P}=\mathcal{I}(\mathbf{x})^{-1}$ is the inverse Fisher information.

The Fisher information matrix in Eq.~\eqref{eq:fisher_general} provides a matrix-valued measure of how informative a selected WP grid point set is for estimating latent luminaire degradation states. To compare different point sets, this matrix must be converted into a scalar design criterion. Because the objective is to reduce the overall uncertainty region of the inverse estimate, this uncertainty is expressed through the confidence ellipsoid induced by the estimator covariance matrix $\mathbf{P}$. For a confidence level $(1-\alpha_{\mathrm{CE}})$, the confidence ellipsoid of $\mathbf{x}$ induced by Eq.~\eqref{eq:cov_inv_fi} is
\begin{equation*}
    \mathcal{E}_{1-\alpha_{\mathrm{CE}}}
    =
    \Bigl\{
        \mathbf{x}\in\mathbb{R}^{J}\ \Big|\
        (\mathbf{x}-\hat{\mathbf{x}})^{\top}\mathbf{P}^{-1}(\mathbf{x}-\hat{\mathbf{x}})
        \le \chi^{2}_{J,\,1-\alpha_{\mathrm{CE}}}
    \Bigr\}.
\end{equation*}

This ellipsoid describes the set of state values that are statistically consistent with the inverse estimate at the prescribed confidence level. More generally, the ellipsoid $\{ \mathbf{x} : (\mathbf{x}-\hat{\mathbf{x}})^{\top}\mathbf{P}^{-1}(\mathbf{x}-\hat{\mathbf{x}})\le \omega\}$ has volume
\begin{equation}
    \operatorname{vol}(\mathcal{E})
    =
    \operatorname{vol}(B_J)\,
    \omega^{J/2}\,
    \sqrt{\det(\mathbf{P})},
    \qquad
    \operatorname{vol}(B_J)=\frac{\pi^{J/2}}{\Gamma\!\left(\tfrac{J}{2}+1\right)}.
    \label{eq:ellipsoid_vol}
\end{equation}

Thus, for a fixed confidence level, reducing the ellipsoid volume is equivalent to reducing $\det(\mathbf{P})$. Because $\mathbf{P}=\mathcal{I}^{-1}$ when the information matrix is nonsingular, reducing $\det(\mathbf{P})$ is equivalent to maximizing $\det(\mathcal{I})$. Therefore, the D-optimal criterion \citep{pukelsheim_optimal_2006} is adopted, and design-stage measurement layout selection is posed as selecting $m$ WP grid points (encoded by $\mathbf{S}$) that maximize
\begin{equation}
    \max_{\mathbf{S}}\ \log\det\!\left(\mathbf{H}^{\top}\boldsymbol{\Omega}^{-1}\mathbf{H}\right)
    \quad
    \text{s.t.}\
    \mathbf{S}\in\{0,1\}^{m\times N}\ \text{selects } m \text{ distinct points},
    \label{eq:dopt_general}
\end{equation}
where $(\mathbf{H},\boldsymbol{\Omega})$ are induced by the selection matrix $\mathbf{S}$ through the definitions above.

\paragraph{Remark (minimum measurement constraint for inverse identifiability).} A fundamental distinction from many measurement selection formulations for field reconstruction is that here the goal is inverse estimation of the $J$-dimensional degradation state $\mathbf{x}$ (one state per luminaire). For the MLE and its covariance to be well defined, $\mathcal{I}(\mathbf{x})=\mathbf{H}^{\top}\boldsymbol{\Omega}^{-1}\mathbf{H}$ must be nonsingular. Because $\boldsymbol{\Omega}$ is positive definite, this requires $\operatorname{rank}(\mathbf{H}) = J$. Since $\mathbf{H}\in\mathbb{R}^{m\times J}$, $\operatorname{rank}(\mathbf{H})\le \min\{m,J\}$, implying the necessary condition
\begin{equation}
    m \ge J.
    \label{eq:k_ge_J}
\end{equation}

If $m<J$, then $\det(\mathcal{I})=0$ and the confidence ellipsoid volume in Eq.~\eqref{eq:ellipsoid_vol} is infinite (unidentifiable directions). Accordingly, the D-optimal criterion in Eq.~\eqref{eq:dopt_general} is meaningful only when the selected WP grid points yield full column rank. This result shows that standalone inverse estimation at a single epoch requires at least as many WP measurements as the number of luminaires in the lighting system. In the following reference layout optimization, the reference layout is constructed with $m=J$ as the minimum identifiable configuration. Because measuring this many WP grid points at every monitoring epoch may still be costly in practice, this minimum configuration later serves as the reference layout on which runtime-stage measurement design is built.

\subsection{Design-stage reference measurement layout optimization}
\label{subsec:offline_opt}

With the design-stage measurement count set to the minimum value required by the identifiability condition in Eq.~\eqref{eq:k_ge_J}, namely $m=J$, the remaining task is to determine which WP grid points to select. The measurement point selection problem therefore reduces to selecting a subset containing exactly $m$ WP grid points under the D-optimal objective. Let $\mathcal{S}\subset\{1,\dots,N\}$ denote the index set of selected points with $|\mathcal{S}|=m$, where $N$ is the total number of candidate WP grid points, and let $\mathbf{S}$ denote the corresponding selection matrix. For any feasible $\mathcal{S}$, the D-optimal objective in Eq.~\eqref{eq:dopt_general} is evaluated as
\begin{equation}
    f(\mathcal{S})
    \triangleq
    \log\det\!\Bigl(\mathbf{H}^{\top}\boldsymbol{\Omega}^{-1}\mathbf{H}\Bigr),
    \qquad
    \mathbf{H}=\mathbf{S}\mathbf{C},\ \
    \boldsymbol{\Omega}=\mathbf{S}\boldsymbol{\Sigma}\mathbf{S}^{\top}+\mathbf{R}.
    \label{eq:dopt_obj_set}
\end{equation}
This fixed-size combinatorial optimization problem is generally nonconvex and nondeterministic polynomial-time hard (NP-hard), so heuristic layout optimization strategies are adopted. Four candidate layouts are considered: the Nadir-point layout, hybrid greedy search (HG) layout, random-initialized genetic algorithm (GA) layout, and nadir-seeded GA layout.

\paragraph{Nadir-point layout.}
The physically interpretable Nadir-point layout places WP measurement points directly beneath luminaire centers. Because $m=J$, this layout selects one WP measurement point at the nadir projection of each luminaire and serves as a transparent benchmark for optimization-based designs.

\paragraph{Hybrid greedy search layout.}
To solve the design-stage WP selection problem with a fixed number of selected points efficiently, the HG procedure is adopted. The method combines randomized greedy construction and beam search to generate a diverse set of promising layouts, followed by deduplication and a limited 1-swap local improvement step. The best layout is retained as the final HG solution according to the D-optimal objective in Eq.~\eqref{eq:dopt_obj_set}. The detailed procedure is summarized in Algorithm~\ref{alg:hg_deployment} in Appendix~\ref{app:hg_sensor_placement}.

\paragraph{Genetic algorithm layouts.}
To further explore the nonconvex selection space, a genetic algorithm is employed under a fixed measurement count constraint. Each individual is encoded as a binary vector $\mathbf{z}\in\{0,1\}^{N}$ satisfying $\mathbf{1}^{\top}\mathbf{z}=m$, which defines the selected set $\mathcal{S}=\{i:z_i=1\}$. The fitness is computed from Eq.~\eqref{eq:dopt_obj_set}. To keep exactly $m$ selected points throughout evolution, a crossover operator based on subsets and a mutation operator based on swaps are used. The former forms offspring by retaining WP grid points shared by both parents and filling the remaining positions from the parental difference set, while the latter exchanges selected and unselected points without changing the number of selected measurement points. Two initialization schemes are considered, producing the Random-initialized GA layout and the Nadir-seeded GA layout. For the Nadir-seeded GA layout, a prescribed fraction of the initial population is generated from the Nadir-point layout and then slightly perturbed by several swaps that keep the number of selected points fixed to enhance diversity; the remaining individuals are generated randomly. The best feasible individual found during evolution is decoded as the GA layout $\mathcal{S}_{\mathrm{GA}}$. The detailed GA procedure is summarized in Algorithm~\ref{alg:ga_deployment} in Appendix~\ref{app:ga_sensor_placement}.

\paragraph{Reference layout.}
Among the four candidate layouts, the configuration with the largest objective value in Eq.~\eqref{eq:dopt_obj_set} is denoted as the reference layout $\mathcal{S}_{\mathrm{ref}}\subset\{1,\dots,N\}$ with $|\mathcal{S}_{\mathrm{ref}}|=m=J$. This layout provides a high-quality set of WP measurement points defined by design-stage measurement layout optimization and serves as the candidate pool for runtime-stage measurement design in \autoref{sec:online_tracking}.

\section{Runtime-stage measurement design and condition-based maintenance optimization}
\label{sec:online_tracking}

The reference layout $\mathcal{S}_{\mathrm{ref}}$ obtained from design-stage measurement layout optimization in \autoref{sec:design_sensor_opt} provides an informative candidate set of WP measurement points for inverse luminaire degradation state estimation. When all $J$ reference points are measured at a single epoch, the resulting observation model satisfies the minimum identifiability requirement derived in Eq.~\eqref{eq:k_ge_J}. However, imposing this full reference measurement requirement at every monitoring visit can still be costly and time-consuming, especially for large-scale lighting systems. The runtime stage therefore keeps $\mathcal{S}_{\mathrm{ref}}$ fixed and uses sequential filtering to reduce the measurement effort per visit. Instead of relying only on instantaneous measurements at one epoch, the filter combines sparse current observations with the degradation dynamics and previously assimilated information. The corresponding runtime-stage measurement design problem is to determine when site visits should occur and which active reference measurement points should be selected at each visit. These decisions define the active reference measurement points at each executed visit and, consequently, the runtime monitoring effort.

The same site visits also provide the information used for CBM. After each filtering update, the posterior degradation state distribution is used to compute posterior predictive failure probabilities, which serve as the risk metric for preventive replacement decisions. Hence, the runtime-stage measurement design is coupled with the CBM risk threshold, forming a budgeted runtime monitoring and CBM policy optimization problem that minimizes expected total downtime and replacement cost subject to a visit--measurement budget.

\subsection{Runtime grid and joint visit--measurement schedule}
\label{subsec:runtime_schedule}

Let $t_0$ denote the initial time with prior state $\mathbf{x}_0:=\mathbf{x}(t_0)$. Let $t_1<t_2<\cdots<t_K$ denote a fixed runtime simulation grid with step size $\delta>0$ (e.g., monthly),
\begin{equation}
    t_k = t_0 + k\delta,\qquad k=1,\ldots,K .
    \label{eq:timegrid}
\end{equation}
Among these grid points, a subset is designated as candidate visit epochs at which a site visit, and therefore WP measurements, may be executed.

\paragraph{Candidate visit grid.}
A periodic candidate visit grid with interval $\Delta=q_{\mathrm{cand}}\delta$ ($q_{\mathrm{cand}}\in\mathbb N_{+}$) is defined as $\mathcal{K}_{\mathrm{cand}}:=\{k\in\{1,\ldots,K\}:\ k \bmod q_{\mathrm{cand}} = 0\}$, so that the first candidate visit occurs at $t_{q_{\mathrm{cand}}}=t_0+\Delta$. Visits are restricted to the candidate grid, i.e., $v_k=0$ for all $k\notin\mathcal{K}_{\mathrm{cand}}$.

\paragraph{Visit indicator on the time grid.}
Let $v_k\in\{0,1\}$ indicate whether a site visit is executed at $t_k$:
\begin{equation}
    v_k=
    \begin{cases}
        1, & \text{a visit is executed at } t_k,\\
        0, & \text{otherwise},
    \end{cases}
    \qquad k=1,\ldots,K.
    \label{eq:vk_def}
\end{equation}
A baseline periodic policy is recovered by setting $v_k\equiv 1$ for all $k\in\mathcal{K}_{\mathrm{cand}}$ and $v_k=0$ otherwise.

\paragraph{Active reference point selection constrained to visits.}
Runtime WP measurement is constrained to the reference layout $\mathcal{S}_{\mathrm{ref}}$ with $|\mathcal{S}_{\mathrm{ref}}|=J$.
At time $t_k$, let $\mathbf{u}_k=[u_{k,1},\ldots,u_{k,J}]^{\top}\in\{0,1\}^{J}$ denote which reference points are actively measured:
\begin{equation}
    u_{k,r}=
    \begin{cases}
        1, & \text{reference point } r \text{ is measured at } t_k,\\
        0, & \text{otherwise},
    \end{cases}
    \qquad r=1,\ldots,J.
    \label{eq:uk_def}
\end{equation}

When $v_k=0$ (no visit), $\mathbf{u}_k=\mathbf{0}$ and no reference point selection is required. Accordingly, the schedule satisfies
\begin{equation}
    u_{k,r}\le v_k,\qquad \forall k,r.
    \label{eq:u_le_v}
\end{equation}

For notational convenience, define the joint schedule vector $\mathbf{z}_k = \bigl[v_k,\ \mathbf{u}_k^{\top}\bigr]\in\{0,1\}^{1+J}$ for $k=1,\ldots,K$, and stack $\{\mathbf{z}_k\}$ to obtain the joint visit--measurement schedule matrix
\begin{equation*}
    \mathbf{Z}
    =
    \begin{bmatrix}
    \mathbf{z}_1^{\top}\\
    \vdots\\
    \mathbf{z}_K^{\top}
    \end{bmatrix}
    \in \{0,1\}^{K\times(1+J)}.
\end{equation*}

A runtime visit--measurement policy can therefore be represented by $\mathbf{Z}$ (or equivalently $(\mathbf{v},\mathbf{U})$), where $\mathbf{U}=[\mathbf{u}_1,\ldots,\mathbf{u}_K]^{\top}\in\{0,1\}^{K\times J}$. The optimization problem developed below seeks an optimal schedule $\mathbf{Z}^{\star}$ jointly with the CBM risk threshold.

Define the active reference measurement point set and active reference measurement count as
\begin{equation*}
\begin{aligned}
    \mathcal{S}_k &:= \{r\in\{1,\ldots,J\}: u_{k,r}=1\},\\
    m_k &:= |\mathcal{S}_k|.
\end{aligned}
\end{equation*}
The set $\mathcal{S}_k$ encodes the active reference measurement points, and $m_k$ gives the active reference measurement count at that epoch. At visit epochs ($v_k=1$), at least one active reference measurement point is measured:
\begin{equation}
    \sum_{r=1}^{J} u_{k,r} \ge v_k,\qquad \forall k.
    \label{eq:atleastone_meas}
\end{equation}
Together, Eq.~\eqref{eq:u_le_v} and Eq.~\eqref{eq:atleastone_meas} enforce that measurements are collected only during executed visits and that every executed visit collects at least one reference-point measurement.

\subsection{State-space and runtime observation models}
\label{subsec:enkf_model}

Given the joint visit--measurement schedule defined above, runtime filtering requires a state evolution model on the runtime grid and an observation model induced by the active reference measurement points. Let $\mathbf{x}_k:=\mathbf{x}(t_k)\in\mathbb{R}^{J}$ denote the degradation state vector at time $t_k$. Under the nonhomogeneous gamma process model in \autoref{subsec:state_illum_model}, the evolution on this runtime grid is
\begin{equation}
    \mathbf{x}_k = \mathbf{x}_{k-1} + \mathbf{g}_k,
    \label{eq:state_gamma_grid}
\end{equation}
where $\mathbf{g}_k\in\mathbb{R}^J_+$ collects the componentwise non-negative gamma process increments over $(t_{k-1},t_k]$.

\paragraph{Reference layout observation model.}
Let $\mathbf{S}_{\mathrm{ref}}\in\{0,1\}^{J\times N}$ denote the selection matrix associated with the reference layout over the $N$ WP grid points. Setting $\mathbf{S}(t)=\mathbf{S}_{\mathrm{ref}}$ in the sparse linear-Gaussian observation model in Eq.~\eqref{eq:obs_lg_t} gives the full reference layout observation model
\begin{equation}
    \mathbf{y}^{\mathrm{ref}}_k
    =
    \mathbf{H}_{\mathrm{ref}}\mathbf{x}_k + \mathbf{d}_{\mathrm{ref}} + \boldsymbol{\nu}^{\mathrm{ref}}_k,
    \qquad
    \boldsymbol{\nu}^{\mathrm{ref}}_k\sim \mathcal{N}(\mathbf{0},\boldsymbol{\Omega}_{\mathrm{ref}}),
    \label{eq:ref_inv_model}
\end{equation}
where $\mathbf{H}_{\mathrm{ref}}=\mathbf{S}_{\mathrm{ref}}\mathbf{C}\in\mathbb{R}^{J\times J}$, $\mathbf{d}_{\mathrm{ref}}=\mathbf{S}_{\mathrm{ref}}\mathbf{b}\in\mathbb{R}^{J}$, and $\boldsymbol{\Omega}_{\mathrm{ref}}=\mathbf{S}_{\mathrm{ref}}\boldsymbol{\Sigma}\mathbf{S}_{\mathrm{ref}}^\top+\mathbf{R}_{\mathrm{ref}}\in\mathbb{S}_{++}^{J}$.

\paragraph{Runtime sparse observation induced by $\mathbf{u}_k$.}
The reference layout observation model above corresponds to measuring all $J$ reference points. Runtime operation is budget-constrained and may therefore use only a subset of the reference measurement points at each executed visit. For an executed visit at $t_k$, write the active set selected by $\mathbf{u}_k$ as $\mathcal{S}_k=\{r_{k,1},\ldots,r_{k,m_k}\}$, where $m_k\le J$ and the indices are given in any fixed ordering. Define the row extraction matrix $\mathbf{M}_k\in\{0,1\}^{m_k\times J}$ as
\begin{equation*}
    \mathbf{M}_k
    \triangleq
    \begin{bmatrix}
        \mathbf{e}_{r_{k,1}}^{\top}\\
        \vdots\\
        \mathbf{e}_{r_{k,m_k}}^{\top}
    \end{bmatrix},
    \qquad
    \mathbf{e}_r\in\mathbb{R}^{J},\quad (\mathbf{e}_r)_{r'}=\mathbb{I}(r'=r),\ r'=1,\ldots,J,
\end{equation*}
so that $\mathbf{M}_k\mathbf{y}^{\mathrm{ref}}_k$ extracts the subset of reference layout measurements that is actually collected.

For $m_k>0$, the resulting runtime measurement model is
\begin{equation}
    \mathbf{y}_k
    =
    \mathbf{H}_k\mathbf{x}_k + \mathbf{d}_k + \boldsymbol{\nu}_k,
    \qquad
    \boldsymbol{\nu}_k\sim \mathcal{N}(\mathbf{0},\boldsymbol{\Omega}_k),
    \label{eq:runtime_obs}
\end{equation}
where $\mathbf{H}_k=\mathbf{M}_k\mathbf{H}_{\mathrm{ref}}\in\mathbb{R}^{m_k\times J}$, $\mathbf{d}_k=\mathbf{M}_k\mathbf{d}_{\mathrm{ref}}\in\mathbb{R}^{m_k}$, and $\boldsymbol{\Omega}_k=\mathbf{M}_k\boldsymbol{\Omega}_{\mathrm{ref}}\mathbf{M}_k^\top\in\mathbb{S}_{++}^{m_k}$. If $v_k=0$, then $\mathbf{u}_k=\mathbf{0}$ and $m_k=0$; in this case, no runtime observation vector is formed and the filter performs forecast-only propagation.

\subsection{Degradation tracking using an EnKF}
\label{subsec:enkf}

The runtime estimator is required to track the joint degradation state of all luminaires from sparse, noisy WP measurements. A standard Kalman filter \cite{rawlings_model_2017, truong-ba_sensor_2025} is not directly suitable because the degradation dynamics follow a gamma process, which produces non-Gaussian, non-negative degradation increments. Particle filters provide a more general sequential Monte Carlo alternative for non-Gaussian state-space models, but they can suffer from severe weight degeneracy in high-dimensional systems \citep{bengtsson_curse--dimensionality_2008, bickel_sharp_2008}. In particular, \citet{bengtsson_curse--dimensionality_2008} show that avoiding degeneracy generally requires the particle population to increase essentially exponentially with the state dimension. This exponential scaling would make a standard particle filter computationally prohibitive for repeated policy evaluation in large-scale LED lighting systems. The EnKF provides a scalable alternative by propagating an ensemble of degradation trajectories and applying corrections based on the Kalman gain rather than importance reweighting \citep{evensen_sequential_1994, evensen_ensemble_2003, katzfuss_understanding_2016}. Although derived under Gaussian assumptions, the EnKF can remain effective for gamma-process degradation tracking because the ensemble represents degradation uncertainty through sampled trajectories \citep{katzfuss_understanding_2016}.

Let $\{\mathbf{x}^{a,(i)}_{k-1}\}_{i=1}^{N_e}$ denote the analysis ensemble available at time $t_{k-1}$. At each runtime grid step, the EnKF first propagates this ensemble through the gamma process degradation model. If a site visit is executed ($v_k=1$), the forecast ensemble is then corrected using the sparse WP measurement vector $\mathbf{y}_k$ defined in Eq.~\eqref{eq:runtime_obs}; otherwise, no measurement update is performed, and the forecast ensemble is used as the analysis ensemble.

\paragraph{Forecast step.}
For each ensemble member, a gamma process degradation increment $\mathbf{g}^{(i)}_k$ is sampled over $(t_{k-1},t_k]$, and the state is propagated as
\begin{equation}
    \mathbf{x}^{f,(i)}_k = \mathbf{x}^{a,(i)}_{k-1} + \mathbf{g}^{(i)}_k,
    \qquad i=1,\ldots,N_e.
    \label{eq:enkf_forecast_runtime}
\end{equation}

\paragraph{Update step (only when $v_k=1$).}
When $v_k=1$, the sparse runtime measurements are assimilated by first projecting each forecast ensemble member through the active observation model in Eq.~\eqref{eq:runtime_obs} to obtain its predicted WP measurement vector:
\begin{equation*}
    \mathbf{y}^{f,(i)}_k = \mathbf{H}_k\mathbf{x}^{f,(i)}_k + \mathbf{d}_k,\qquad i=1,\ldots,N_e,
\end{equation*}
with forecast sample means $\bar{\mathbf{x}}^f_k=N_e^{-1}\sum_{i=1}^{N_e}\mathbf{x}^{f,(i)}_k$ and $\bar{\mathbf{y}}^f_k=N_e^{-1}\sum_{i=1}^{N_e}\mathbf{y}^{f,(i)}_k$.

The corresponding forecast state and predicted measurement anomaly matrices are
\begin{equation*}
    \mathbf{A}_k=
    \begin{bmatrix}
        (\mathbf{x}^{f,(1)}_k-\bar{\mathbf{x}}^f_k)^\top\\
        \vdots\\
        (\mathbf{x}^{f,(N_e)}_k-\bar{\mathbf{x}}^f_k)^\top
    \end{bmatrix}
    \in\mathbb{R}^{N_e\times J},
    \qquad
    \mathbf{B}_k=
    \begin{bmatrix}
        (\mathbf{y}^{f,(1)}_k-\bar{\mathbf{y}}^f_k)^\top\\
        \vdots\\
        (\mathbf{y}^{f,(N_e)}_k-\bar{\mathbf{y}}^f_k)^\top
    \end{bmatrix}
    \in\mathbb{R}^{N_e\times m_k}.
\end{equation*}

The empirical cross-covariance between states and predicted measurements and the predicted measurement covariance are then estimated as
\begin{equation*}
    \mathbf{P}^{xy}_k=\frac{1}{N_e-1}\mathbf{A}_k^\top \mathbf{B}_k,
    \qquad
    \mathbf{P}^{yy}_k=\frac{1}{N_e-1}\mathbf{B}_k^\top \mathbf{B}_k + \boldsymbol{\Omega}_k,
\end{equation*}
which gives the ensemble Kalman gain
\begin{equation}
    \mathbf{K}_k=\mathbf{P}^{xy}_k(\mathbf{P}^{yy}_k)^{-1}.
\end{equation}

In the stochastic EnKF update, perturbed observations are generated as
\begin{equation*}
    \mathbf{y}^{(i,\mathrm{pert})}_k
    =
    \mathbf{y}_k+\boldsymbol{\xi}^{(i)}_k,
    \qquad 
    \boldsymbol{\xi}^{(i)}_k
    \sim\mathcal{N}(\mathbf{0},\boldsymbol{\Omega}_k),
\end{equation*}
and each ensemble member is updated by
\begin{equation}
    \mathbf{x}^{a,(i)}_k
    =
    \mathbf{x}^{f,(i)}_k
    +
    \mathbf{K}_k\Bigl(\mathbf{y}^{(i,\mathrm{pert})}_k-\mathbf{y}^{f,(i)}_k\Bigr),
    \qquad i=1,\ldots,N_e.
    \label{eq:enkf_update_runtime}
\end{equation}

If no visit is executed ($v_k=0$), no measurement update is performed, and the forecast ensemble becomes the analysis ensemble:
\begin{equation*}
    \mathbf{x}^{a,(i)}_k = \mathbf{x}^{f,(i)}_k,\qquad i=1,\ldots,N_e.
\end{equation*}

\paragraph{Posterior predictive failure probability.}
For a nonterminal visit epoch $t_k$ with $v_k=1$, let
\begin{equation}
    \kappa(k) := \min\{\ell>k: v_\ell=1\}
    \label{eq:kappa_def}
\end{equation}
denote the index of the next scheduled visit. This lookahead calculation is not required for terminal visits with no subsequent scheduled visit. Given the analysis ensemble $\{\mathbf{x}^{a,(i)}_{k}\}_{i=1}^{N_e}$ after assimilating measurements at $t_k$, a posterior predictive ensemble at $t_{\kappa(k)}$ is generated by propagating each member forward:
\begin{equation}
    \mathbf{x}^{f,(i)}_{\kappa(k)\mid k}
    =
    \mathbf{x}^{a,(i)}_{k}
    +
    \mathbf{g}^{(i)}_{k\to \kappa(k)},
    \qquad i=1,\ldots,N_e,
    \label{eq:jump_forecast}
\end{equation}
where $\mathbf{g}^{(i)}_{k\to \kappa(k)}$ denotes the aggregated gamma process increments over $(t_k,t_{\kappa(k)}]$ (equivalently, the sum of the intermediate increments on the runtime grid).

Let $x_{\mathrm{th}}$ denote the degradation failure threshold. Under the $L_{70}$ criterion and the definition in Eq.~\eqref{eq:LO}, $x_{\mathrm{th}}=0.3$. For luminaire $j$, let $\hat p_{j,k}$ denote the posterior predictive failure probability of exceeding this threshold before the next visit, conditional on all measurements up to $t_k$. Because the gamma process degradation state is nondecreasing, this event before the next visit is equivalent to exceeding the threshold by $t_{\kappa(k)}$. The probability is therefore estimated from the propagated ensemble as
\begin{equation}
    \hat p_{j,k}
    \triangleq
    \Pr\!\left(x_{j,\kappa(k)} > x_{\mathrm{th}} \mid \mathbf{y}_{1:k}\right)
    \approx
    \frac{1}{N_e}\sum_{i=1}^{N_e}\mathbb{I}\!\left(x^{f,(i)}_{j,\kappa(k)\mid k}>x_{\mathrm{th}}\right),
    \qquad j=1,\ldots,J.
    \label{eq:phat_def}
\end{equation}

\subsection{Budgeted monitoring and risk-based maintenance optimization}
\label{subsec:budget_runtime_design}

The posterior predictive failure probabilities in Eq.~\eqref{eq:phat_def} provide the risk metric for runtime maintenance decisions. At each nonterminal visit epoch, they are used to determine whether individual luminaires should be preventively replaced to avoid failure before the next visit. This budgeted runtime-stage measurement design and CBM policy formulation therefore balances monitoring effort, preventive replacement, and failed-luminaire downtime by minimizing expected total downtime and replacement cost subject to a visit--measurement budget.

\paragraph{Risk-based preventive replacement rule.}
Let $p_{\mathrm{th}}\in(0,1)$ denote the risk threshold applied to the posterior predictive failure probability. The probability $\hat p_{j,k}$ is evaluated at visit epochs with a subsequent scheduled visit; at terminal visit epochs with no subsequent scheduled visit, no lookahead risk is computed and $\hat p_{j,k}$ is set to zero for notational convenience. For luminaire $j$, the replacement decision made at visit time $t_k$ is
\begin{equation}
    d_{j,k} = v_k\,\mathbb{I}\!\left(\hat p_{j,k} > p_{\mathrm{th}}\right),
    \qquad j=1,\ldots,J.
    \label{eq:replace_rule}
\end{equation}
Thus, replacement can occur only when a site visit is scheduled ($v_k=1$), and only for luminaires whose predicted probability of exceeding the failure threshold before the next visit is larger than $p_{\mathrm{th}}$.

When $d_{j,k}=1$, luminaire $j$ is replaced at time $t_k$. Operationally, this replacement restores the luminaire to an ``as-good-as-new'' condition. In the Monte Carlo rollout, this is represented by resetting both the filtering ensemble and the simulated true state:
\begin{equation}
    x^{a,(i)}_{j,k}\leftarrow 0,\ \forall i,
    \qquad\text{and}\qquad
    x^{\mathrm{true}}_{j,k}\leftarrow 0,
    \qquad \text{if } d_{j,k}=1.
    \label{eq:reset_rule}
\end{equation}

\paragraph{Total downtime and replacement cost.}
Preventive maintenance (PM) actions are implemented as replacements and counted through the indicator
\begin{equation*}
    \mathrm{PM}_{j,k}
    =
    v_k\,\mathbb{I}\!\left(\hat p_{j,k} > p_{\mathrm{th}}\right)
    =d_{j,k},
    \qquad j=1,\ldots,J,
\end{equation*}
which is identical to the decision variable in Eq.~\eqref{eq:replace_rule}. The runtime grid in Eq.~\eqref{eq:timegrid} is used consistently for state propagation, WP measurement opportunities, filtering updates, preventive replacement decisions, and downtime accounting. Downtime is accumulated over the simulation intervals rather than on a separate grid. Because consecutive simulation epochs are spaced by $\delta$, downtime is expressed in units of one simulation step. Let $\mathrm{DT}_{j,k}$ denote the failed-luminaire downtime accrued by luminaire $j$ over the interval $(t_{k-1},t_k]$, expressed in units of $\delta$:
\begin{equation}
    \mathrm{DT}_{j,k}
    =
    \frac{1}{\delta}\int_{t_{k-1}}^{t_k}
    \mathbb{I}\!\left(x^{\mathrm{true}}_{j}(t)>x_{\mathrm{th}}\right)\,dt,
    \qquad j=1,\ldots,J,\ k=1,\ldots,K.
    \label{eq:downtime_def}
\end{equation}
Thus, $0\le\mathrm{DT}_{j,k}\le1$: $\mathrm{DT}_{j,k}=0$ if luminaire $j$ is not failed during $(t_{k-1},t_k]$, $\mathrm{DT}_{j,k}=1$ if it is failed throughout the interval, and intermediate values are used when the simulated failure crossing occurs within the interval. If preventive replacement is performed at $t_k$, the reset in Eq.~\eqref{eq:reset_rule} is then applied for subsequent propagation of the true state. Therefore, $\mathrm{PM}_{j,k}$ represents PM actions implemented as replacements and triggered by applying the risk threshold to posterior predictive failure probabilities at visit epochs, whereas $\mathrm{DT}_{j,k}$ represents interval downtime counted on the common runtime grid.

For diagnostic reporting, PM events are further classified using the simulated true degradation trajectory. A PM event is classified as true positive PM (TP-PM) if the replaced luminaire would have crossed the failure threshold before the next visit under the counterfactual case without replacement at $t_k$; otherwise, it is classified as false positive PM (FP-PM). This decomposition is used only for simulation-based reporting and is not available to the monitoring and CBM policy.

The total downtime and replacement cost function, denoted by $C_{\mathrm{DR}}$, is defined by aggregating failed-luminaire downtime and preventive replacements with their corresponding unit costs:
\begin{equation}
    C_{\mathrm{DR}}(p_{\mathrm{th}},\mathbf{Z})
    =
    c_{\mathrm{DT}}\sum_{k=1}^{K}\sum_{j=1}^{J}\mathrm{DT}_{j,k}
    +
    c_{\mathrm{PM}}\sum_{k=1}^{K}\sum_{j=1}^{J}\mathrm{PM}_{j,k}.
    \label{eq:CDR_def}
\end{equation}
Here, $c_{\mathrm{DT}}$ is the normalized downtime cost per failed luminaire per simulation interval of length $\delta$, and $c_{\mathrm{PM}}$ is the unit cost per preventive replacement.

\paragraph{Visit--measurement cost under a budget.}
Runtime monitoring also incurs visit--measurement cost. Let $c_{\mathrm{visit}}$ be the fixed cost per executed site visit and $c_{\mathrm{meas}}$ the cost per measured reference point. The total visit--measurement cost induced by schedule $\mathbf{Z}$ is
\begin{equation}
    C_{\mathrm{insp}}(\mathbf{Z})
    =
    c_{\mathrm{visit}}\sum_{k=1}^{K} v_k
    +
    c_{\mathrm{meas}}\sum_{k=1}^{K}\sum_{r=1}^{J} u_{k,r}.
    \label{eq:Cinsp_def}
\end{equation}

A total monitoring budget $B_{\mathrm{mon}}$ is imposed:
\begin{equation}
    C_{\mathrm{insp}}(\mathbf{Z})\le B_{\mathrm{mon}}.
    \label{eq:budget_constraint}
\end{equation}

\paragraph{Budgeted runtime-stage measurement design and CBM policy problem (general form).}
The general problem couples runtime-stage measurement design, represented by the visit--measurement schedule $\mathbf{Z}$ (equivalently $(\mathbf{v},\mathbf{U})$), with the CBM risk threshold $p_{\mathrm{th}}$. The objective is to minimize the expected total downtime and replacement cost subject to the visit--measurement budget and schedule consistency constraints:
\begin{equation}
\begin{aligned}
    \min_{p_{\mathrm{th}}\in(0,1),\ \mathbf{Z}\in\{0,1\}^{K\times(1+J)}}\quad
    & \mathbb{E}\!\left[C_{\mathrm{DR}}(p_{\mathrm{th}},\mathbf{Z})\right] \\
    \text{s.t.}\quad
    & C_{\mathrm{insp}}(\mathbf{Z})\le B_{\mathrm{mon}}, \\
    & u_{k,r}\le v_k,
    && k=1,\ldots,K,\ r=1,\ldots,J, \\
    & \sum_{r=1}^{J}u_{k,r}\ge v_k,
    && k=1,\ldots,K, \\
    & v_k=0,
    && k\notin\mathcal{K}_{\mathrm{cand}}.
\end{aligned}
    \label{eq:budgeted_opt}
\end{equation}

\subsection{Tractable policy parameterization and sequential grid search with paired screening}
\label{sec:solution_framework}

The general formulation in Eq.~\eqref{eq:budgeted_opt} defines a budgeted runtime monitoring and CBM policy through the visit--measurement schedule $\mathbf{Z}=(\mathbf{v},\mathbf{U})$ and the risk threshold $p_{\mathrm{th}}$. For a fixed pair $(p_{\mathrm{th}},\mathbf{Z})$, the expected total downtime and replacement cost cannot be evaluated analytically because it depends on stochastic degradation trajectories, measurement noise, EnKF updates, and threshold-triggered maintenance actions. We therefore evaluate candidate policies through Monte Carlo (MC) simulation.

\paragraph{Stochastic policy evaluation.}

For a candidate schedule $\mathbf{Z}$ and threshold $p_{\mathrm{th}}$, one stochastic rollout requires (i) sampling degradation trajectories and measurement noise, (ii) updating latent states through EnKF at visit epochs, (iii) applying the replacement rule in Eq.~\eqref{eq:replace_rule}, and (iv) accumulating the resulting total downtime and replacement cost. Because $d_{j,k}=v_k\mathbb{I}(\hat p_{j,k}>p_{\mathrm{th}})$ is threshold-triggered, small perturbations around $p_{\mathrm{th}}$ can switch replacement actions and cause discontinuous changes in the realized total downtime and replacement cost. The expected total downtime and replacement cost is estimated by Monte Carlo averaging,
\begin{equation}
\bar{C}_{\mathrm{DR}}\bigl(p_{\mathrm{th}},\mathbf{Z}\bigr)
=
\frac{1}{N_{\mathrm{MC}}}\sum_{w=1}^{N_{\mathrm{MC}}}C_{\mathrm{DR}}^{(w)}\bigl(p_{\mathrm{th}},\mathbf{Z}\bigr)
\approx
\mathbb{E}\!\left[C_{\mathrm{DR}}\bigl(p_{\mathrm{th}},\mathbf{Z}\bigr)\right],
\label{eq:mc_est_runtime}
\end{equation}
Here, $N_{\mathrm{MC}}$ is the number of MC replications.

Direct optimization of Eq.~\eqref{eq:budgeted_opt} over the full decision matrix is computationally prohibitive because $\mathbf{Z}$ encodes high-dimensional, coupled visit and measurement decisions and the simulation-based objective is non-smooth and noisy. These features motivate the periodic parameterization and sequential screening strategy developed below.

\paragraph{Periodic visit--measurement policy parameterization.}
To obtain a tractable parameterization of runtime-stage measurement design and CBM decisions, we restrict the general decision space in Eq.~\eqref{eq:budgeted_opt} to periodic visit--measurement policies indexed by
\begin{equation*}
    \boldsymbol{\theta}
    :=
    \bigl(t_{\mathrm{first}},\Delta_{\mathrm{visit}},n_{\mathrm{zone}},p_{\mathrm{th}}\bigr),
\end{equation*}
where $t_{\mathrm{first}}$ is the first visit time, $\Delta_{\mathrm{visit}}$ is the periodic visit interval, $n_{\mathrm{zone}}$ is the zone-count parameter for the runtime reference layout partition, and $p_{\mathrm{th}}$ is the risk threshold. Under the zone cyclic rule introduced below, $n_{\mathrm{zone}}$ also determines the active reference measurement count per executed visit.

Let $\{t^{\mathrm{cand}}_q\}_{q=1}^{Q}$ denote the ordered candidate visit epochs induced by $\mathcal{K}_{\mathrm{cand}}$, where $Q=|\mathcal{K}_{\mathrm{cand}}|$. Write $t_{\mathrm{first}}=t^{\mathrm{cand}}_{q_0}$ and $\Delta_{\mathrm{visit}}=h\Delta$ with integer $q_0\in\{1,\ldots,Q\}$ and $h\in\mathbb{N}_{+}$. The visit indicator on the candidate grid is then
\begin{equation}
    v_q(\boldsymbol{\theta})
    =
    \mathbb{I}\!\left(q\ge q_0,\ (q-q_0)\bmod h=0\right),
    \qquad q=1,\ldots,Q.
    \label{eq:periodic_visit_indicator}
\end{equation}
This policy parameterization substantially reduces search dimensionality while retaining explicit control of when to visit through $(t_{\mathrm{first}},\Delta_{\mathrm{visit}})$, how to partition the reference layout through $n_{\mathrm{zone}}$, and which active reference measurement points are selected through the zone cyclic selection rule.

\paragraph{Reference layout partition for zone cyclic measurements.}
Given the visit sequence in Eq.~\eqref{eq:periodic_visit_indicator}, the remaining task is to specify the measurement matrix $\mathbf{U}$, i.e., which active reference measurement points are selected at each executed visit. For runtime scheduling, let $\mathcal{R}_{\mathrm{ref}}:=\{1,\ldots,J\}$ denote the index set of the $J$ reference points in the design-stage reference layout. These reference point indices are partitioned into $n_{\mathrm{zone}}$ nonempty disjoint zones, where $n_{\mathrm{zone}}$ denotes the number of zones in the runtime partition:
\begin{equation*}
    \mathcal{R}_{\mathrm{ref}}
    =
    \bigcup_{g=1}^{n_{\mathrm{zone}}}\mathcal{G}_g,
    \qquad
    \mathcal{G}_g\cap\mathcal{G}_{g'}=\varnothing\ \ (g\neq g').
\end{equation*}

Each zone $\mathcal{G}_g$ contains $n_g:=|\mathcal{G}_g|$ reference points. A fixed cyclic ordering $\mathcal{G}_g=(r_{g,1},r_{g,2},\ldots,r_{g,n_g})$, $g=1,\ldots,n_{\mathrm{zone}}$, is assigned within each zone, where $r_{g,\ell}$ denotes the $\ell$-th reference point in zone $g$.

\paragraph{Zone cyclic measurement rule.}
At each executed visit, exactly one reference point is measured from each zone; therefore, the active reference measurement count per executed visit is $n_{\mathrm{zone}}$. Let $V_q^{\mathrm{cum}}:=\sum_{\ell=1}^{q} v_\ell(\boldsymbol{\theta})$ denote the cumulative number of executed visits up to candidate epoch $q$. If $v_q(\boldsymbol{\theta})=1$, the selected position in zone $g$ follows the cyclic rule
\begin{equation}
    \ell_g(q)
    =
    1+\bigl((V_q^{\mathrm{cum}}-1)\bmod n_g\bigr),
    \qquad g=1,\ldots,n_{\mathrm{zone}}.
    \label{eq:zone_cyclic_pointer}
\end{equation}

The set of active reference measurement points selected at epoch $t^{\mathrm{cand}}_q$ is therefore
\begin{equation}
    \mathcal{S}_q=
    \{\,r_{g,\ell_g(q)}:\ g=1,\ldots,n_{\mathrm{zone}}\,\},
    \qquad \text{if } v_q(\boldsymbol{\theta})=1,
    \label{eq:Sq_zone_cyclic_aosga}
\end{equation}
and $\mathcal{S}_q=\varnothing$ when no visit is executed.

\paragraph{Mapping to the runtime grid.}
Let $k(q)$ denote the corresponding index on the runtime grid defined in Eq.~\eqref{eq:timegrid}, so that $t_{k(q)}=t^{\mathrm{cand}}_q$. The decoded periodic policy is mapped to the runtime-grid visit--measurement variables in Eqs.~\eqref{eq:vk_def}--\eqref{eq:uk_def} as
\begin{equation*}
    v_{k(q)} = v_q(\boldsymbol{\theta}),
    \qquad
    u_{k(q),r} = \mathbb{I}(r\in\mathcal{S}_q).
\end{equation*}

For epochs that are not candidates, $v_k=0$ and $\mathbf{u}_k=\mathbf{0}$. Hence the runtime visit--measurement decisions satisfy Eqs.~\eqref{eq:u_le_v}--\eqref{eq:atleastone_meas} by construction.

\paragraph{Budget-feasible periodic plans.}

Under the zone cyclic measurement rule, the visit--measurement cost in Eq.~\eqref{eq:Cinsp_def} simplifies to
\begin{equation}
    C_{\mathrm{insp}}(\boldsymbol{\theta})
    =
    \sum_{q=1}^{Q} v_q(\boldsymbol{\theta})\left(c_{\mathrm{visit}}+n_{\mathrm{zone}}c_{\mathrm{meas}}\right).
    \label{eq:Cinsp_aosga}
\end{equation}

Therefore, the budget constraint in Eq.~\eqref{eq:budget_constraint} is equivalent to the visit count bound
\begin{equation}
    \sum_{q=1}^{Q} v_q(\boldsymbol{\theta})
    \le
    V_{\max}^{\mathrm{zone}},
    \qquad
    V_{\max}^{\mathrm{zone}}
    :=
    \left\lfloor
    \frac{B_{\mathrm{mon}}}{c_{\mathrm{visit}}+n_{\mathrm{zone}}c_{\mathrm{meas}}}
    \right\rfloor.
    \label{eq:Vmax_zone_aosga}
\end{equation}

The periodic decoder maps each budget-feasible parameter tuple to an executable runtime schedule,
\begin{equation*}
    \boldsymbol{\theta}
    \ \mapsto\
    \mathbf{Z}(\boldsymbol{\theta})
    \equiv
    \bigl(\mathbf{v}(\boldsymbol{\theta}),\mathbf{U}(\boldsymbol{\theta})\bigr),
\end{equation*}
thereby enforcing the visit--measurement constraints and budget feasibility while maintaining a low-dimensional search space.

\paragraph{Sequential grid search with paired screening.}
With the above restriction, periodic visit--measurement and CBM policy optimization is performed over the low-dimensional parameter vector $\boldsymbol{\theta}$ using sequential grid search with paired screening (SGS-PS). The reduced problem is
\begin{equation}
\min_{\boldsymbol{\theta}\in\Theta_{\mathrm{feas}}}
\mathbb{E}\!\left[C_{\mathrm{DR}}\bigl(p_{\mathrm{th}},\mathbf{Z}(\boldsymbol{\theta})\bigr)\right],
\label{eq:periodic_runtime_opt}
\end{equation}
where $p_{\mathrm{th}}$ is the risk threshold component of $\boldsymbol{\theta}$, and $\Theta_{\mathrm{feas}}$ is defined by the periodic structure in Eq.~\eqref{eq:periodic_visit_indicator}, the zone cyclic measurement rule, and the budget feasibility condition in Eq.~\eqref{eq:Vmax_zone_aosga}. In implementation, SGS-PS enumerates candidate parameter tuples, decodes them into periodic schedules $\mathbf{Z}(\boldsymbol{\theta})$, and evaluates them using the Monte Carlo estimator in Eq.~\eqref{eq:mc_est_runtime}.

Because each evaluation of $C_{\mathrm{DR}}$ requires simulating stochastic degradation paths, noisy WP measurements, EnKF updates, and threshold-triggered maintenance actions over the full planning horizon, a sufficiently large Monte Carlo replication budget is needed to obtain statistically reliable estimates. Even under the periodic policy parameterization, the parameter search grid can still generate many feasible candidates. Applying the full replication budget to every feasible candidate would therefore be computationally expensive. Accordingly, SGS-PS combines exhaustive enumeration over the low-dimensional parameter grid with sequential paired screening. Candidates are first evaluated with a small number of paired replications, the current incumbent is identified from the estimated mean total downtime and replacement cost, and pairwise cost differences relative to the incumbent are used to eliminate candidates that are statistically inferior. Surviving candidates receive additional replications in later stages, so simulation effort is concentrated on policies that remain competitive. The final survivor set is reported as the retained set of near-optimal periodic policies. Full pseudocode is given in Algorithm~\ref{alg:seq_grid_runtime} in Appendix~\ref{app:seq_screening_runtime}, and the instance-specific grids, parameter enumeration strategy, and replication settings are provided in the case study section.

\section{Case study and results}
\label{sec:case_study_settings}
This section instantiates the proposed two-stage monitoring design and budgeted CBM framework in an office-zone case study. It first defines the application geometry, luminaire configuration, degradation model calibration, cost assumptions, and baseline simulation settings. It then presents the design-stage reference measurement layout selection, runtime-stage measurement design and budgeted CBM policy optimization, and sensitivity analyses for hitting time uncertainty and monitoring budget.

\subsection{Case description and baseline settings}
\begin{figure}[pos=!htbp]
    \centering
    \begin{subfigure}[b]{0.47\textwidth}
        \centering
        \includegraphics[width=\textwidth,height=4.0cm,keepaspectratio]{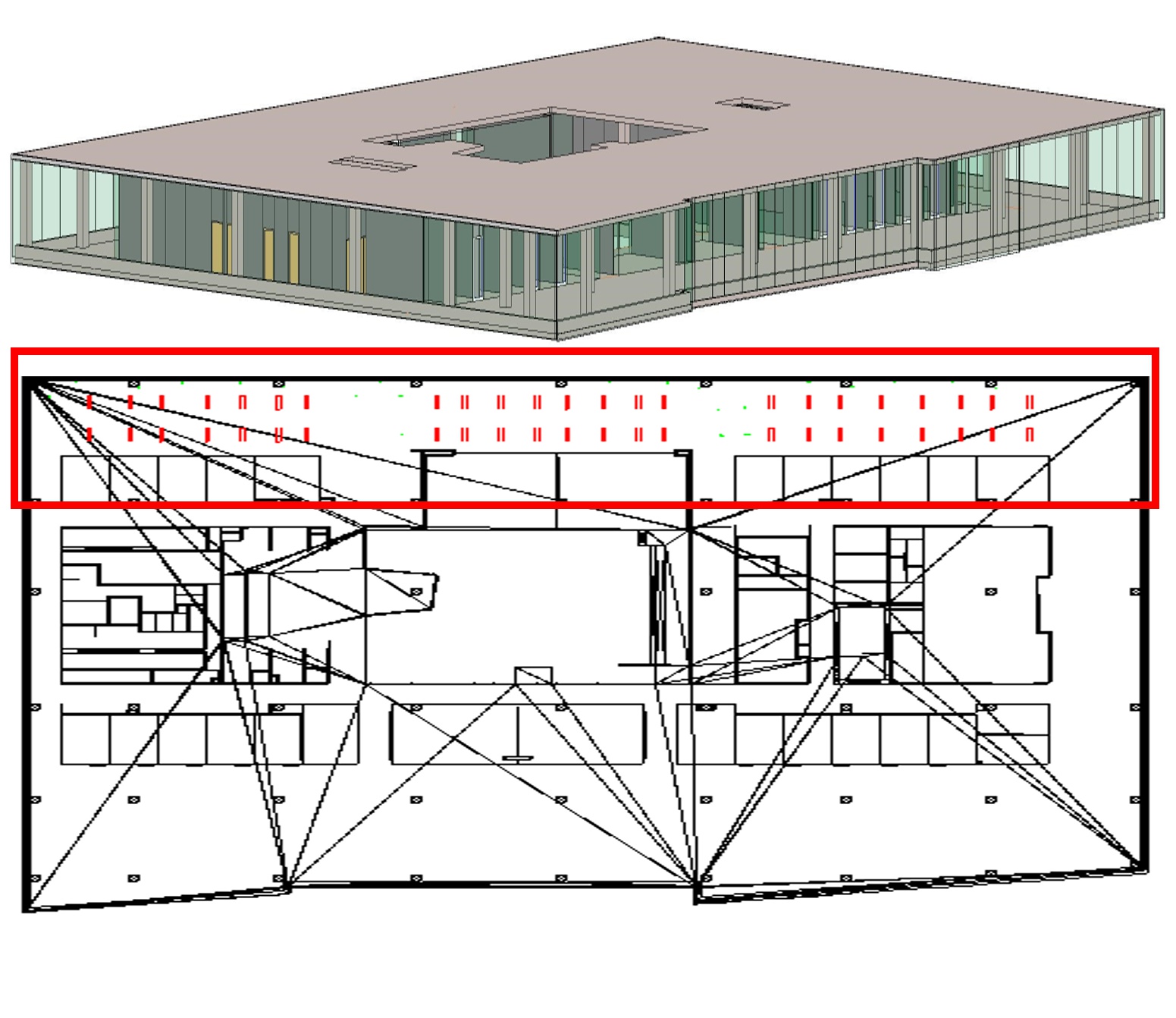}
        \caption{Office level containing Zone~1}
        \textit{Note:} The area in the red box is Zone~1.
        \label{fig:RLDFCS}
    \end{subfigure}
    \hfill
    \begin{subfigure}[b]{0.47\textwidth}
        \centering
        \includegraphics[width=\textwidth,height=4.0cm,keepaspectratio]{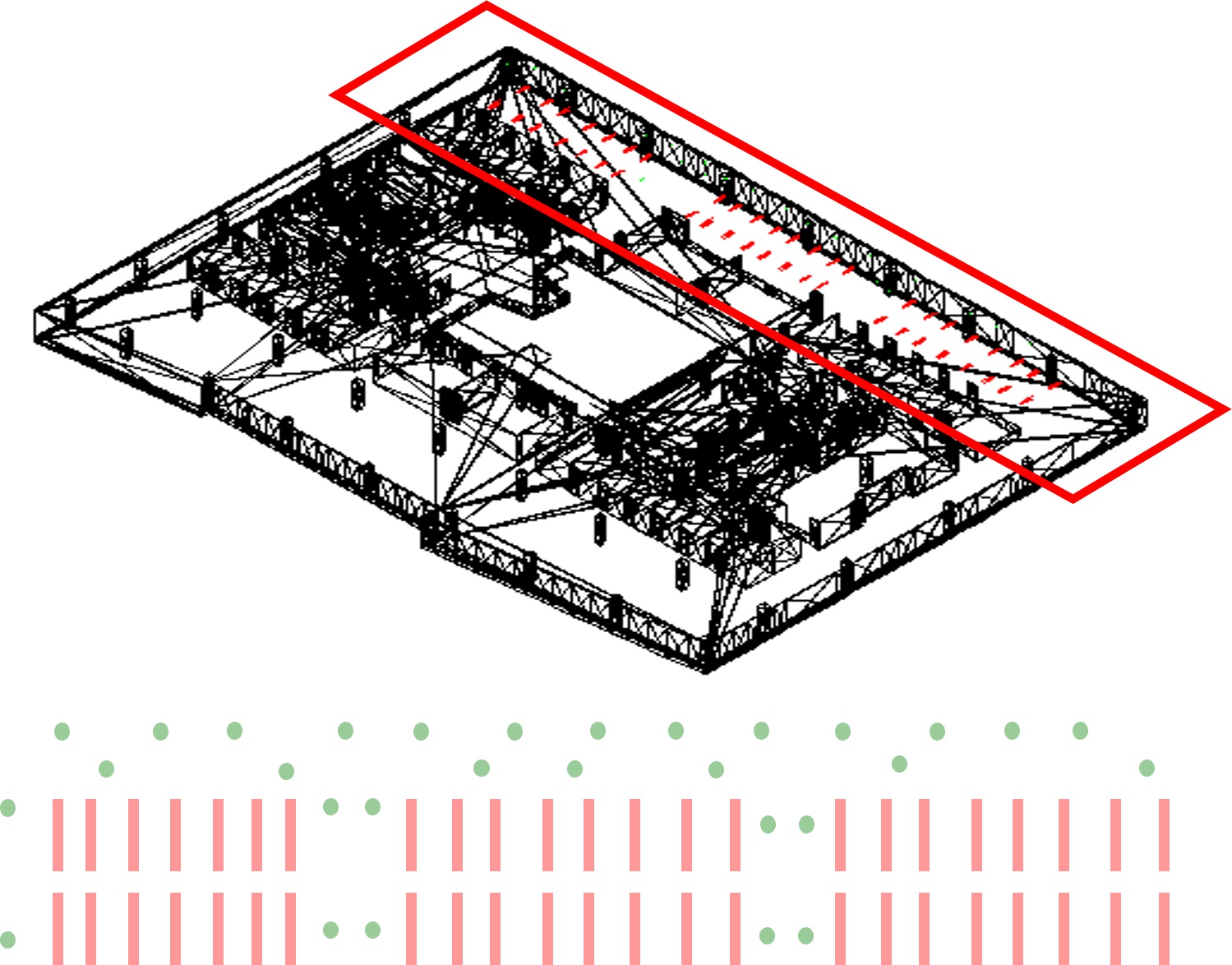}
        \caption{LED lighting system layout}
        \textit{Note:} Red rectangles denote B7 and green circles denote D13 luminaires.
        \label{fig:DLLS}
    \end{subfigure}
    \caption{Application case and LED lighting system layout, reproduced from \citep{shi_gamma_2026}.}
    \label{fig:DACL}
\end{figure}

The proposed framework is evaluated using Zone~1 of a Queensland University of Technology office building. As illustrated in Fig.~\ref{fig:DACL}, this zone supports typical office activities, including writing, reading, and data processing tasks. The room has a floor-plan length of 65.38~m, a width of 6.80~m, and a ceiling height of 3.55~m. For illuminance assessment, the WP is placed 0.80~m above the finished floor and covers a 57.73~m $\times$ 4.80~m rectangular area.

The lighting system comprises $J=76$ LED luminaires, including 46 B7 and 30 D13 units (Fig.~\ref{fig:DLLS}), and is assumed to operate 12~hours per day over a 50-year operational horizon, corresponding to a typical building lifetime. Feasible WP measurement points are defined on a 0.5~m grid, giving 1428 candidate points in total and satisfying the maximum spacing requirement in BS EN~12464-1:2021 \citep{the_british_standards_institution_light_2021}. Together with the Radiance scene settings documented in our previous work \citep{shi_gamma_2026}, this luminaire layout and WP grid define the state-to-measurement data used to construct the linear-Gaussian surrogate observation model. Runtime simulation uses a monthly grid ($\delta=1$ month), with candidate visits allowed at every monthly grid point ($q_{\mathrm{cand}}=1$, so $\Delta=\delta=1$ month), yielding $Q=600$ candidate visit epochs over the horizon. The normalized cost settings follow the visit--measurement and downtime--replacement cost definitions in \autoref{sec:online_tracking}, and the monitoring budget levels correspond to the monitoring budget (MB) scenario labels MB-25, MB-50 (base), and MB-100 used in the runtime analysis. Table~\ref{tab:PFSM} summarizes the main case study settings.

\begin{table}[width=1\linewidth,cols=2,pos=h]
    \caption{Settings specific to the case study used to instantiate the proposed method.}
    \label{tab:PFSM}
    \resizebox{\columnwidth}{!}{%
        \begin{tabular*}{\tblwidth}{@{} LL @{} }
            \toprule
            \textbf{Parameter} & \textbf{Value} \\
            \midrule
            Zone~1 length, width, height (m) & 65.38, 6.80, 3.55 \\
            WP grid spacing (m) and total number of points & 0.5, 1428 \\
            Number of B7, D13 luminaires & 46, 30 \\
            Horizon & $T=50$ years \\
            Simulation grid & $\delta=1$ (month) \\
            Candidate visit grid interval & $q_{\mathrm{cand}}=1$, $\Delta=\delta=1$ month \\
            Downtime accounting interval & $\delta=1$ month \\
            Candidate visit epochs & $Q=600$ \\
            WP candidate grid and reference layout size & $N=1428$, $m=J=76$ \\
            Photometer reading noise standard deviation & $\sigma_{\mathrm{read}}=6.48~\mathrm{lx}$ \\
            Normalized visit and measurement unit costs & $c_{\mathrm{meas}}=1$, $c_{\mathrm{visit}}=50$ \\
            Normalized downtime and preventive replacement unit costs & $c_{\mathrm{DT}}=c_{\mathrm{PM}}=1$ \\
            Monitoring-budget scenario labels (normalized cost units) & MB-25, MB-50 (base), MB-100: $B_{\mathrm{mon}}\in\{6300,12600,25200\}$\\
            \bottomrule
        \end{tabular*}%
    }
\end{table}

Because LM-84-type luminaire-level degradation data are unavailable for the case study luminaires, the degradation-model posterior is calibrated using LM-80 LED-package data and extrapolated to the service reference temperature of $45^\circ\mathrm{C}$, following our previous work \citep{ies_lm84_2020,iesna_approved_2008,shi_gamma_2026}.

\subsection{Design-stage measurement layout selection results}
This subsection reports the design-stage reference measurement layout selection results for the methods described in \autoref{subsec:offline_opt}. The reference layout is constructed with the minimum identifiable measurement count, $m=J=76$, as required by the minimum identifiable configuration stated in Eq.~\eqref{eq:k_ge_J}. Therefore, all candidate layouts are compared at the same measurement count and under the same linear-Gaussian observation model, with performance evaluated using the D-optimal criterion. The compared layouts are the Nadir-point layout, Hybrid greedy search layout, Random-initialized GA layout, and Nadir-seeded GA layout.

\begin{figure}[pos=!htbp]
    \centering
    \begin{subfigure}[b]{0.48\textwidth}
        \centering
        \includegraphics[width=\textwidth]{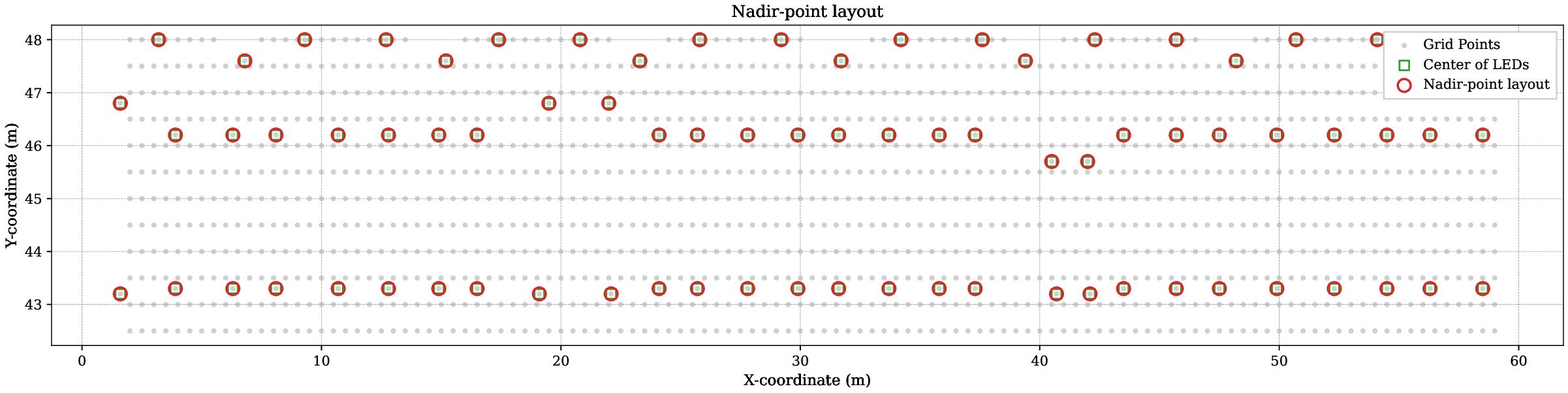}
        \caption{Nadir-point layout}
        \label{fig:deployment_nadir}
    \end{subfigure}
    \hfill
    \begin{subfigure}[b]{0.48\textwidth}
        \centering
        \includegraphics[width=\textwidth]{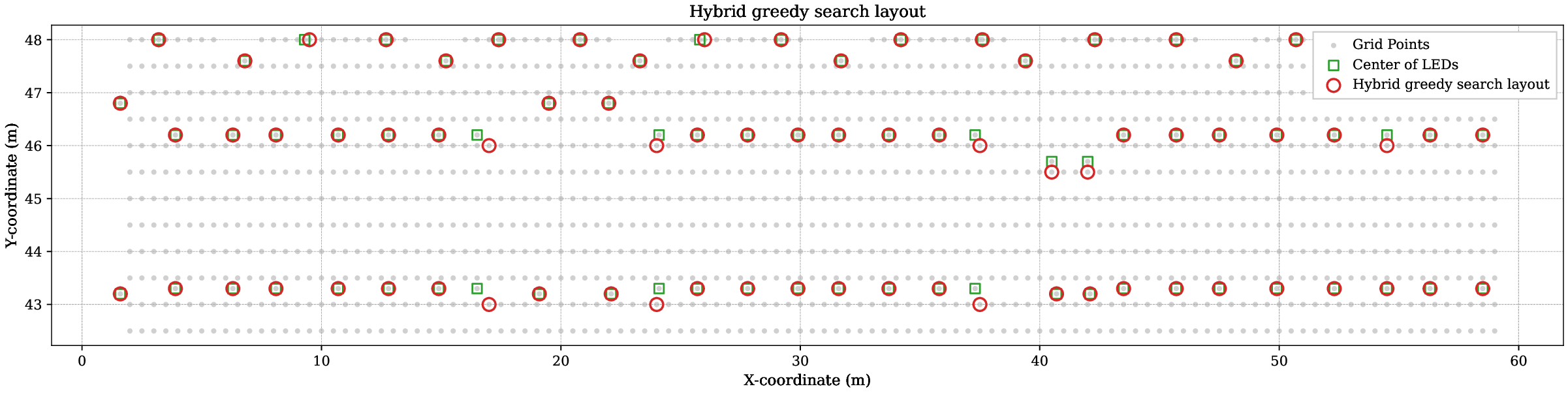}
        \caption{Hybrid greedy search layout}
        \label{fig:deployment_hg}
    \end{subfigure}

    \begin{subfigure}[b]{0.48\textwidth}
        \centering
        \includegraphics[width=\textwidth]{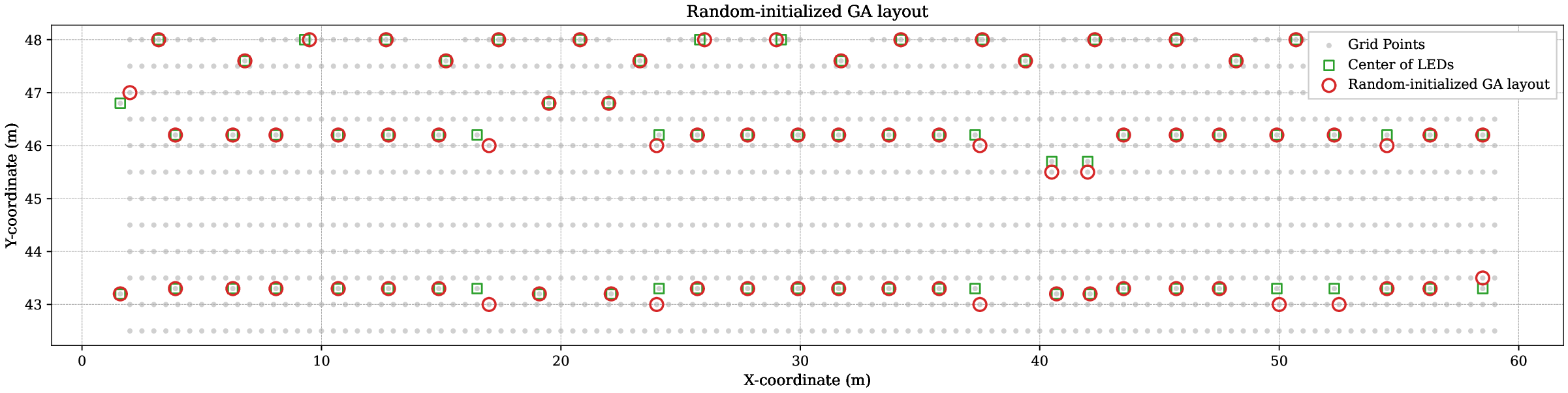}
        \caption{Random-initialized GA layout}
        \label{fig:deployment_ga_random}
    \end{subfigure}
    \hfill
    \begin{subfigure}[b]{0.48\textwidth}
        \centering
        \includegraphics[width=\textwidth]{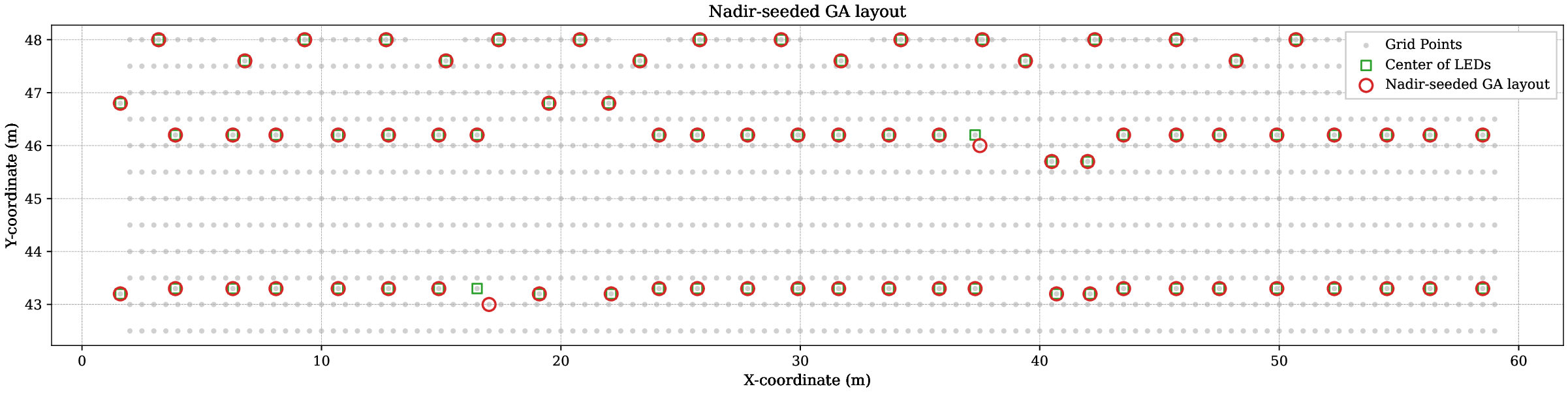}
        \caption{Nadir-seeded GA layout}
        \label{fig:deployment_ga_bench}
    \end{subfigure}
    \caption{Comparison of design-stage measurement layouts for the four candidate methods.}
    \label{fig:deployment_layout_compare}
\end{figure}

The comparative performance is summarized in Table~\ref{tab:deployment_compare}, which reports the D-optimal objective $\log\det(\mathcal{I})$ defined in Eq.~\eqref{eq:dopt_obj_set}. Among the four candidates, the Hybrid greedy search layout achieves the highest value ($562.550461$), indicating the largest Fisher-information determinant and, equivalently, the smallest generalized inverse-estimation uncertainty volume. Fig.~\ref{fig:deployment_layout_compare} provides a qualitative comparison of the resulting measurement point layouts. The Nadir-point layout places measurement points directly beneath luminaire centers, reflecting a local one-to-one benchmark. By contrast, the Hybrid greedy search layout and the two GA layouts distribute selected points more broadly across the working plane. This broader distribution is consistent with the globally coupled observation model, in which each WP measurement reflects contributions from multiple luminaires rather than only the nearest luminaire.

\begin{table}
    \caption{Comparison of D-optimal objective values for design-stage reference layout candidates.}
    \label{tab:deployment_compare}
    \centering
    \begin{tabular}{l c}
        \toprule
        \textbf{Method} & \makecell[c]{$\boldsymbol{\log\det(\mathcal{I})}$\\(higher is better)} \\
        \midrule
        Nadir-point layout & $562.242649$ \\
        Hybrid greedy search layout & $562.550461$ \\
        Random-initialized GA layout & $562.046456$ \\
        Nadir-seeded GA layout & $562.312133$ \\
        \bottomrule
    \end{tabular}
\end{table}

Based on this quantitative comparison, the HG layout is selected as the reference layout $\mathcal{S}_{\mathrm{ref}}$ for subsequent runtime-stage measurement design. This layout contains $m=J=76$ WP measurement points, satisfies the single-epoch identifiability requirement, and defines the fixed reference set from which active reference measurement points are selected during runtime monitoring. In this way, the design-stage D-optimal layout is carried into the budgeted runtime monitoring and CBM optimization while avoiding runtime measurement selection over all $N=1428$ WP grid points.

\subsection{Runtime-stage measurement design and budgeted CBM performance}

\subsubsection{Search setup}
\begin{figure}[pos=!htbp]
    \centering
    \begin{subfigure}[b]{0.48\textwidth}
        \centering
        \includegraphics[width=\textwidth,height=4.0cm,keepaspectratio]{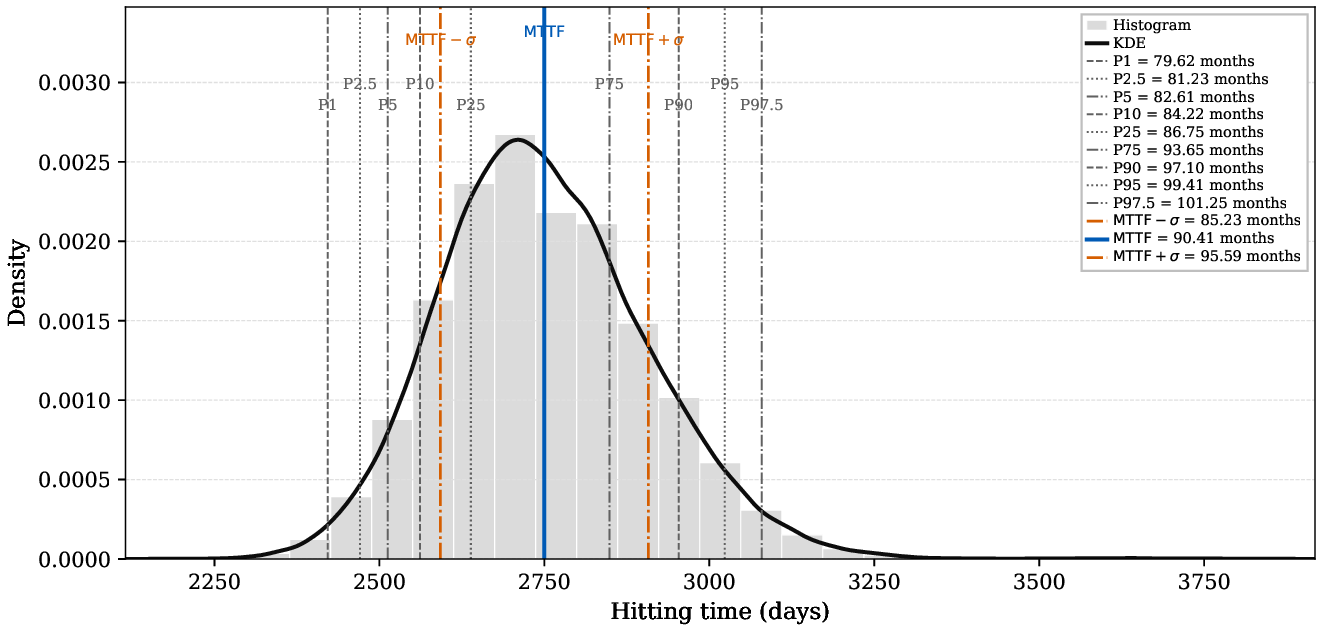}
        \caption{Baseline uncertainty setting}
        \label{fig:baseline_hitting_time_pdf_all_percentiles}
    \end{subfigure}
    \hfill
    \begin{subfigure}[b]{0.48\textwidth}
        \centering
        \includegraphics[width=\textwidth,height=4.0cm,keepaspectratio]{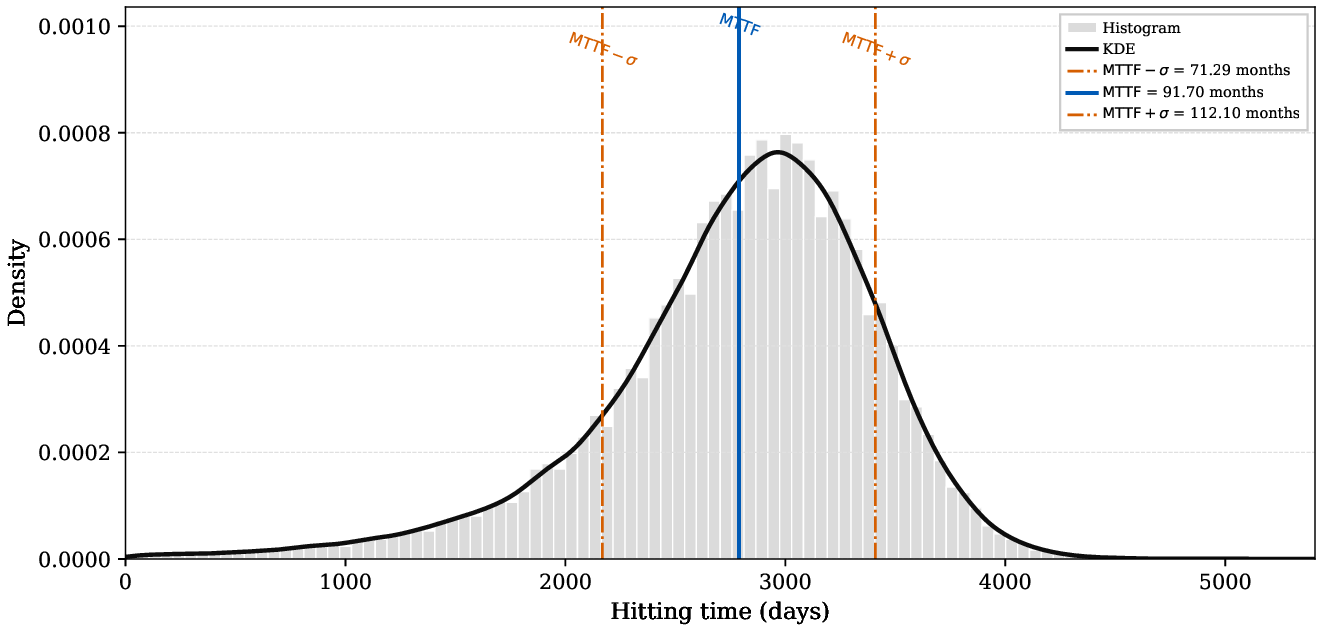}
        \caption{Enlarged uncertainty setting}
        \label{fig:selected_c150_pdf_own_xlim}
    \end{subfigure}
    \caption{Posterior predictive distributions of the luminaire degradation hitting time under the baseline and enlarged hitting-time uncertainty settings. The baseline percentiles are used to inform the search range for the first visit and periodic visit interval grids, while the enlarged distribution supports the uncertainty sensitivity analysis.}
    \label{fig:hitting_time_distribution_comparison}
\end{figure}

Building on the baseline settings in Table~\ref{tab:PFSM}, runtime-stage measurement design and CBM policy optimization are implemented using SGS-PS, as summarized in Algorithm~\ref{alg:seq_grid_runtime} in Appendix~\ref{app:seq_screening_runtime}. The search grids and screening settings are summarized in Table~\ref{tab:runtime_search_setup}. To define physically meaningful ranges for the visit schedule variables, posterior samples of the degradation model parameters are propagated through the gamma process model to estimate the posterior predictive distribution of the first hitting time to the luminaire failure threshold. As shown in Fig.~\ref{fig:baseline_hitting_time_pdf_all_percentiles}, the resulting hitting time distribution is concentrated around 80--100 months under the case study service condition, with the 1st, 50th, and 95th percentiles equal to approximately 80, 90, and 100 months, respectively. Therefore, the first visit grid is defined from 3 months to 102 months using a 3-month step: the lower bound allows early information collection before appreciable degradation, whereas the upper bound is the first quarterly grid point beyond the 95th posterior hitting time percentile. The periodic visit interval grid uses the same range and resolution to cover policies from quarterly monitoring to very sparse periodic visits. For each fixed $(p_{\mathrm{th}},n_{\mathrm{zone}})$ pair, this gives $|\mathcal{T}_{\mathrm{first}}|=|\mathcal{D}|=34$ and hence $|\Theta_0|=34^2=1156$ periodic visit plans. The active reference measurement count grid is $\mathcal{N}_{\mathrm{zone}}=\{4,8,\ldots,76\}$, where $n_{\mathrm{zone}}=76$ corresponds to measuring the full design-stage reference layout at each visit. The risk threshold grid is non-uniform and contains $|\mathcal{P}_{\mathrm{th}}|=52$ candidate values. Thus, the outer grid contains $52\times19=988$ fixed $(p_{\mathrm{th}},n_{\mathrm{zone}})$ combinations, each screened over 1156 periodic visit plans.

\begin{table}[width=1\linewidth,cols=4,pos=h]
    \caption{SGS-PS search grids and screening settings for runtime-stage measurement design and CBM policy optimization.}
    \label{tab:runtime_search_setup}
    \resizebox{\columnwidth}{!}{%
        \begin{tabular*}{\tblwidth}{@{} LLLL @{} }
            \toprule
            \textbf{Item} & \textbf{Value} & \textbf{Step or rule} & \textbf{Count} \\
            \midrule
            First visit time & $\mathcal{T}_{\mathrm{first}}=\{3,6,\ldots,102\}$ months & 3 months & 34 \\
            Periodic visit interval & $\mathcal{D}=\{3,6,\ldots,102\}$ months & 3 months & 34 \\
            Active reference measurement count & $\mathcal{N}_{\mathrm{zone}}=\{4,8,\ldots,76\}$ & 4 reference points & 19 \\
            Risk threshold & \makecell[l]{$\mathcal{P}_{\mathrm{th}}=\{0.001a:a=1,\ldots,10\}$\\$\cup\{0.01a:a=2,\ldots,30\}$\\$\cup\{0.05a:a=7,\ldots,19\}$} & non-uniform & 52 \\
            \makecell[l]{Periodic visit plans per fixed\\$(p_{\mathrm{th}},n_{\mathrm{zone}})$} & $|\Theta_0|=34^2=1156$ & paired screening & 1156 \\
            Fixed outer grid combinations & $|\mathcal{P}_{\mathrm{th}}|\,|\mathcal{N}_{\mathrm{zone}}|=52\times19=988$ & exhaustive outer grid & 988 \\
            EnKF ensemble size & $N_e=1000$ ensemble members & fixed & -- \\
            Screening significance level & $\alpha_{\mathrm{sv}}=0.05$ & one-sided paired screening & -- \\
            MC replication schedule and limit & \makecell[l]{$\{N_{\mathrm{MC},s}\}_{s=1}^{4}=(32,128,512,2048)$,\\$N_{\mathrm{MC},\max}=2048$} & fourfold schedule & 4 stages \\
            Incumbent patience & $N_{\mathrm{patience}}=8$ screening stages & early stopping & -- \\
            \bottomrule
        \end{tabular*}%
    }
\end{table}

Policy values are estimated using the EnKF and SGS-PS settings in Table~\ref{tab:runtime_search_setup}. The Monte Carlo replication schedule follows the standard error scaling $O(N_{\mathrm{MC}}^{-1/2})$: increasing the number of replications by a factor of four approximately halves the standard error of the sample mean. At each SGS-PS stage, the incumbent is the currently best policy, defined as the candidate with the lowest stage mean total downtime and replacement cost.

The visit, measurement, failure, replacement, and budget values in Table~\ref{tab:PFSM} are expressed in normalized cost units. Each measurement at an active reference measurement point is assigned unit cost, $c_{\mathrm{meas}}=1$, and a site visit is assigned $c_{\mathrm{visit}}=50$, corresponding to 50 times the per-point measurement cost. Because absolute monitoring costs can be project-specific, the monitoring budget is interpreted as a normalized visit--measurement budget in this case study. Under the zone cyclic measurement rule, a visit with active reference measurement count $n_{\mathrm{zone}}$ costs $c_{\mathrm{visit}}+n_{\mathrm{zone}}c_{\mathrm{meas}}=50+n_{\mathrm{zone}}$, so each budget level can be converted into a budget-implied visit count bound through Eq.~\eqref{eq:Vmax_zone_aosga}. The base budget setting is $B_{\mathrm{mon}}=12600$; under the full reference measurement count $n_{\mathrm{zone}}=76$, this corresponds to $V_{\max}^{\mathrm{zone}}=\lfloor12600/126\rfloor=100$ visits, as shown in Table~\ref{tab:budget_sensitivity_setup}. This base budget is 50\% of the reference budget defined by the most measurement-intensive periodic plan considered in the runtime-stage measurement design grid, with $t_{\mathrm{first}}=3$ months, $\Delta_{\mathrm{visit}}=3$ months, and $n_{\mathrm{zone}}=76$. This plan executes 200 visits over the 50-year horizon, and each visit costs $c_{\mathrm{visit}}+76c_{\mathrm{meas}}=126$, giving a total monitoring cost of $25200$. The other monitoring budget scenarios are defined as 25\% and 100\% of the same reference budget, as summarized in Table~\ref{tab:budget_sensitivity_setup}.

\begin{table}[width=1\linewidth,cols=4,pos=h]
    \caption{Monitoring budget sensitivity settings.}
    \label{tab:budget_sensitivity_setup}
    \resizebox{\columnwidth}{!}{%
        \begin{tabular*}{\tblwidth}{@{} LLLL @{} }
            \toprule
            \textbf{Scenario ID} & \textbf{Budget level} & \textbf{Monitoring budget} & \textbf{Maximum visits for $n_{\mathrm{zone}}=76$} \\
            \midrule
            MB-25 & 25\% & $6300$ & 50 \\
            MB-50 (base) & 50\% & $12600$ & 100 \\
            MB-100 & 100\% & $25200$ & 200 \\
            \bottomrule
        \end{tabular*}%
    }
\end{table}

The normalized downtime and preventive replacement unit costs are set equal, $c_{\mathrm{DT}}=c_{\mathrm{PM}}=1$, so that $c_{\mathrm{DT}}$ is interpreted as the normalized downtime cost per failed luminaire per simulation interval and the base objective treats failed-luminaire downtime and preventive replacements symmetrically. Because $\delta=1$ month in this case study, downtime cost is accumulated in failed-luminaire months. The base budget is used for the main policy landscape and performance presentation; subsequent analyses examine the effects of active reference measurement count, hitting time uncertainty, and monitoring budget.

Before evaluating the runtime monitoring and CBM policies, the state-tracking behavior of the EnKF was illustrated over a 10-year diagnostic horizon. Fig.~\ref{fig:app_enkf_tracking} in Appendix~\ref{app:enkf_tracking} compares the true degradation trajectories, EnKF estimates, and observation-only estimates for four randomly selected luminaires. The comparison illustrates that the EnKF tracks the degradation states more closely than estimates based only on the observation model.

\subsubsection{Representative policy landscape for $n_{\mathrm{zone}}=16$}
To illustrate the runtime policy landscape before comparing all active reference measurement counts, this subsection fixes $n_{\mathrm{zone}}=16$ under the base monitoring budget MB-50. SGS-PS is applied separately for each active reference measurement count $n_{\mathrm{zone}}$. For a fixed pair $(n_{\mathrm{zone}},p_{\mathrm{th}})$, it screens the candidate periodic visit schedules defined by the first visit time $t_{\mathrm{first}}$ and the periodic visit interval $\Delta_{\mathrm{visit}}$. Therefore, for each value of $p_{\mathrm{th}}$, the output is a retained set of near-optimal visit schedules and their estimated mean total downtime and replacement cost. In some cases, SGS-PS retains multiple schedules for the same $(n_{\mathrm{zone}},p_{\mathrm{th}})$ pair, indicating that these schedules are not statistically distinguishable under the paired screening criterion and the maximum replication limit $N_{\mathrm{MC},\max}=2048$. For plotting and discussion, the retained schedule with the lowest estimated mean total downtime and replacement cost is reported as the representative policy for that $p_{\mathrm{th}}$. For $n_{\mathrm{zone}}=16$, the base budget implies $V_{\max}^{\mathrm{zone}}=\lfloor12600/(50+16)\rfloor=190$ visits. The same procedure is applied to all other values of $n_{\mathrm{zone}}$.
\begin{figure}[pos=!htbp]
    \centering
    \begin{subfigure}[b]{0.49\textwidth}
        \centering
        \includegraphics[width=\textwidth,height=5.8cm,keepaspectratio]{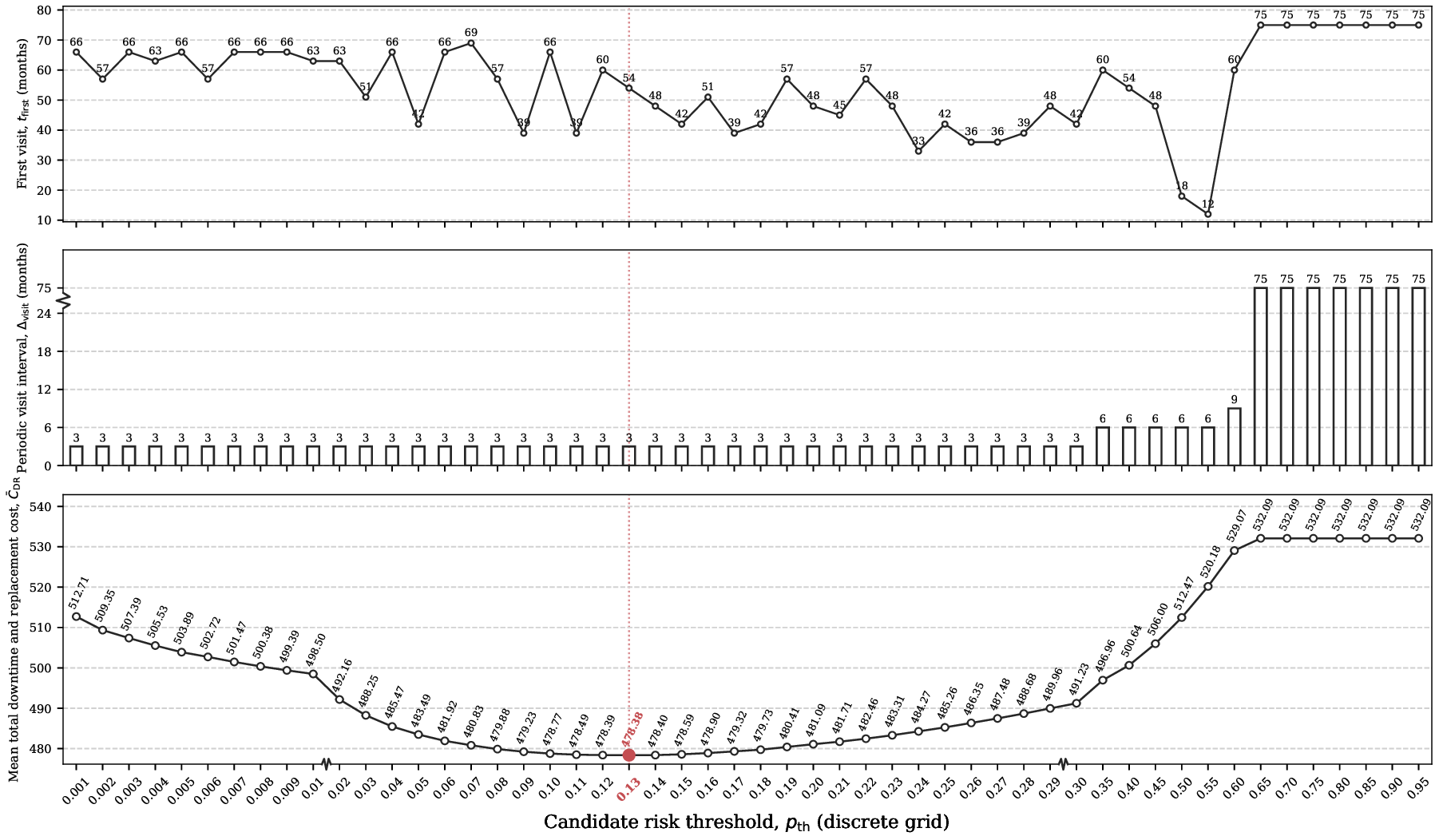}
        \caption{Representative policy for $n_{\mathrm{zone}}=16$}
        \label{fig:winner_policy_vs_thresholds_B12600_nz16}
    \end{subfigure}
    \hfill
    \begin{subfigure}[b]{0.49\textwidth}
        \centering
        \includegraphics[width=\textwidth,height=5.8cm,keepaspectratio]{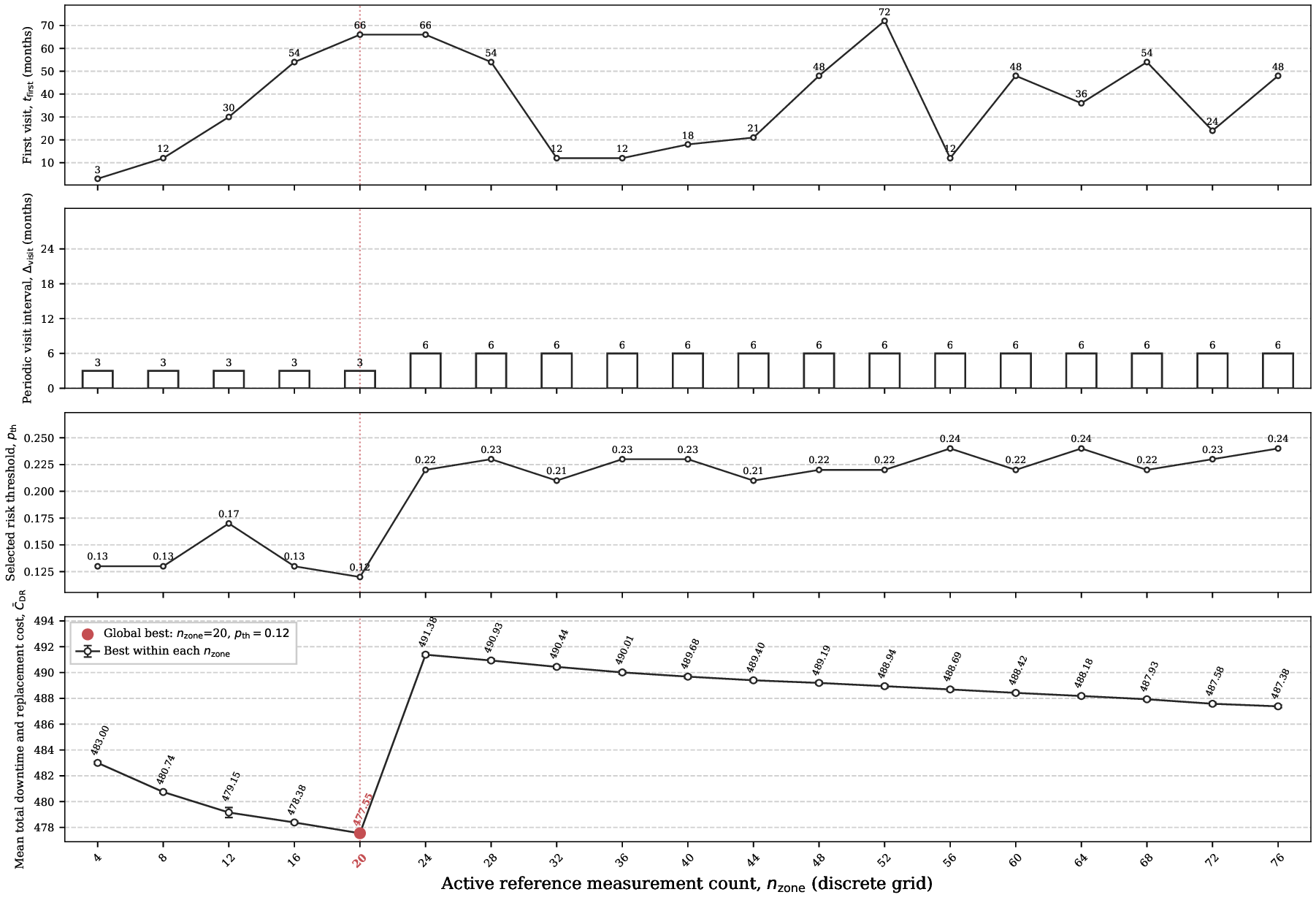}
        \caption{Best retained policies across $n_{\mathrm{zone}}$}
        \label{fig:global_best_across_nz_B12600}
    \end{subfigure}
    \caption{Runtime-stage measurement design and CBM policy results under the baseline hitting-time uncertainty setting and the base monitoring budget MB-50. The panels compare the representative threshold-dependent policy for $n_{\mathrm{zone}}=16$ with the best retained policies across active reference measurement counts.}
    \label{fig:baseline_runtime_policy_summary}
\end{figure}

As shown in Fig.~\ref{fig:winner_policy_vs_thresholds_B12600_nz16}, the three panels share the same risk threshold axis $p_{\mathrm{th}}$. The bottom panel reports the estimated mean total downtime and replacement cost $\bar{C}_{\mathrm{DR}}$; the middle panel shows the representative periodic visit interval $\Delta_{\mathrm{visit}}$; and the top panel shows the representative first visit time $t_{\mathrm{first}}$. For $n_{\mathrm{zone}}=16$ under MB-50, the estimated total downtime and replacement cost decreases from $\bar{C}_{\mathrm{DR}}=512.71$ at $p_{\mathrm{th}}=0.001$ to its minimum at $p_{\mathrm{th}}=0.13$, where $\bar{C}_{\mathrm{DR}}=478.38$. The corresponding representative schedule has $t_{\mathrm{first}}=54$ months and $\Delta_{\mathrm{visit}}=3$ months. The cost then increases as the threshold becomes more conservative and reaches a plateau at high thresholds of approximately $\bar{C}_{\mathrm{DR}}=532.09$ for $p_{\mathrm{th}}\ge0.65$. This trend reflects the threshold-dependent balance between premature preventive replacement and failed-luminaire downtime. When $p_{\mathrm{th}}$ is very small, preventive replacement is triggered easily, producing unnecessary replacement cost. Increasing the threshold initially suppresses these premature replacements and reduces the estimated total downtime and replacement cost. Beyond the optimum, however, replacement becomes harder to trigger, and delayed replacement and failed-luminaire downtime increasingly offset the savings from fewer premature replacements.

The representative periodic visit interval also changes with the risk threshold. The selected representative policy uses the shortest tested interval, $\Delta_{\mathrm{visit}}=3$ months, over the low and intermediate threshold range up to $p_{\mathrm{th}}=0.30$. It then uses $\Delta_{\mathrm{visit}}=6$ months for $p_{\mathrm{th}}=0.35$--$0.55$, $\Delta_{\mathrm{visit}}=9$ months at $p_{\mathrm{th}}=0.60$, and $\Delta_{\mathrm{visit}}=75$ months for $p_{\mathrm{th}}\ge0.65$. Thus, frequent quarterly monitoring is retained around the best risk threshold region, whereas very conservative replacement rules are paired with sparse monitoring because additional visits provide fewer useful intervention opportunities.

The top panel in Fig.~\ref{fig:winner_policy_vs_thresholds_B12600_nz16} shows that the selected first visit time varies less systematically across $p_{\mathrm{th}}$ than the periodic visit interval. This weaker pattern follows from the role of the monitoring budget and from the fact that $t_{\mathrm{first}}$ mainly acts as a phase parameter once feasibility is satisfied. In this study, the budget acts as a feasibility constraint rather than as a requirement to exhaust all available resources. For fixed values of the monitoring budget $B_{\mathrm{mon}}$, the active reference measurement count $n_{\mathrm{zone}}$, and the periodic visit interval $\Delta_{\mathrm{visit}}$, the budget mainly imposes a lower bound on the feasible first visit time $t_{\mathrm{first}}$: shorter periodic visit intervals require a later first visit to keep the total number of visits within the budget. For example, when $\Delta_{\mathrm{visit}}=3$ months, the visit limit $V_{\max}^{\mathrm{zone}}=190$ requires $t_{\mathrm{first}}>30$ months, or at least 33 months on the 3-month grid. Once this feasibility condition is satisfied, later first visit times remain feasible even if part of the monitoring budget is unused. The selected $t_{\mathrm{first}}$ is then determined by its interaction with the degradation hitting time distribution, the periodic visit interval, the finite decision horizon, and simulation variability.

\subsubsection{Preventive replacement and downtime cost components}

Fig.~\ref{fig:F_winner_PM_downtime_cost_components_vs_pth_discrete_B12600_nz16} further decomposes the total downtime and replacement cost trend in Fig.~\ref{fig:winner_policy_vs_thresholds_B12600_nz16} into simulation-diagnostic TP-PM cost, FP-PM cost, and downtime cost, while the overlaid curve reports the estimated mean total downtime and replacement cost $\bar{C}_{\mathrm{DR}}$. This decomposition shows that the minimum is obtained by reducing premature preventive replacement without allowing failed-luminaire downtime to grow too much. At the most aggressive threshold, $p_{\mathrm{th}}=0.001$, $\bar{C}_{\mathrm{DR}}$ is about $512.71$ and is dominated by FP-PM cost, about $485.78$, with TP-PM cost about $26.90$ and negligible downtime cost. At the best representative threshold, $p_{\mathrm{th}}=0.13$, $\bar{C}_{\mathrm{DR}}$ decreases to about $478.38$; FP-PM cost falls to about $248.25$, while TP-PM cost and downtime cost increase to about $222.38$ and $7.76$, respectively.

Within the quarterly monitoring regime, the threshold effect is especially clear because the selected visit interval is fixed. As $p_{\mathrm{th}}$ increases from $0.001$ to $0.30$, FP-PM cost decreases from about $486$ to about $148$, while TP-PM cost increases from about $27$ to about $315$ and downtime cost increases from essentially zero to about $29$. Thus, increasing the risk threshold initially improves performance by suppressing premature replacements. Beyond the minimum, however, the remaining reduction in FP-PM cost is outweighed by the growth in TP-PM and downtime costs, so $\bar{C}_{\mathrm{DR}}$ starts to increase.

The component redistribution is not completely smooth because the representative policy is re-optimized at each risk threshold. The most visible changes occur when the selected periodic visit interval $\Delta_{\mathrm{visit}}$ switches between interval regimes, rather than from a smooth change in $p_{\mathrm{th}}$ alone. Such switches change both the number of decision epochs and the prediction horizon before the next visit used by the PM rule. For example, after the transition from the quarterly regime to a regime with a longer interval around $p_{\mathrm{th}}=0.35$, the policy has higher TP-PM cost and lower FP-PM cost; the components are approximately $401$ for TP-PM cost, $72$ for FP-PM cost, and $24$ for downtime cost. At more conservative thresholds, downtime cost continues to increase, reaching about $50$ at $p_{\mathrm{th}}=0.60$. In the plateau at high thresholds, $p_{\mathrm{th}}\ge0.65$, the selected representative policy produces a clustered preventive replacement pattern: replacements are still triggered by posterior predictive probabilities at visit epochs, but sparse monitoring and long lookahead horizons make many luminaires satisfy the trigger at the same visit. In this region, $\bar{C}_{\mathrm{DR}}$ is about $532.09$ and is almost entirely TP-PM cost, with negligible FP-PM and downtime costs.

\subsubsection{Value of additional measurement information}

\begin{figure}[pos=!htbp]
    \centering
    \begin{subfigure}{0.90\textwidth}
        \centering
        \includegraphics[width=\textwidth,height=4.2cm,keepaspectratio]{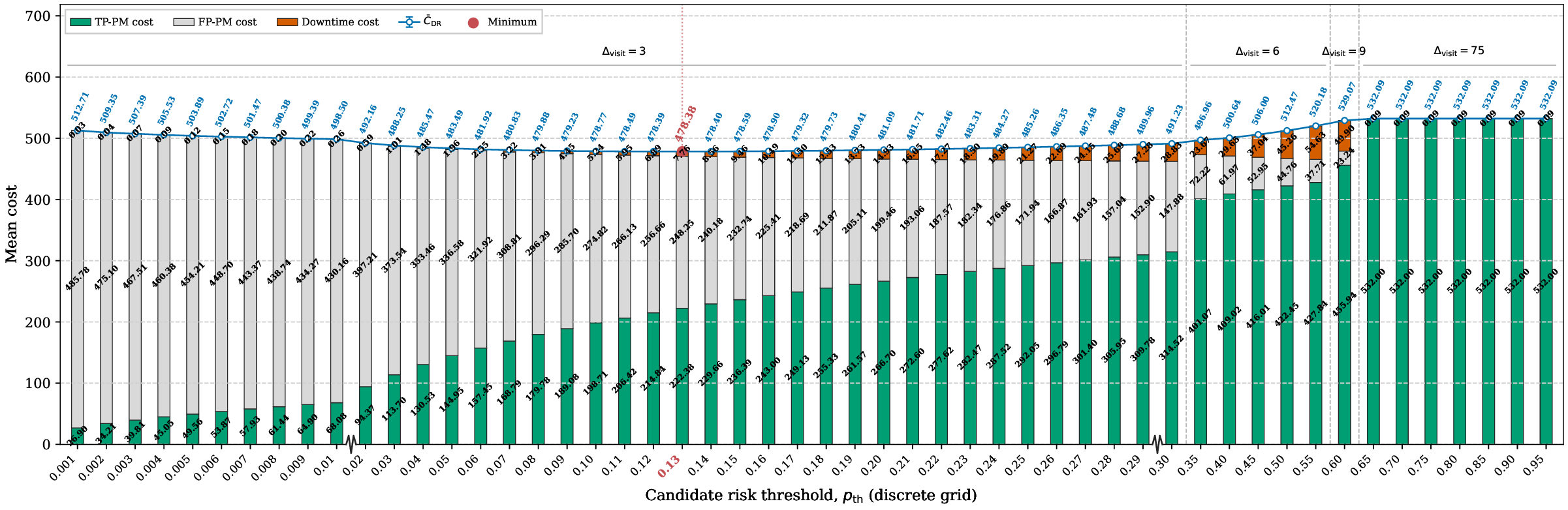}
        \caption{$n_{\mathrm{zone}}=16$}
        \label{fig:F_winner_PM_downtime_cost_components_vs_pth_discrete_B12600_nz16}
    \end{subfigure}

    \vspace{0.6em}

    \begin{subfigure}{0.90\textwidth}
        \centering
        \includegraphics[width=\textwidth,height=4.2cm,keepaspectratio]{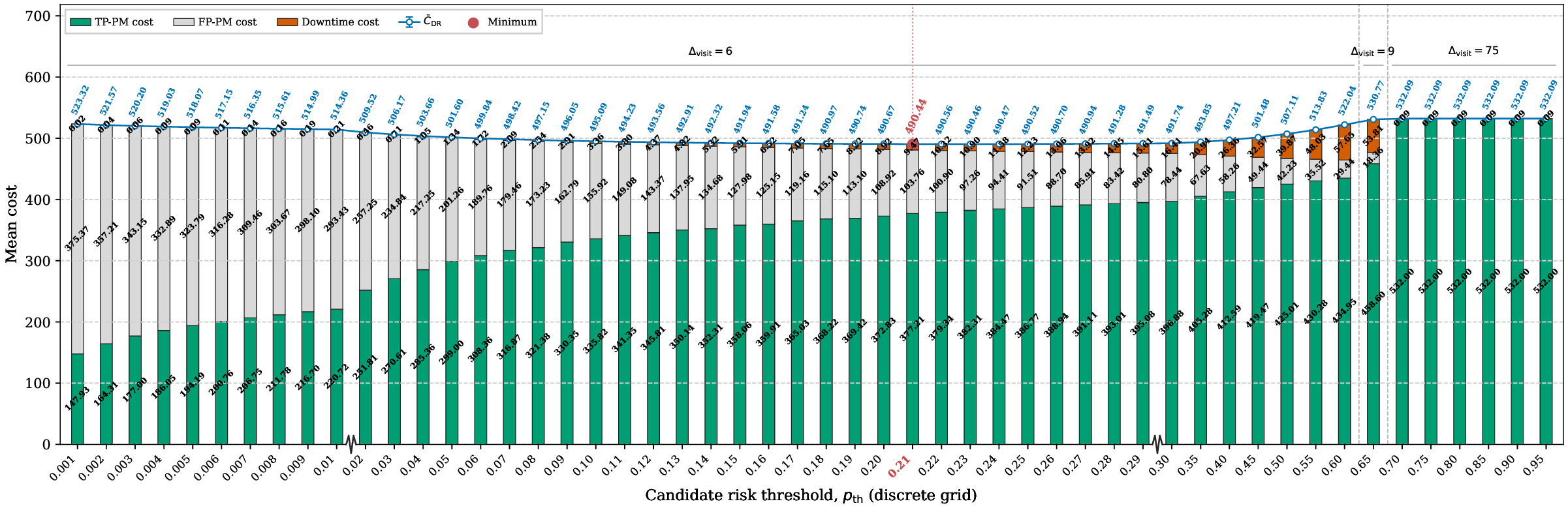}
        \caption{$n_{\mathrm{zone}}=32$}
        \label{fig:F_winner_PM_downtime_cost_components_vs_pth_discrete_B12600_nz32}
    \end{subfigure}
    \caption{Downtime and replacement cost components of the representative runtime policies under the base monitoring budget MB-50, comparing active reference measurement counts $n_{\mathrm{zone}}=16$ and $n_{\mathrm{zone}}=32$.}
    \label{fig:F_winner_PM_downtime_cost_components_vs_pth_discrete_B12600_nz16_nz32}
\end{figure}

The preceding cost component analysis focuses on the threshold-dependent tradeoff for a fixed active reference measurement count, $n_{\mathrm{zone}}=16$. To assess the role of measurement information in the resulting CBM policy, this subsection compares that baseline with $n_{\mathrm{zone}}=32$ under the same base monitoring budget. Since $n_{\mathrm{zone}}$ determines the number of active reference measurement points assimilated at each visit, increasing it changes the information available to the EnKF update and, consequently, the posterior predictive failure probabilities entering the PM rule. Fig.~\ref{fig:F_winner_PM_downtime_cost_components_vs_pth_discrete_B12600_nz32} presents the corresponding diagnostic cost component decomposition, which is used to examine whether additional spatial measurement information is associated with fewer premature FP-PM decisions and less failed-luminaire downtime.

Comparing Figs.~\ref{fig:F_winner_PM_downtime_cost_components_vs_pth_discrete_B12600_nz16} and~\ref{fig:F_winner_PM_downtime_cost_components_vs_pth_discrete_B12600_nz32} illustrates the value of additional active reference measurement points, although the comparison must be interpreted together with the selected visit schedule because the monitoring budget couples spatial information with visit frequency. The comparison is clearest where the retained representative schedules use the same six-month visit interval. At $p_{\mathrm{th}}=0.35$, increasing the active reference measurement count from $n_{\mathrm{zone}}=16$ to $n_{\mathrm{zone}}=32$ reduces $\bar{C}_{\mathrm{DR}}$ from about 496.96 to about 493.85. This reduction is associated with decreases in FP-PM cost, from about 72.22 to 67.63, and downtime cost, from about 23.67 to 20.94, while TP-PM cost increases slightly from about 401.07 to 405.28. A similar pattern appears at $p_{\mathrm{th}}=0.45$, where $\bar{C}_{\mathrm{DR}}$ decreases from about 506.00 to about 501.48, with lower FP-PM and downtime costs and a slightly higher TP-PM cost. When the visit frequency is comparable, this pattern is consistent with additional measurement information improving state estimation and shifting replacement decisions away from premature FP-PM and failed-luminaire downtime toward more timely TP-PM.

The $n_{\mathrm{zone}}=32$ cost component landscape also preserves the same threshold tradeoff observed for $n_{\mathrm{zone}}=16$. In Fig.~\ref{fig:F_winner_PM_downtime_cost_components_vs_pth_discrete_B12600_nz32}, increasing $p_{\mathrm{th}}$ initially reduces FP-PM cost, and $\bar{C}_{\mathrm{DR}}$ decreases from about 523.32 at $p_{\mathrm{th}}=0.001$ to about 490.44 at $p_{\mathrm{th}}=0.21$. Beyond this region, downtime cost grows as replacement becomes more conservative, and $\bar{C}_{\mathrm{DR}}$ increases. In the plateau at high thresholds, $p_{\mathrm{th}}\ge0.70$, the same clustered preventive replacement outcome appears, with $\bar{C}_{\mathrm{DR}}$ about 532.09 and almost entirely attributable to TP-PM cost. Thus, additional measurement information is beneficial when policies have similar visit frequencies, but under the fixed monitoring budget it does not by itself dominate the effect of changing the visit interval; this budget tradeoff is summarized across all $n_{\mathrm{zone}}$ in the next subsection.

\subsubsection{Best active reference measurement count under the monitoring budget}
The preceding analyses examine threshold-dependent behavior for selected active reference measurement counts. The base budget results are now aggregated over all $n_{\mathrm{zone}}\in\mathcal{N}_{\mathrm{zone}}$ to identify the active reference measurement count that gives the lowest estimated total downtime and replacement cost. For each fixed $n_{\mathrm{zone}}$, the retained feasible policies are first compared over the selected $t_{\mathrm{first}}$, $\Delta_{\mathrm{visit}}$, and $p_{\mathrm{th}}$ values obtained from SGS-PS. The policy with the lowest estimated mean total downtime and replacement cost is then reported as the best retained policy for that active reference measurement count. Comparing these best retained policies across all $n_{\mathrm{zone}}$ values gives the best active reference measurement count under the base monitoring budget.

Fig.~\ref{fig:global_best_across_nz_B12600} compares the best retained policy for each active reference measurement count after selecting $t_{\mathrm{first}}$, $\Delta_{\mathrm{visit}}$, and $p_{\mathrm{th}}$ for that fixed $n_{\mathrm{zone}}$. The main cost pattern is shown by the bottom panel. Within the $\Delta_{\mathrm{visit}}=3$ months regime, increasing $n_{\mathrm{zone}}$ from 4 to 20 reduces the estimated total downtime and replacement cost from about 483.00 to the overall minimum of about 477.55 at $n_{\mathrm{zone}}=20$ and $p_{\mathrm{th}}=0.12$. This reduction is consistent with the benefit of additional active reference measurement information when the visit frequency is unchanged. When $n_{\mathrm{zone}}$ increases from 20 to 24, the selected interval changes from $\Delta_{\mathrm{visit}}=3$ to $\Delta_{\mathrm{visit}}=6$ months, and the estimated total downtime and replacement cost increases to about 491.38. This discontinuity reflects the budget-induced tradeoff between spatial information and temporal update frequency: because each visit costs $50+n_{\mathrm{zone}}$, increasing the active reference measurement count reduces the number of affordable visits and makes quarterly monitoring less favorable under the fixed budget.

Within the $\Delta_{\mathrm{visit}}=6$ months regime, the estimated total downtime and replacement cost decreases gradually from about 491.38 at $n_{\mathrm{zone}}=24$ to about 487.38 at $n_{\mathrm{zone}}=76$. However, this improvement remains above the best quarterly policy, indicating that the additional spatial information only partly compensates for the lower update frequency. Thus, under MB-50, the best active reference measurement count is the intermediate value $n_{\mathrm{zone}}=20$, rather than the smallest or largest tested value. The selected policies indicate two regimes: a quarterly monitoring regime in which additional active reference measurement points improve the retained policy while temporal update frequency is preserved, and a six-month monitoring regime in which the marginal value of additional spatial information is limited by less frequent state updates. The selected $t_{\mathrm{first}}$ varies irregularly because it mainly acts as a phase parameter of the periodic schedule. The selected risk threshold is generally lower in the quarterly regime, about 0.12--0.17, and higher in the six-month regime, about 0.21--0.24, but it fluctuates within each regime rather than following a monotone trend.

\subsection{Sensitivity to hitting time uncertainty}

Under the case study service condition, the resulting posterior predictive hitting time distribution is relatively concentrated, with a standard deviation of 5.18 months, as shown in Fig.~\ref{fig:baseline_hitting_time_pdf_all_percentiles}. This narrow distribution is consistent with the controlled laboratory conditions under which LM-80 data are collected, whereas field operation can involve additional uncertainty from environmental variation, installation conditions, usage patterns, and unit-to-unit heterogeneity. This sensitivity analysis therefore examines how the optimized runtime-stage measurement design and CBM policy respond when the hitting time distribution is made wider. The hitting time standard deviation is increased to approximately four times its baseline value, from 5.18 months to 20.41 months, while approximately preserving the mean hitting time, which changes only slightly from 90.41 months to 91.07 months. The resulting wider distribution is shown in Fig.~\ref{fig:selected_c150_pdf_own_xlim}. The analysis is conducted under the base monitoring budget MB-50, so that the effect of hitting time uncertainty is separated from the monitoring budget sensitivity analysis.

\begin{figure}[pos=!htbp]
    \centering
    \begin{subfigure}[b]{0.49\textwidth}
        \centering
        \includegraphics[width=\textwidth,height=5.8cm,keepaspectratio]{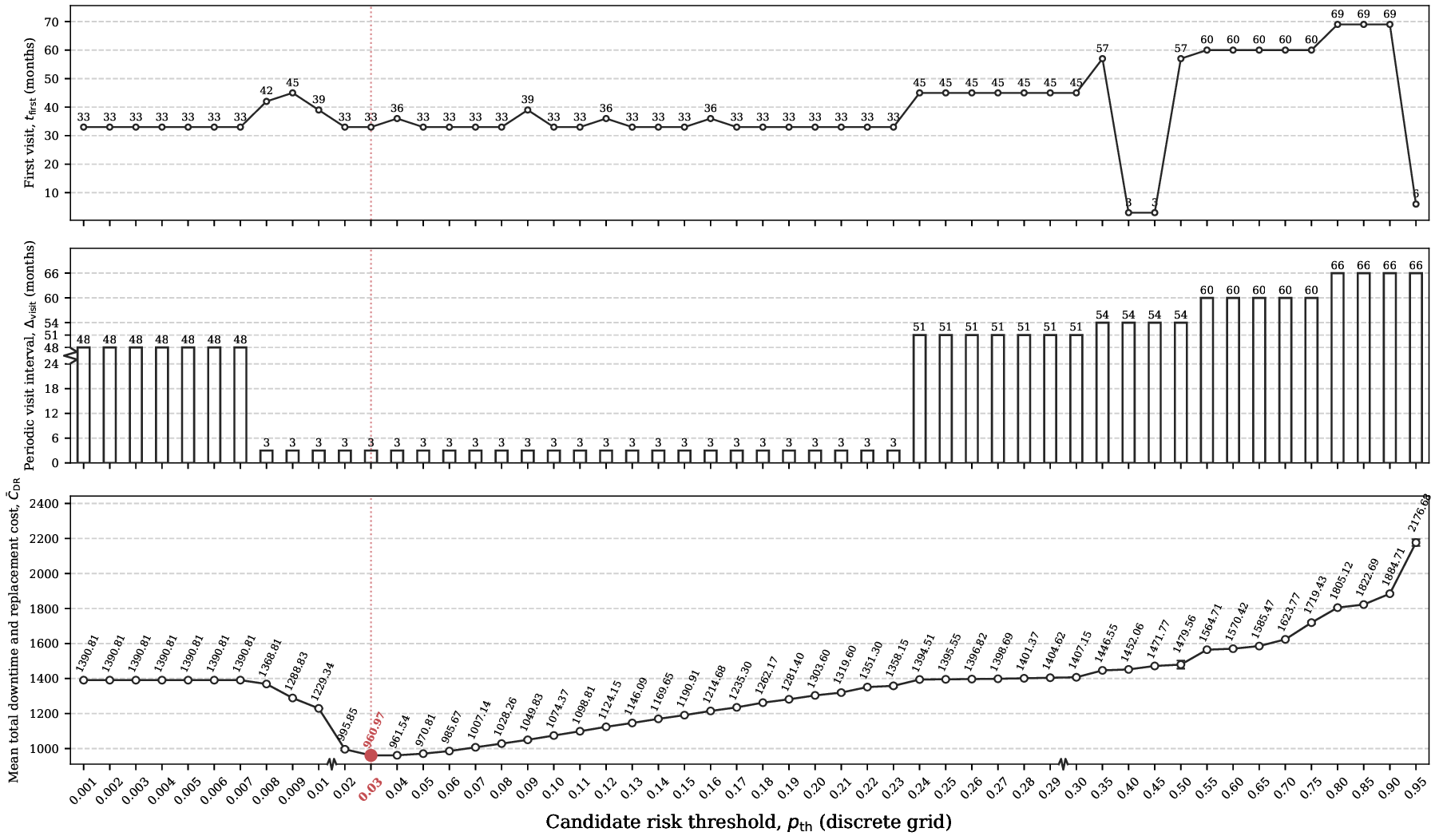}
        \caption{Representative policy for $n_{\mathrm{zone}}=16$}
        \label{fig:winner_policy_vs_thresholds_B12600_c150_nz16}
    \end{subfigure}
    \hfill
    \begin{subfigure}[b]{0.49\textwidth}
        \centering
        \includegraphics[width=\textwidth,height=5.8cm,keepaspectratio]{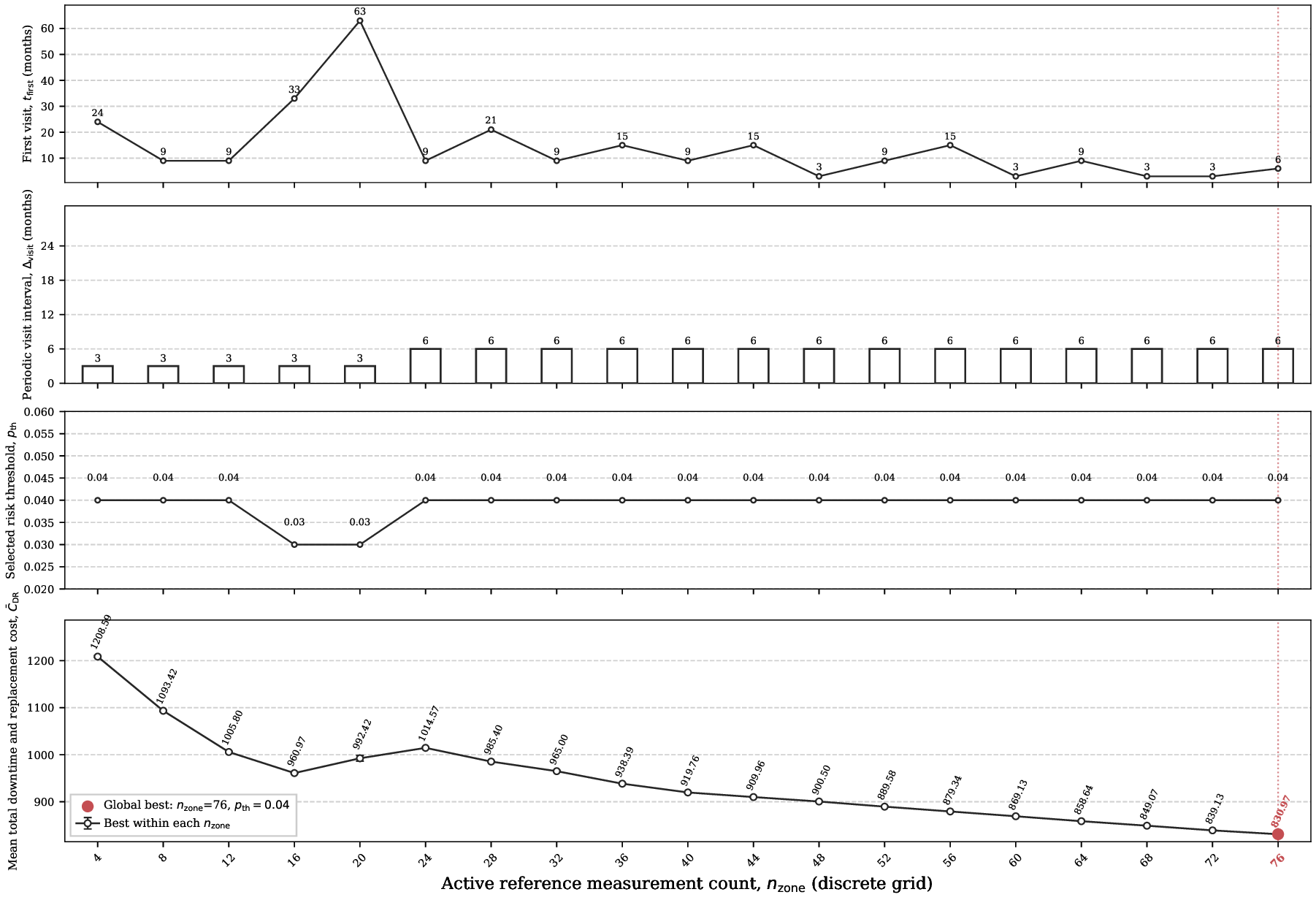}
        \caption{Best retained policies across $n_{\mathrm{zone}}$}
        \label{fig:global_best_across_nz_B12600_c150}
    \end{subfigure}
    \caption{Runtime-stage measurement design and CBM policy results under enlarged hitting-time uncertainty and the base monitoring budget MB-50. The panels compare the representative threshold-dependent policy for $n_{\mathrm{zone}}=16$ with the best retained policies across active reference measurement counts.}
    \label{fig:enlarged_uncertainty_runtime_policy_summary}
\end{figure}

Fig.~\ref{fig:winner_policy_vs_thresholds_B12600_c150_nz16} shows that the representative policy landscape for $n_{\mathrm{zone}}=16$ changes substantially when the hitting time uncertainty is enlarged. Compared with the baseline result in Fig.~\ref{fig:winner_policy_vs_thresholds_B12600_nz16}, the best representative threshold shifts from $p_{\mathrm{th}}=0.13$ to a more aggressive value of $p_{\mathrm{th}}=0.03$. The corresponding estimated total downtime and replacement cost increases markedly, from about 478.38 to about 960.97. Very small thresholds, $p_{\mathrm{th}}=0.001$--0.008, give much higher costs of about 1368.81--1390.81, indicating excessive preventive replacement under the wider hitting time distribution. As the threshold increases above the optimum, the cost rises again, reaching about 2176.68 at $p_{\mathrm{th}}=0.95$. Because the mean hitting time is nearly unchanged in this sensitivity setting, the higher cost mainly reflects the greater difficulty of timing preventive replacements when failure time predictions are less concentrated, rather than faster degradation.

The selected visit schedule also differs from the baseline case. In Fig.~\ref{fig:winner_policy_vs_thresholds_B12600_nz16}, the baseline representative policies use $\Delta_{\mathrm{visit}}=3$ months near the optimum and select longer intervals mainly at higher risk thresholds. Under enlarged uncertainty, the selected interval becomes more strongly threshold dependent. Very small thresholds, $p_{\mathrm{th}}\le0.007$, select a long interval of $\Delta_{\mathrm{visit}}=48$ months; the policy then switches to quarterly monitoring for $p_{\mathrm{th}}=0.008$--0.23, including the threshold with the minimum cost at $p_{\mathrm{th}}=0.03$; and more conservative thresholds select sparse intervals, with $\Delta_{\mathrm{visit}}=51$ months for $p_{\mathrm{th}}=0.24$--0.30 and stepwise increases to 54--66 months for $p_{\mathrm{th}}\ge0.35$. This pattern indicates a stronger interaction between the risk threshold and the visit schedule when failure time prediction is less certain. The enlarged uncertainty therefore changes not only the cost scale, but also the structure of the representative policy landscape.

Fig.~\ref{fig:global_best_across_nz_B12600_c150} summarizes the best retained policy for each active reference measurement count under enlarged hitting time uncertainty. Compared with the baseline result in Fig.~\ref{fig:global_best_across_nz_B12600}, the selected periodic visit intervals follow a similar two-regime structure: the best policies use $\Delta_{\mathrm{visit}}=3$ months for smaller active reference measurement counts and $\Delta_{\mathrm{visit}}=6$ months for larger active reference measurement counts. The selected first visit time remains irregular, again reflecting its role as a phase parameter of the periodic schedule rather than a monotone decision variable. The selected risk thresholds are lower than in the baseline case, with $p_{\mathrm{th}}=0.04$ for most active reference measurement counts and $p_{\mathrm{th}}=0.03$ at $n_{\mathrm{zone}}=16$ and 20, indicating that a more aggressive replacement trigger is preferred when failure time prediction is less concentrated.

The most visible effect of enlarged hitting time uncertainty is the substantial increase in the estimated mean total downtime and replacement cost. For example, at $n_{\mathrm{zone}}=20$, the estimated total downtime and replacement cost increases from about 477.55 in the baseline setting to about 992.42 under enlarged uncertainty. At $n_{\mathrm{zone}}=76$, the corresponding cost increases from about 487.38 to about 830.97. Under the setting with enlarged uncertainty, the best retained policy occurs at $n_{\mathrm{zone}}=76$, with $p_{\mathrm{th}}=0.04$, $\Delta_{\mathrm{visit}}=6$ months, and an estimated total downtime and replacement cost of about 830.97. This differs from the baseline optimum at $n_{\mathrm{zone}}=20$, $p_{\mathrm{th}}=0.12$, and $\Delta_{\mathrm{visit}}=3$ months. The contrast reflects a shift in the relative value of temporal and spatial information. In the baseline setting, where hitting time uncertainty is small, the EnKF predictions are sufficiently informative that preserving frequent quarterly updates is more valuable than further increasing the active reference measurement count once doing so forces a six-month visit interval; although the cost decreases again as $n_{\mathrm{zone}}$ increases within the six-month regime, it does not recover the best quarterly policy. Under enlarged uncertainty, the filter has less predictive concentration between visits, so additional active reference measurement points provide more useful correction information. Consequently, the total cost generally decreases as $n_{\mathrm{zone}}$ increases, apart from local changes caused by the jointly selected $t_{\mathrm{first}}$ and $\Delta_{\mathrm{visit}}$. The cost component diagnostics in Figs.~\ref{fig:app_global_best_components_B12600} and~\ref{fig:app_global_best_components_B12600_c150} in Appendix~\ref{app:nzone_component_diagnostics} support this interpretation by showing that the larger active reference measurement counts reduce failed-luminaire downtime and FP-PM cost under enlarged uncertainty.

\subsubsection{Block-replacement benchmark comparison}

For comparison, a block-replacement benchmark is evaluated under the same two hitting time uncertainty settings. The benchmark uses the same periodic timing grid but does not assimilate WP measurements, does not update latent luminaire degradation states using the EnKF, and does not apply a risk threshold. It therefore represents a purely time-based policy characterized by a periodic replacement interval. Detailed cost component curves for the benchmark are provided in Fig.~\ref{fig:app_block_replacement_components} in Appendix~\ref{app:block_replacement_diagnostics}; the main comparison is summarized in Table~\ref{tab:block_replacement_comparison}.

\begin{table}[width=1\linewidth,cols=6,pos=h]
    \caption{Comparison between the block-replacement benchmark and the best retained budgeted CBM policy under the base monitoring budget MB-50.}
    \label{tab:block_replacement_comparison}
    \resizebox{\columnwidth}{!}{%
        \begin{tabular*}{\tblwidth}{@{} LLLLLL @{} }
            \toprule
            \textbf{Hitting time uncertainty} & \textbf{Block replacement interval} & \textbf{Block cost} & \textbf{CBM policy} & \textbf{CBM cost} & \textbf{Cost reduction} \\
            \midrule
            Baseline & 75 months & 532.00 & \makecell[l]{$t_{\mathrm{first}}=66$ months\\$\Delta_{\mathrm{visit}}=3$ months\\$n_{\mathrm{zone}}=20$\\$p_{\mathrm{th}}=0.12$} & 477.55 & 10.2\% \\
            Enlarged & 51 months & 1395.42 & \makecell[l]{$t_{\mathrm{first}}=6$ months\\$\Delta_{\mathrm{visit}}=6$ months\\$n_{\mathrm{zone}}=76$\\$p_{\mathrm{th}}=0.04$} & 830.97 & 40.5\% \\
            \bottomrule
        \end{tabular*}%
    }
\end{table}

Table~\ref{tab:block_replacement_comparison} shows that the proposed budgeted CBM framework improves on the block-replacement benchmark in both uncertainty settings, reducing the estimated total downtime and replacement cost by about 10.2\% in the baseline setting and about 40.5\% under enlarged hitting time uncertainty. The improvement arises from the joint policy optimization: the visit schedule, active reference measurement count, and risk threshold are selected together under the monitoring budget, allowing measurement information and replacement timing to adapt to the uncertainty level.

\subsection{Sensitivity to monitoring budget}

This subsection returns to the baseline hitting time uncertainty setting and examines the effect of changing the monitoring budget. The base budget result in Fig.~\ref{fig:global_best_across_nz_B12600} is used as the reference. Two additional budget levels are considered: the tighter MB-25 setting with $B_{\mathrm{mon}}=6300$ and the full reference MB-100 setting with $B_{\mathrm{mon}}=25200$. Because the budget-implied visit count bound in Eq.~\eqref{eq:Vmax_zone_aosga} depends on $n_{\mathrm{zone}}$, changing the monitoring budget affects the feasible tradeoff between visit frequency and the number of active reference measurement points per visit. The sensitivity analysis therefore compares the best retained policy for each $n_{\mathrm{zone}}$, rather than a fixed-$n_{\mathrm{zone}}$ threshold landscape.

\begin{figure}[pos=!htbp]
    \centering
    \begin{subfigure}[b]{0.49\textwidth}
        \centering
        \includegraphics[width=\textwidth,height=6.0cm,keepaspectratio]{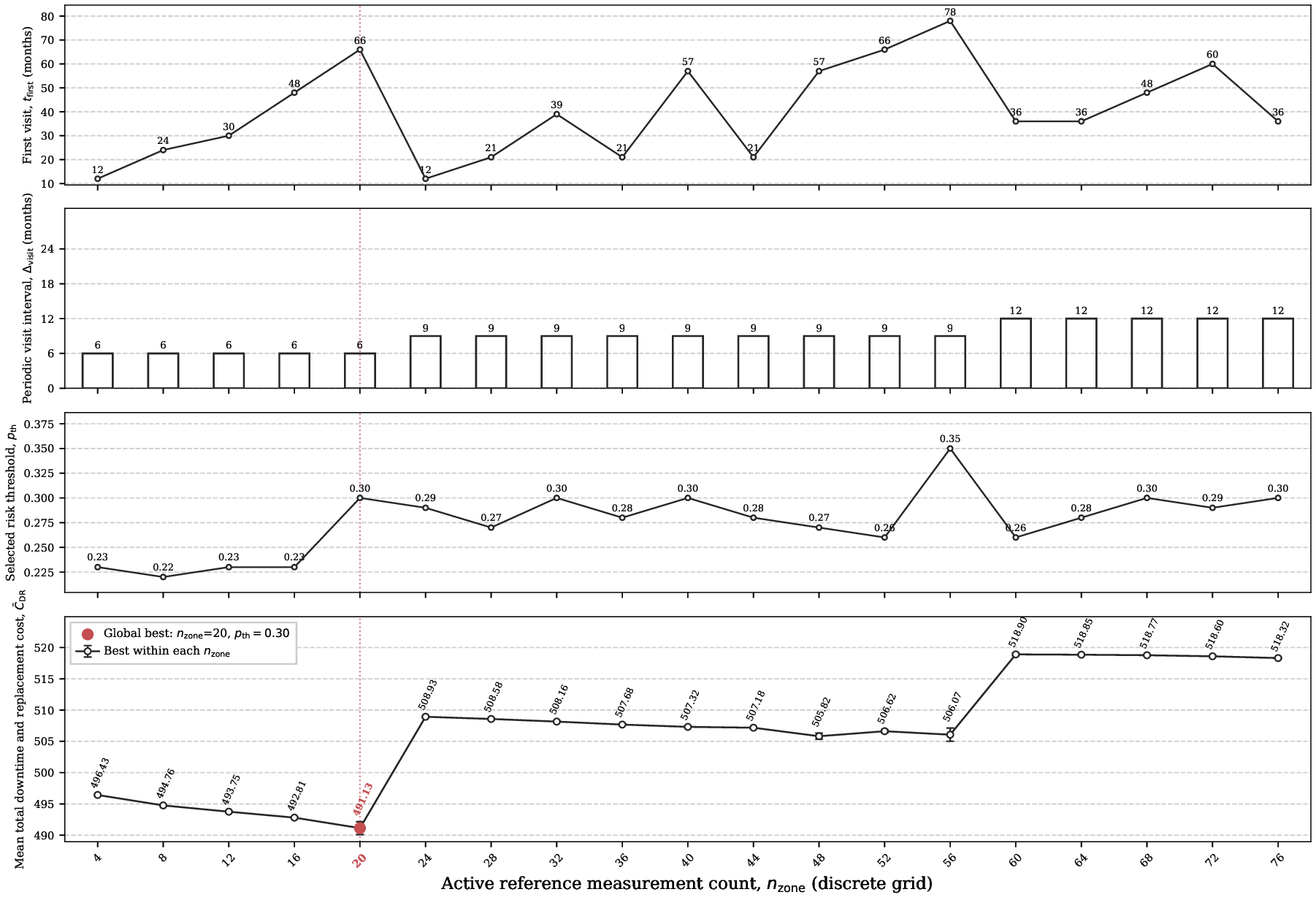}
        \caption{MB-25}
        \label{fig:global_best_across_nz_B6300}
    \end{subfigure}
    \hfill
    \begin{subfigure}[b]{0.49\textwidth}
        \centering
        \includegraphics[width=\textwidth,height=6.0cm,keepaspectratio]{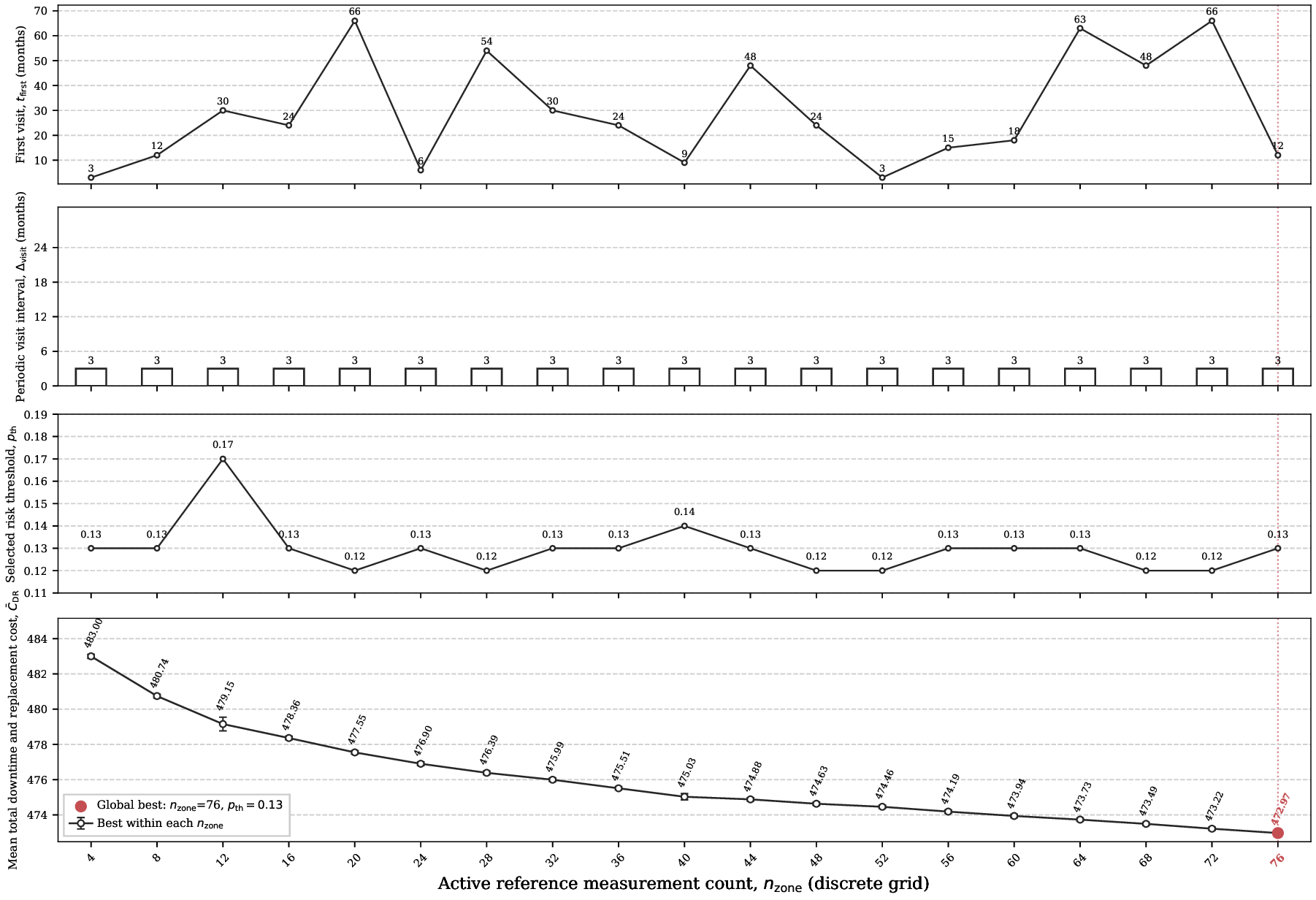}
        \caption{MB-100}
        \label{fig:global_best_across_nz_B25200}
    \end{subfigure}
    \caption{Best retained runtime-stage measurement design and CBM policy for each active reference measurement count $n_{\mathrm{zone}}$ under the tighter MB-25 and full-reference MB-100 monitoring budgets. Each panel group shows the selected first visit time, periodic visit interval, risk threshold, and estimated mean total downtime and replacement cost.}
    \label{fig:global_best_across_nz_budget_sensitivity}
\end{figure}

Fig.~\ref{fig:global_best_across_nz_B6300} shows that the tighter budget shifts the best retained policies toward longer periodic visit intervals. For smaller active reference measurement counts, $n_{\mathrm{zone}}=4$--20, the selected policies use $\Delta_{\mathrm{visit}}=6$ months, whereas the corresponding best policies under MB-50 use $\Delta_{\mathrm{visit}}=3$ months up to $n_{\mathrm{zone}}=20$. As $n_{\mathrm{zone}}$ increases, the per-visit monitoring cost rises and the budget forces still longer intervals: $\Delta_{\mathrm{visit}}=9$ months for intermediate active reference measurement counts and $\Delta_{\mathrm{visit}}=12$ months for the largest values of $n_{\mathrm{zone}}$. The overall best policy under MB-25 remains at $n_{\mathrm{zone}}=20$, with $p_{\mathrm{th}}=0.30$, $\Delta_{\mathrm{visit}}=6$ months, and an estimated total downtime and replacement cost of about 491.13, compared with the MB-50 optimum of about 477.55 at $n_{\mathrm{zone}}=20$ and $p_{\mathrm{th}}=0.12$. Thus, when the budget is halved relative to the base setting, the selected optimum keeps the same active reference measurement count but uses a longer visit interval. The resulting increase in the estimated total downtime and replacement cost reflects the reduced ability to update the state estimate and trigger preventive replacements at appropriate times under less frequent monitoring.

Under the full reference budget, the budget constraint no longer forces a tradeoff between increasing the active reference measurement count and maintaining quarterly visits for any tested value of $n_{\mathrm{zone}}$. As shown in Fig.~\ref{fig:global_best_across_nz_B25200}, the best retained policy uses $\Delta_{\mathrm{visit}}=3$ months for every value of $n_{\mathrm{zone}}$. The estimated total downtime and replacement cost decreases as the active reference measurement count increases, from about 483.00 at $n_{\mathrm{zone}}=4$ to the minimum of about 472.97 at $n_{\mathrm{zone}}=76$. Accordingly, the benefit of additional spatial measurement information is no longer offset by a reduction in temporal update frequency. The selected risk thresholds lie between $p_{\mathrm{th}}=0.12$ and 0.17, while the selected first visit time remains irregular because it primarily acts as a phase parameter of the periodic schedule. The lowest estimated mean under MB-100 occurs at the full reference measurement count $n_{\mathrm{zone}}=76$, with $p_{\mathrm{th}}=0.13$ and $\Delta_{\mathrm{visit}}=3$ months. Compared with the MB-50 optimum at $n_{\mathrm{zone}}=20$, the MB-100 results therefore favor using the complete reference measurement layout when quarterly monitoring can be maintained.

Overall, the budget sensitivity results support the interpretation from Fig.~\ref{fig:global_best_across_nz_B12600}. Under a fixed monitoring budget, the best retained policies first preserve sufficiently frequent visits and then use the remaining budget to collect additional active reference measurement points. When the budget is tight, increasing $n_{\mathrm{zone}}$ beyond the range that permits the shortest feasible visit interval forces longer intervals, and the additional spatial information does not compensate for the loss of temporal update frequency. Under the base budget, this tradeoff produces an interior optimum at $n_{\mathrm{zone}}=20$. When the budget is sufficiently large to maintain quarterly visits for all active reference measurement counts, additional measurement information becomes beneficial across the grid and the optimum moves to the full reference measurement count at $n_{\mathrm{zone}}=76$. Thus, the qualitative allocation logic of the proposed budgeted monitoring framework remains consistent across budget levels in the baseline uncertainty setting: temporal update frequency is prioritized when it is budget-limited, and spatial measurement information is exploited once the visit frequency is not compromised.

\section{Conclusion}
\label{sec:conclusion}

This paper proposed a two-stage monitoring design and budgeted condition-based maintenance (CBM) framework for LED lighting systems. At the design stage, a Radiance-based linear-Gaussian observation model and an identifiability-constrained D-optimal design were used to select an informative reference layout for inverse estimation of latent luminaire degradation states. At the runtime stage, an ensemble Kalman filter assimilated sparse measurements from this layout together with nonhomogeneous gamma process predictions, and posterior predictive failure probabilities were used to guide preventive replacement decisions. Under a monitoring budget, the evaluated periodic policy jointly selected visit timing, the active reference measurement count per visit, and the risk threshold, thereby capturing the tradeoff among temporal monitoring frequency, spatial measurement information, and replacement timing.

The office-zone case study demonstrated the proposed workflow from reference layout selection to runtime monitoring and CBM policy optimization. Under the base monitoring budget, the best retained policy used $n_{\mathrm{zone}}=20$, $p_{\mathrm{th}}=0.12$, and $\Delta_{\mathrm{visit}}=3$ months, indicating that frequent state updates were prioritized before allocating additional budget to active reference measurements. Compared with the block-replacement benchmark, the best retained budgeted CBM policy reduced the estimated total downtime and replacement cost by 10.2\% under baseline hitting time uncertainty and by 40.5\% under enlarged hitting time uncertainty. The sensitivity analyses further showed that higher hitting time uncertainty increased the value of condition information, favoring lower risk thresholds and larger active reference measurement counts. The budget analysis showed that visit frequency is prioritized under restrictive monitoring budgets, whereas the complete reference measurement layout becomes beneficial once quarterly visits can be maintained.

Future work should validate the framework using field measurements from operating lighting systems to calibrate the observation model, degradation process, and cost parameters under real conditions. Further extensions may consider adaptive or event-triggered monitoring, visit-specific active measurement selection, and multi-zone or building-scale maintenance planning with shared resources and heterogeneous operating environments.

\section*{Acknowledgments}
This research was funded by the Australian Research Council (ARC) through the ARC Research Hub for Resilient and Intelligent Infrastructure Systems (RIIS) (IH210100048). The authors also appreciate the collaboration and support from their industry partner, Fredon, and Queensland University of Technology. We acknowledge support from the Queensland University of Technology (QUT) through its provision of subsidised access to specialised research infrastructure, including high-performance computing (HPC) and the research virtual desktop infrastructure (rVDI), and the eResearch team at QUT for providing expertise in the enablement of this project.

\section*{Declaration of competing interest}
The authors declare that they have no known competing financial interests or personal relationships that could have appeared to influence the work reported in this paper.

\section*{Data availability}
The data supporting the findings of this study are available from the corresponding author upon reasonable request. The building geometry and industry-related information are subject to access restrictions from the collaborating partner.

\printcredits

\bibliographystyle{model1-num-names}

\bibliography{references}

\appendix
\numberwithin{equation}{section}
\numberwithin{figure}{section}
\numberwithin{table}{section}

\section{Appendix}
\subsection{Details of surrogate model parameter estimation}
\label{app:surrogate_est_details}

A Sobol sequence is used to generate quasi-random samples over the luminaire degradation state space, with each component in $[0,1]$, providing more space-filling coverage than standard pseudo-random sampling \citep{sobol_construction_2011}.

Let $\mathcal{D}_{\mathrm{train}}=\{(\mathbf{x}_i,\mathbf{E}_i)\}_{i=1}^{N_s}$ denote the training dataset generated by Radiance, where $\mathbf{x}_i\in\mathbb{R}^{J}$ is a sampled state vector and $\mathbf{E}_i\in\mathbb{R}^{N}$ is the corresponding WP illuminance field vector. Stacking samples yields
\begin{equation}
    \mathbf{Y} =
    \begin{bmatrix}
    \mathbf{E}_1^{\top}\\
    \vdots\\
    \mathbf{E}_{N_s}^{\top}
    \end{bmatrix}
    \in\mathbb{R}^{N_s\times N},
    \qquad
    \mathbf{X} =
    \begin{bmatrix}
    \mathbf{x}_1^{\top}\\
    \vdots\\
    \mathbf{x}_{N_s}^{\top}
    \end{bmatrix}
    \in\mathbb{R}^{N_s\times J}.
\end{equation}

Using an augmented design matrix $\tilde{\mathbf{X}}=[\mathbf{X}\ \ \mathbf{1}_{N_s}] \in \mathbb{R}^{N_s\times(J+1)}$ and an augmented parameter matrix $\tilde{\mathbf{C}}=[\mathbf{C}\ \ \mathbf{b}]\in\mathbb{R}^{N\times(J+1)}$, the surrogate model in Eq.~\eqref{eq:sr1} can be written compactly as
\begin{equation}
    \mathbf{Y} = \tilde{\mathbf{X}}\,\tilde{\mathbf{C}}^{\top} + \mathbf{E}_{\mathrm{res}},
    \label{eq:sr_mat}
\end{equation}
where $\mathbf{E}_{\mathrm{res}}\in\mathbb{R}^{N_s\times N}$ denotes residuals. OLS is used to estimate $\tilde{\mathbf{C}}$:
\begin{equation}
    \hat{\tilde{\mathbf{C}}}^{\top}
    = \left(\tilde{\mathbf{X}}^{\top}\tilde{\mathbf{X}}\right)^{-1}\tilde{\mathbf{X}}^{\top}\mathbf{Y}.
    \label{eq:ols_solution}
\end{equation}

The residual covariance is estimated by the sample covariance of OLS residuals:
\begin{equation}
    \hat{\boldsymbol{\Sigma}}
    = \frac{1}{N_s-(J+1)}\,\mathbf{E}_{\mathrm{res}}^{\top}\mathbf{E}_{\mathrm{res}},
    \qquad
    \mathbf{E}_{\mathrm{res}}:=\mathbf{Y}-\tilde{\mathbf{X}}\hat{\tilde{\mathbf{C}}}^{\top}.
    \label{eq:sigma_hat}
\end{equation}

\subsection{Hybrid greedy procedure for design-stage WP measurement point selection}
\label{app:hg_sensor_placement}

\begin{algorithm}[H]
\caption{Hybrid greedy search for WP measurement point selection with a fixed number of selected points}
\label{alg:hg_deployment}
\scriptsize
\setstretch{0.75}
\begin{algorithmic}[1]
\Require Candidate WP grid point set $\{1,\dots,N\}$, target number of selected points $m=J$, greedy parameter $\alpha$, number of greedy constructions $N_{\mathrm{grasp}}$, number of retained greedy seeds $N_{\mathrm{seed}}$, beam width $B$, number of polished candidates $N_{\mathrm{polish}}$, maximum local-search rounds $N_{\mathrm{swap}}$ ($N_{\mathrm{swap}}=50$ in this study)
\Ensure Hybrid greedy search layout $\mathcal{S}_{\mathrm{HG}}$

\State For any nonempty set $\mathcal{S}$, let $\mathbf{H}_{\mathcal{S}}=\mathbf{S}_{\mathcal{S}}\mathbf{C}$ and $\boldsymbol{\Omega}_{\mathcal{S}}=\mathbf{S}_{\mathcal{S}}\boldsymbol{\Sigma}\mathbf{S}_{\mathcal{S}}^{\top}+\sigma_{\mathrm{read}}^2\mathbf{I}_{|\mathcal{S}|}$
\State Compute $\boldsymbol{\Omega}_{\mathcal{S}}=\mathbf{L}_{\mathcal{S}}\mathbf{L}_{\mathcal{S}}^{\top}$ and $\mathbf{W}_{\mathcal{S}}=\mathbf{L}_{\mathcal{S}}^{-1}\mathbf{H}_{\mathcal{S}}$
\State Define the construction score
\Statex \hspace{1.5em}$\displaystyle g(\mathcal{S})=\begin{cases}
\log\det\!\left(\mathbf{W}_{\mathcal{S}}\mathbf{W}_{\mathcal{S}}^{\top}\right), & 1\le |\mathcal{S}|<m,\\
\log\det\!\left(\mathbf{W}_{\mathcal{S}}^{\top}\mathbf{W}_{\mathcal{S}}\right), & |\mathcal{S}|=m,
\end{cases}$
\Statex \hspace{1.5em}where the complete-layout score at $|\mathcal{S}|=m=J$ equals the D-optimal objective in Eq.~\eqref{eq:dopt_obj_set}

\State $\mathcal{P}_{\mathrm{grasp}} \gets \emptyset$
\For{$a = 1,\dots,N_{\mathrm{grasp}}$}
    \State $\mathcal{S} \gets \emptyset$
    \While{$|\mathcal{S}| < m$}
        \ForAll{$j \in \{1,\dots,N\}\setminus\mathcal{S}$}
            \State compute $q_j\gets g(\mathcal{S}\cup\{j\})$
        \EndFor
        \State discard candidates with non-finite scores and compute $q_{\min}\gets\min_j q_j$ and $q_{\max}\gets\max_j q_j$
        \State $q_{\mathrm{RCL}}\gets q_{\max}-\alpha(q_{\max}-q_{\min})$ and $\mathcal{R}\gets\{j:q_j\ge q_{\mathrm{RCL}}\}$
        \State randomly select one candidate from $\mathcal{R}$ and add it to $\mathcal{S}$
    \EndWhile
    \State add $\mathcal{S}$ to $\mathcal{P}_{\mathrm{grasp}}$
\EndFor
\State deduplicate $\mathcal{P}_{\mathrm{grasp}}$
\State retain the $N_{\mathrm{seed}}$ greedy solutions with the largest complete-layout scores in $\mathcal{P}_{\mathrm{grasp}}$

\State initialize beam $\mathcal{B}_0 \gets \{\emptyset\}$
\For{$t=1,\dots,m$}
    \State $\mathcal{C}_t \gets \emptyset$
    \ForAll{$\mathcal{S}\in\mathcal{B}_{t-1}$}
        \ForAll{$j\in\{1,\dots,N\}\setminus\mathcal{S}$}
            \State add candidate $(\mathcal{S}\cup\{j\},g(\mathcal{S}\cup\{j\}))$ to $\mathcal{C}_t$
        \EndFor
    \EndFor
    \State deduplicate candidates in $\mathcal{C}_t$ and discard candidates with non-finite scores
    \State retain the $B$ candidates with the largest scores as $\mathcal{B}_t$
\EndFor
\State let $\mathcal{P}_{\mathrm{beam}} \gets \mathcal{B}_m$

\State form pooled candidate set $\mathcal{P} \gets \mathcal{P}_{\mathrm{grasp}} \cup \mathcal{P}_{\mathrm{beam}}$
\State deduplicate $\mathcal{P}$
\State rank candidates in $\mathcal{P}$ in descending order of their complete-layout scores
\State retain the best $N_{\mathrm{polish}}$ candidates as $\mathcal{P}_{\mathrm{polish}}$

\ForAll{$\mathcal{S} \in \mathcal{P}_{\mathrm{polish}}$}
    \State $r\gets 0$, \texttt{improved}$\gets\texttt{true}$
    \While{\texttt{improved} and $r<N_{\mathrm{swap}}$}
        \State \texttt{improved}$\gets\texttt{false}$, $r\gets r+1$
        \ForAll{$i\in\mathcal{S}$ in randomized order}
            \ForAll{$j\in\{1,\dots,N\}\setminus\mathcal{S}$ in randomized order}
                \State $\mathcal{S}'\gets(\mathcal{S}\setminus\{i\})\cup\{j\}$
                \If{$g(\mathcal{S}')>g(\mathcal{S})$}
                    \State $\mathcal{S}\gets\mathcal{S}'$, \texttt{improved}$\gets\texttt{true}$; accept the first improvement and restart the \textbf{while} loop
                \EndIf
            \EndFor
        \EndFor
    \EndWhile
\EndFor

\State \Return the polished WP grid point layout with the largest complete-layout score as $\mathcal{S}_{\mathrm{HG}}$
\end{algorithmic}
\end{algorithm}

\subsection{Genetic algorithm for design-stage WP measurement point selection}
\label{app:ga_sensor_placement}

\begin{algorithm}[H]
\caption{Genetic algorithm for WP measurement point selection optimization with a fixed number of selected points}
\label{alg:ga_deployment}
\scriptsize
\setstretch{0.75}
\begin{algorithmic}[1]
\Require Candidate WP grid point set $\{1,\dots,N\}$, target number of selected points $m=J$, population size $P$, maximum generations $G_{\max}$, tournament size $K_{\mathrm{tour}}$, elite count $E$, crossover probability $p_c$, mutation probability $p_m$, mutation-swap range $[r_{\mathrm{mut}}^{\min},r_{\mathrm{mut}}^{\max}]$
\Require Initialization mode (random or nadir-seeded), Nadir-point layout $\mathcal{S}_{\mathrm{nadir}}$ when required, nadir-seeded fraction $\rho_{\mathrm{nadir}}$, seed-swap range $[r_{\mathrm{seed}}^{\min},r_{\mathrm{seed}}^{\max}]$, diversity threshold $d_{\min}$, restart fraction $\rho_{\mathrm{restart}}$, saturation limit $G_{\mathrm{sat}}$
\Ensure GA layout $\mathcal{S}_{\mathrm{GA}}$

\State Encode each candidate WP measurement point layout as a binary vector $\mathbf{z}\in\{0,1\}^{N}$ satisfying $\mathbf{1}^{\top}\mathbf{z}=m$

\If{random initialization is used}
    \For{$i=1,\dots,P$}
        \State Generate one random feasible individual with exactly $m$ selected WP grid points
        \State Add it to the initial population $\mathcal{X}^{(0)}$
    \EndFor
\Else
    \State $P_{\mathrm{nadir}} \gets \lfloor \rho_{\mathrm{nadir}}P \rceil$, \quad $P_{\mathrm{rand}} \gets P-P_{\mathrm{nadir}}$
    \For{$i=1,\dots,P_{\mathrm{nadir}}$}
        \State Independently initialize $\mathcal{S}\gets\mathcal{S}_{\mathrm{nadir}}$ and draw $r\in\{r_{\mathrm{seed}}^{\min},\dots,r_{\mathrm{seed}}^{\max}\}$
        \State Remove $r$ points sampled without replacement from $\mathcal{S}_{\mathrm{nadir}}$ and add $r$ points sampled without replacement from its complement
        \State Convert the perturbed set $\mathcal{S}$ into a binary vector and add it to $\mathcal{X}^{(0)}$
    \EndFor
    \For{$i=1,\dots,P_{\mathrm{rand}}$}
        \State Generate one random feasible individual with exactly $m$ selected WP grid points
        \State Add it to $\mathcal{X}^{(0)}$
    \EndFor
\EndIf

\ForAll{individuals $\mathbf{z}\in\mathcal{X}^{(0)}$}
    \State Decode $\mathcal{S}=\{i:z_i=1\}$
    \State Evaluate fitness $f(\mathcal{S})$ using Eq.~\eqref{eq:dopt_obj_set}
\EndFor
\State Set $\mathbf{z}_{\mathrm{best}}$ to the highest-fitness initial individual, $g\gets0$, and $n_{\mathrm{stall}}\gets0$

\While{$g<G_{\max}\ \text{and}\ n_{\mathrm{stall}}<G_{\mathrm{sat}}$}
    \State $g\gets g+1$
    \State Select $P/2$ mating parents by tournament selection with tournament size $K_{\mathrm{tour}}$
    \State Generate offspring using subset crossover with probability $p_c$, preserving $\mathbf{1}^{\top}\mathbf{z}=m$
    \State With probability $p_m$, mutate an offspring using $r\in\{r_{\mathrm{mut}}^{\min},\dots,r_{\mathrm{mut}}^{\max}\}$ selected--unselected swaps
    \State Form $\mathcal{X}^{(g)}$ from the offspring while retaining the best $E$ elite individuals, and evaluate all required fitness values
    \State Compute the unique-individual ratio $d_g=|\operatorname{unique}(\mathcal{X}^{(g)})|/P$
    \If{$d_g<d_{\min}$}
        \State Protect the $E$ elite individuals and replace the worst $\lfloor\rho_{\mathrm{restart}}P\rfloor$ non-elite individuals with random feasible individuals
        \State Evaluate the replacement individuals
    \EndIf
    \If{the best fitness in $\mathcal{X}^{(g)}$ exceeds the global best fitness}
        \State Update $\mathbf{z}_{\mathrm{best}}$ and set $n_{\mathrm{stall}}\gets0$
    \Else
        \State $n_{\mathrm{stall}}\gets n_{\mathrm{stall}}+1$
    \EndIf
\EndWhile

\State Decode $\mathcal{S}_{\mathrm{GA}}=\{i:z_{\mathrm{best},i}=1\}$
\State \Return $\mathcal{S}_{\mathrm{GA}}$
\end{algorithmic}
\end{algorithm}

\subsection{Sequential grid search with paired screening for runtime-stage measurement design and CBM policy optimization}
\label{app:seq_screening_runtime}

\begin{algorithm}[H]
\caption{Sequential grid search with paired screening (SGS-PS) for periodic monitoring policies}
\label{alg:seq_grid_runtime}
\scriptsize
\setstretch{0.75}
\begin{algorithmic}[1]
\Require first visit grid $\mathcal{T}_{\mathrm{first}}$, periodic visit interval grid $\mathcal{D}$, fixed $(p_{\mathrm{th}},n_{\mathrm{zone}})$, Monte Carlo (MC) replication schedule $\{N_{\mathrm{MC},s}\}_{s=1}^{S}$, maximum replication limit $N_{\mathrm{MC},\max}$, screening significance level $\alpha_{\mathrm{sv}}$, incumbent patience $N_{\mathrm{patience}}$
\Ensure retained set $\Theta_{\mathrm{final}}$, final replication count $N_{\mathrm{final}}$, stop reason

\State For fixed $(p_{\mathrm{th}},n_{\mathrm{zone}})$, construct candidate set $\Theta_0=\{(t_{\mathrm{first}},\Delta_{\mathrm{visit}}): t_{\mathrm{first}}\in\mathcal{T}_{\mathrm{first}},\Delta_{\mathrm{visit}}\in\mathcal{D}\}$
\State Build periodic plan $\mathbf{Z}(\theta)$ for each $\theta\in\Theta_0$
\State Remove candidates whose periodic plans violate the monitoring budget
\State Initialize cache $\mathcal{L}(\theta)\leftarrow[\,]$, \texttt{prevIncumbent}$\leftarrow\varnothing$, \texttt{stagnantStages}$\leftarrow 0$

\For{$s=1,2,\dots,S$}
    \ForAll{$\theta\in\Theta_{s-1}$}
        \State $i_{\mathrm{start}}\leftarrow |\mathcal{L}(\theta)|$
        \If{$i_{\mathrm{start}}<N_{\mathrm{MC},s}$}
            \State Simulate additional replications $i=i_{\mathrm{start}}+1,\dots,N_{\mathrm{MC},s}$ and append outcomes to $\mathcal{L}(\theta)$
        \EndIf
        \State Compute stage mean $\hat\mu_s(\theta)=\frac{1}{N_{\mathrm{MC},s}}\sum_{i=1}^{N_{\mathrm{MC},s}}C_{\mathrm{DR}}^{(i)}\bigl(p_{\mathrm{th}},\mathbf{Z}(\theta)\bigr)$
    \EndFor

    \State Set incumbent $\theta_s^{\star}\in\arg\min_{\theta\in\Theta_{s-1}}\hat\mu_s(\theta)$
    \If{$\theta_s^{\star}=\texttt{prevIncumbent}$}
        \State \texttt{stagnantStages}$\leftarrow$\texttt{stagnantStages}$+1$
    \Else
        \State \texttt{stagnantStages}$\leftarrow 0$
    \EndIf
    \State \texttt{prevIncumbent}$\leftarrow\theta_s^{\star}$

    \State Initialize survivor set $\Theta_s\leftarrow\{\theta_s^{\star}\}$
    \ForAll{$\theta\in\Theta_{s-1}$, $\theta\neq\theta_s^{\star}$}
        \State Form paired differences $D_i(\theta)=C_{\mathrm{DR}}^{(i)}\bigl(p_{\mathrm{th}},\mathbf{Z}(\theta)\bigr)-C_{\mathrm{DR}}^{(i)}\bigl(p_{\mathrm{th}},\mathbf{Z}(\theta_s^{\star})\bigr)$, $i=1,\dots,N_{\mathrm{MC},s}$
        \State Compute $\bar D_s(\theta)$, $SE_s(\theta)$, and $t_s(\theta)=\bar D_s(\theta)/SE_s(\theta)$
        \State Let $t_{\mathrm{crit}}=t_{1-\alpha_{\mathrm{sv}},\,N_{\mathrm{MC},s}-1}$
        \If{$t_s(\theta)\le t_{\mathrm{crit}}$}
            \State Retain $\theta$: $\Theta_s\leftarrow\Theta_s\cup\{\theta\}$
        \Else
            \State Eliminate $\theta$ as statistically inferior to the incumbent
        \EndIf
    \EndFor

    \If{$|\Theta_s|=1$}
        \State stop reason $\leftarrow$ single survivor; \textbf{break}
    \ElsIf{$N_{\mathrm{MC},s}\ge N_{\mathrm{MC},\max}$}
        \State stop reason $\leftarrow$ $N_{\mathrm{MC},\max}$ reached; \textbf{break}
    \ElsIf{\texttt{stagnantStages}$\ge N_{\mathrm{patience}}$}
        \State stop reason $\leftarrow$ incumbent stable; \textbf{break}
    \EndIf
\EndFor

\State $\Theta_{\mathrm{final}}\leftarrow\Theta_s$, \quad $N_{\mathrm{final}}\leftarrow N_{\mathrm{MC},s}$
\State \Return $\Theta_{\mathrm{final}}$, $N_{\mathrm{final}}$, stop reason
\end{algorithmic}
\end{algorithm}

\subsection{EnKF state-tracking diagnostics}
\label{app:enkf_tracking}

\begin{figure}[pos=!htbp]
    \centering
    \begin{subfigure}[b]{0.49\textwidth}
        \centering
        \includegraphics[width=\textwidth]{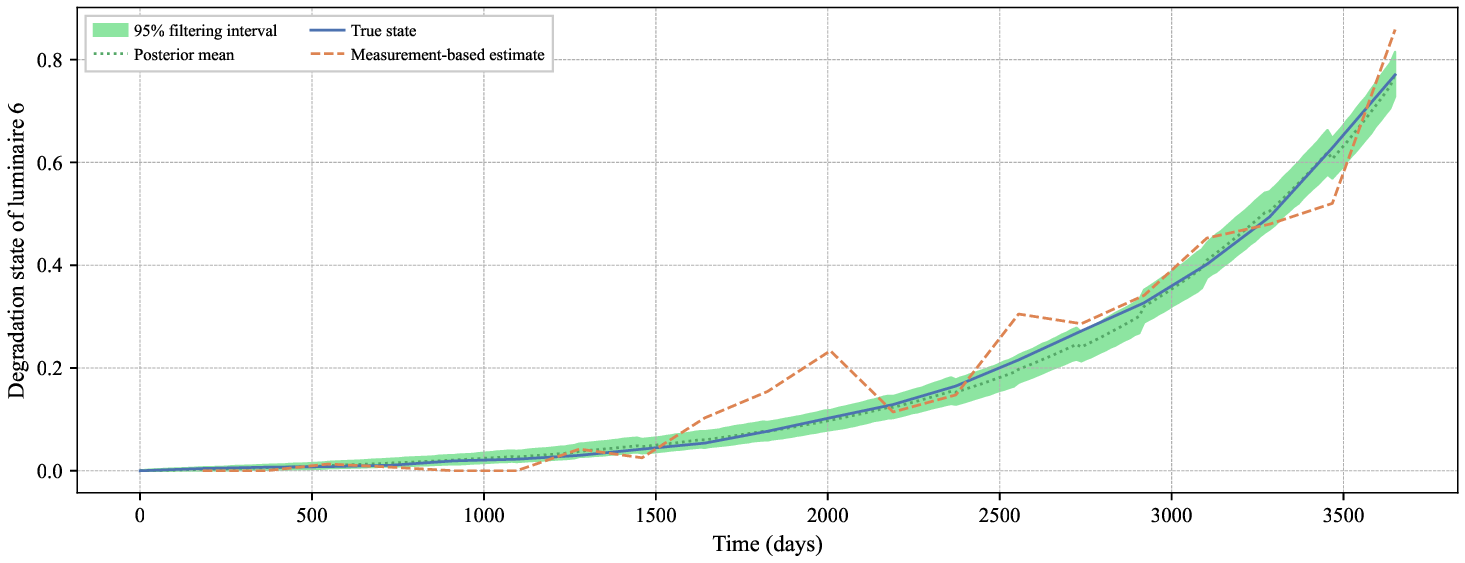}
        \caption{Luminaire 6}
        \label{fig:app_enkf_tracking_led06}
    \end{subfigure}
    \hfill
    \begin{subfigure}[b]{0.49\textwidth}
        \centering
        \includegraphics[width=\textwidth]{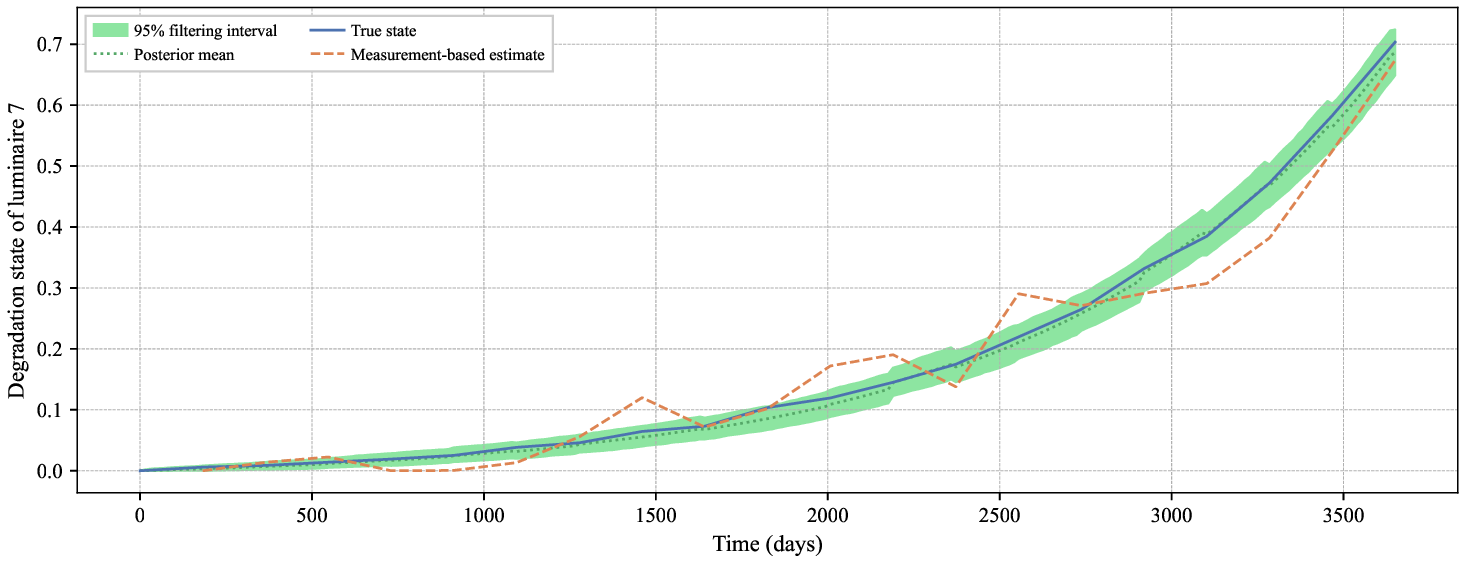}
        \caption{Luminaire 7}
        \label{fig:app_enkf_tracking_led07}
    \end{subfigure}

    \begin{subfigure}[b]{0.49\textwidth}
        \centering
        \includegraphics[width=\textwidth]{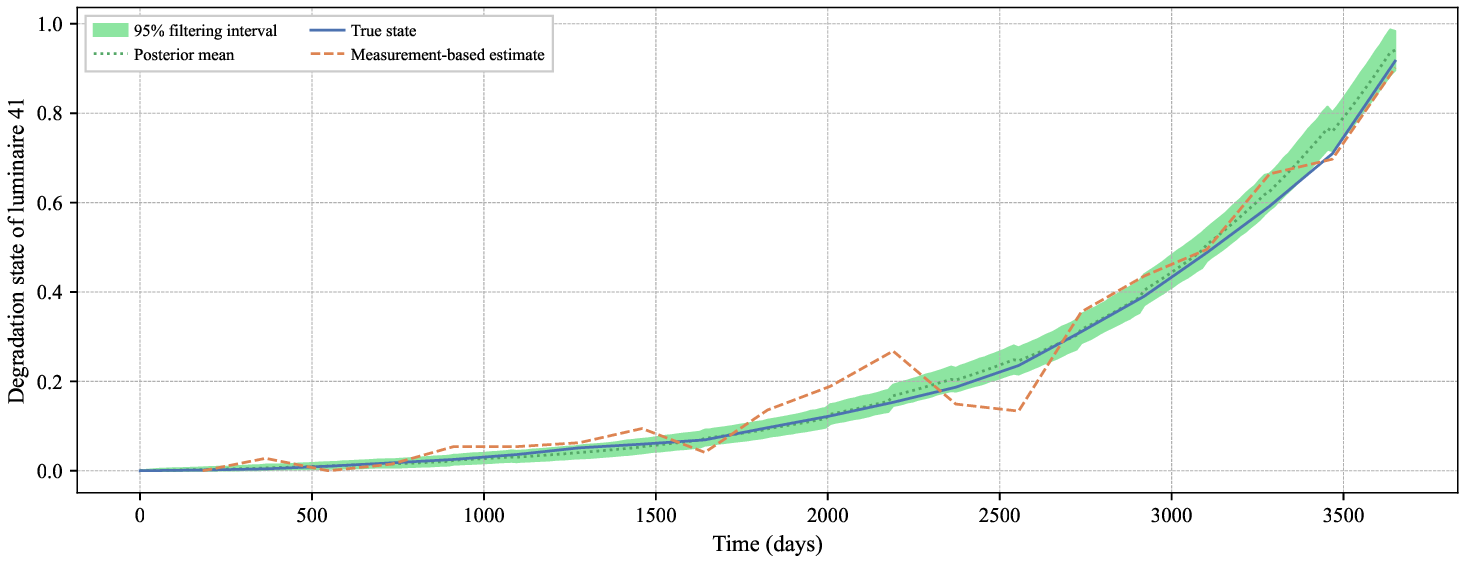}
        \caption{Luminaire 41}
        \label{fig:app_enkf_tracking_led41}
    \end{subfigure}
    \hfill
    \begin{subfigure}[b]{0.49\textwidth}
        \centering
        \includegraphics[width=\textwidth]{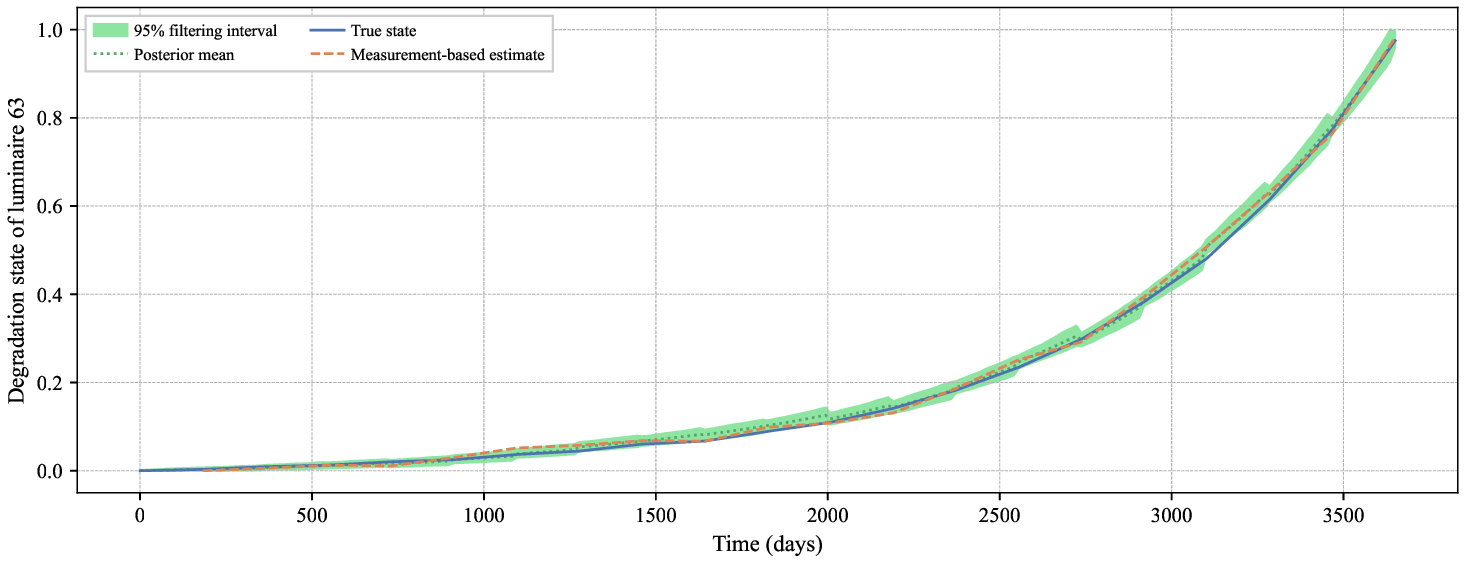}
        \caption{Luminaire 63}
        \label{fig:app_enkf_tracking_led63}
    \end{subfigure}
    \caption{EnKF state-tracking diagnostics for four randomly selected luminaires under an illustrative 10-year simulation scenario. The blue solid lines denote the true degradation states. The green dashed lines and shaded regions denote the EnKF ensemble means and pointwise 95\% ensemble intervals, respectively. The yellow lines denote the corresponding observation-only estimates obtained without either degradation-model propagation or sequential filtering.}
    \label{fig:app_enkf_tracking}
\end{figure}

Figure~\ref{fig:app_enkf_tracking} compares the EnKF and observation-only degradation-state estimates with the true trajectories for four randomly selected luminaires under an illustrative simulation scenario. The diagnostic illustrates state-tracking behavior independently of the subsequent CBM decisions.

\subsection{Cost-component diagnostics for active reference measurement count sensitivity}
\label{app:nzone_component_diagnostics}

\begin{figure}[pos=!htbp]
    \centering
    \begin{subfigure}[b]{0.49\textwidth}
        \centering
        \includegraphics[width=\textwidth,height=4.6cm,keepaspectratio]{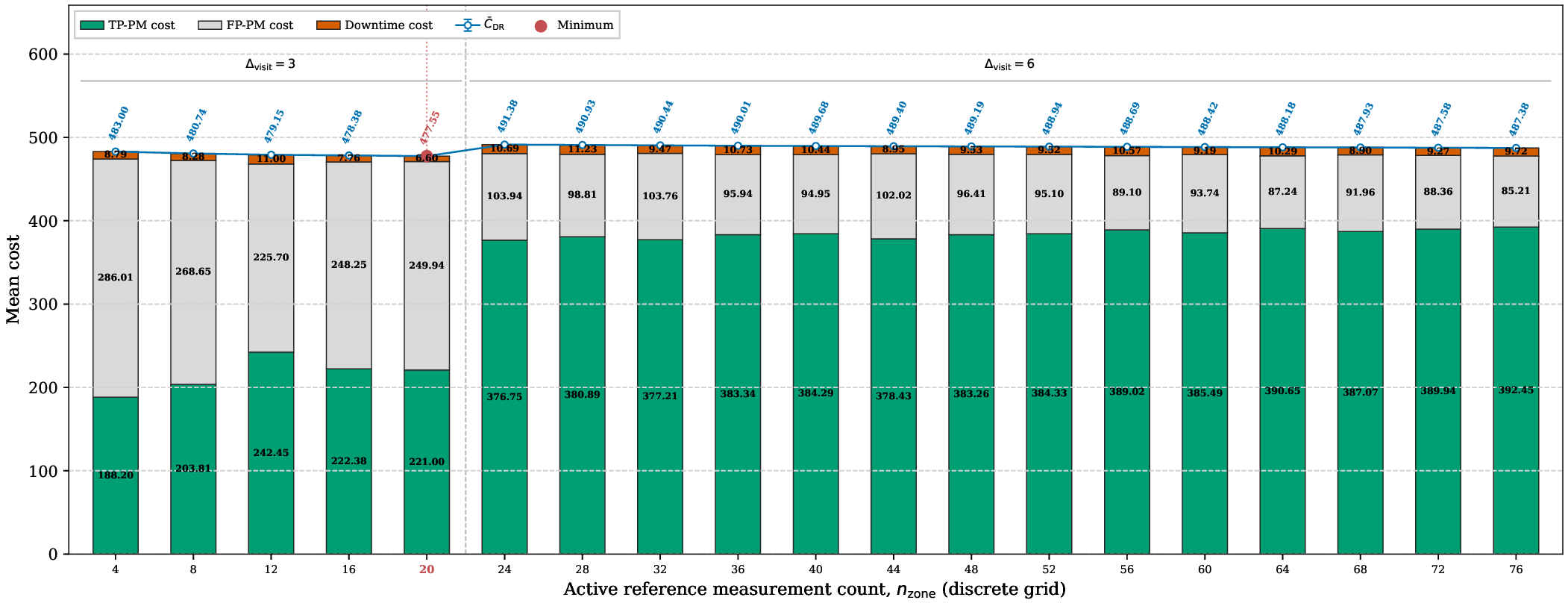}
        \caption{Baseline hitting-time uncertainty}
        \label{fig:app_global_best_components_B12600}
    \end{subfigure}
    \hfill
    \begin{subfigure}[b]{0.49\textwidth}
        \centering
        \includegraphics[width=\textwidth,height=4.6cm,keepaspectratio]{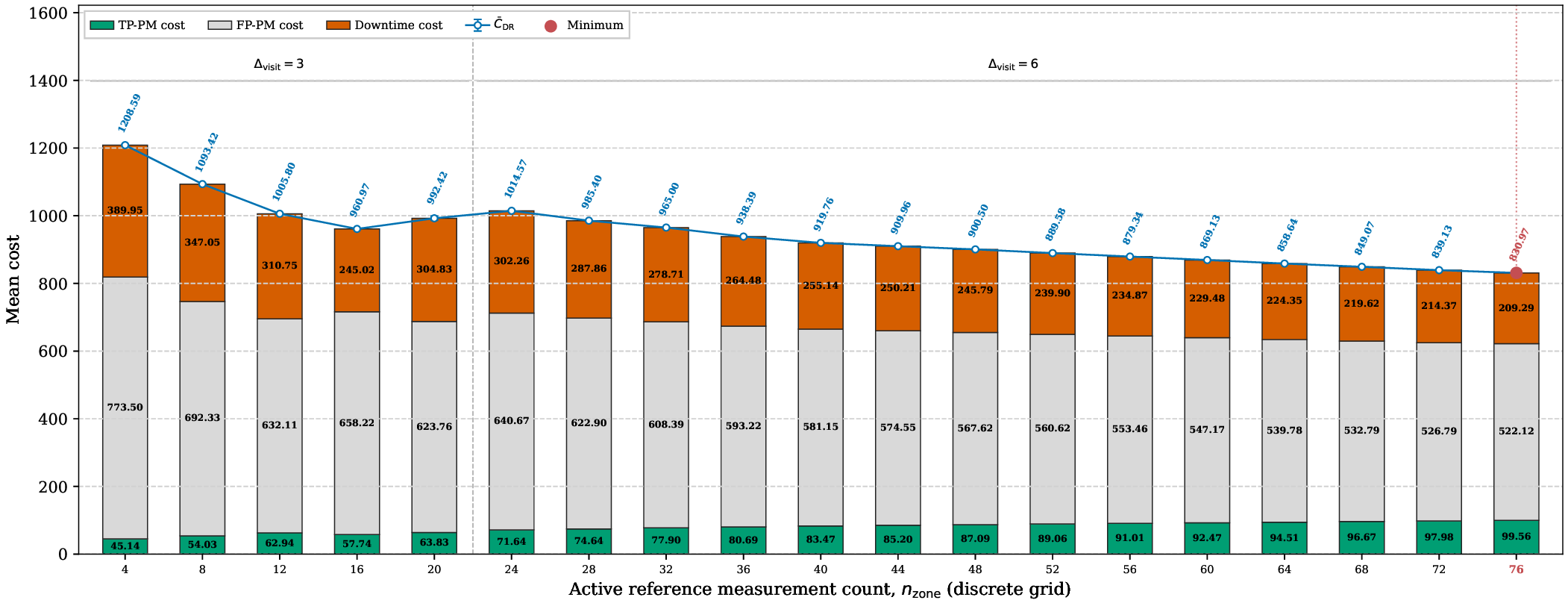}
        \caption{Enlarged hitting-time uncertainty}
        \label{fig:app_global_best_components_B12600_c150}
    \end{subfigure}
    \caption{Diagnostic downtime and replacement cost components of the best retained policy for each active reference measurement count under the baseline and enlarged hitting-time uncertainty settings and the base monitoring budget MB-50.}
    \label{fig:app_global_best_components_B12600_uncertainty_compare}
\end{figure}

Figure~\ref{fig:app_global_best_components_B12600_uncertainty_compare} decomposes the downtime and preventive replacement costs of the best retained policy at each active reference measurement count under the baseline and enlarged hitting-time uncertainty settings.

\subsection{Block-replacement benchmark diagnostics}
\label{app:block_replacement_diagnostics}

\begin{figure}[pos=!htbp]
    \centering
    \begin{subfigure}{0.49\textwidth}
        \centering
        \includegraphics[width=\textwidth]{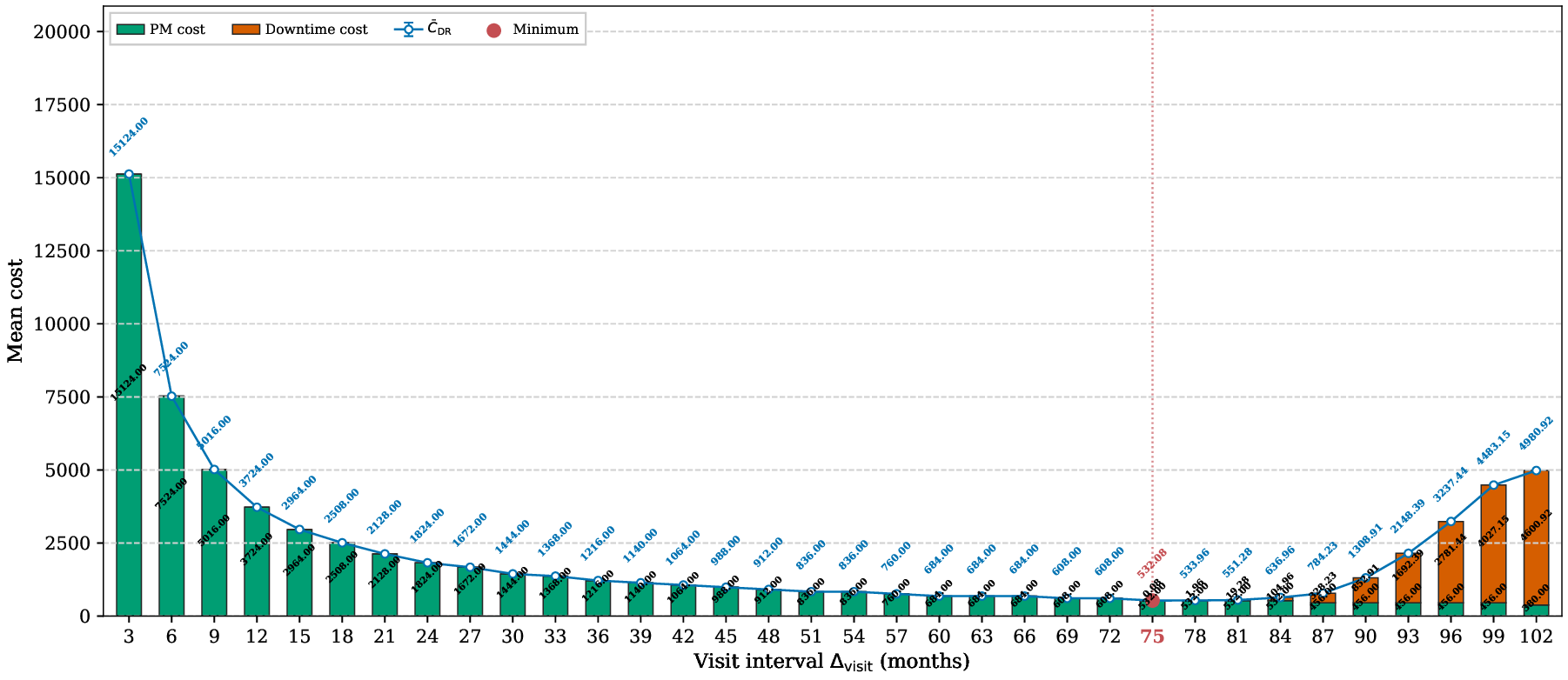}
        \caption{Baseline hitting-time uncertainty}
        \label{fig:app_block_replacement_varc1}
    \end{subfigure}
    \hfill
    \begin{subfigure}{0.49\textwidth}
        \centering
        \includegraphics[width=\textwidth]{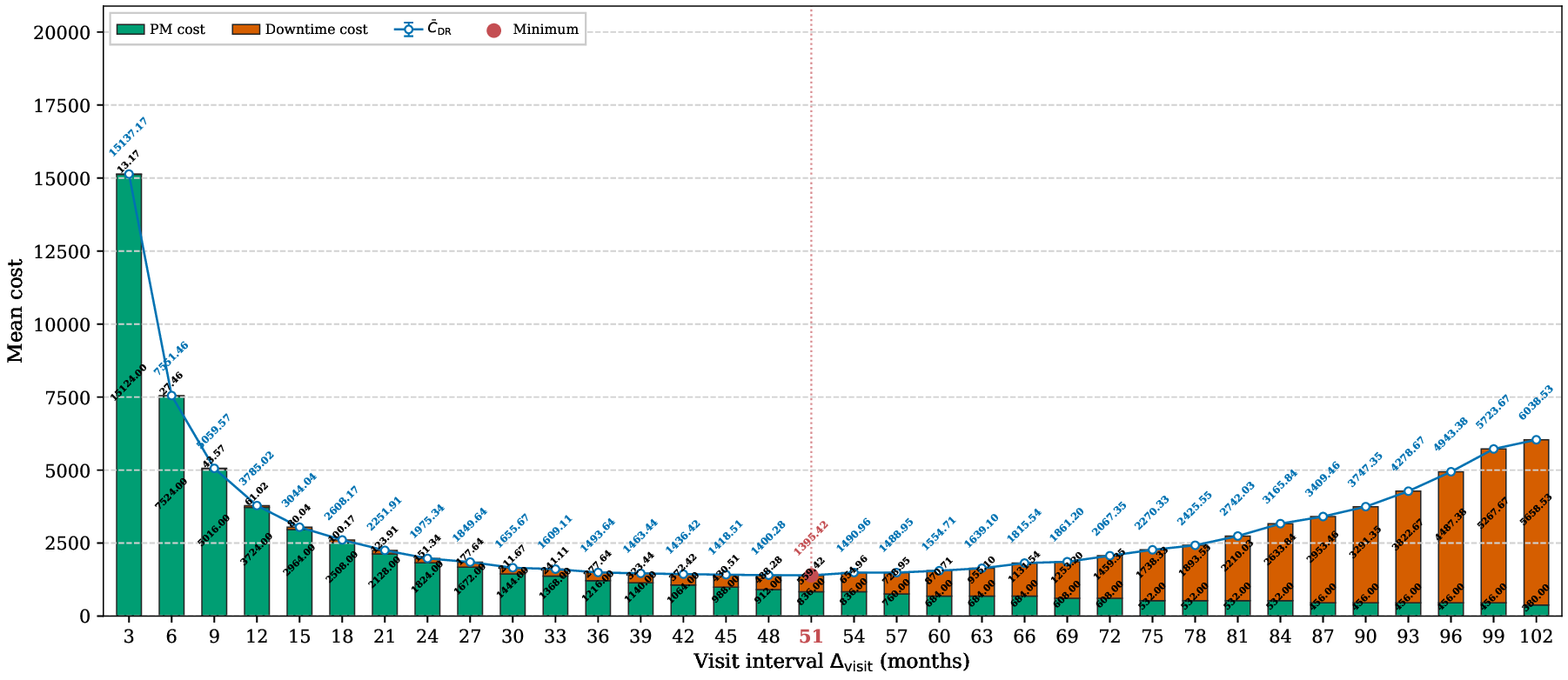}
        \caption{Enlarged hitting-time uncertainty}
        \label{fig:app_block_replacement_varc150}
    \end{subfigure}
    \caption{Diagnostic downtime and replacement cost components of the block-replacement benchmark under the two hitting-time uncertainty settings. The benchmark uses the same periodic timing grid but does not assimilate WP measurements or apply risk-threshold-based CBM decisions.}
    \label{fig:app_block_replacement_components}
\end{figure}

Figure~\ref{fig:app_block_replacement_components} decomposes the corresponding costs of the block-replacement benchmark across the periodic interval grid under the same two uncertainty settings.

\clearpage


\end{document}